\documentclass{CUP-JNL-AAS}

\usepackage{amsmath, nccmath}
\usepackage{graphicx}
\usepackage{float}
\usepackage{subcaption}
\usepackage{dsfont}
\usepackage{bm}
\usepackage{listings}
\usepackage{tabularx}
\newcolumntype{C}{>{\centering\arraybackslash}X}
\usepackage{url, hyperref}

\newcommand{\tabref}[1]{Table~\ref{#1}}
\newcommand{\figref}[1]{Figure~\ref{#1}}

\begin{document}

\lefttitle{Cancer mortality projection}
\righttitle{A. Ar{\i}k, A.J.G. Cairns and G. Streftaris}

\papertitle{Article}

%\jnlPage{1}{8}
%\jnlDoiYr{2020}
%\doival{10.1017/xxxxx}

\title{Cancer mortality projection: disparities, COVID-19, and late diagnosis impact}

\begin{authgrp}
\author{A. Ar{\i}k, A.J.G. Cairns, G. Streftaris}
\affiliation{School of Mathematical and Computer Sciences, Heriot-Watt University, and Maxwell Institute for Mathematical
Sciences, Edinburgh EH14 4AS, UK\\
$^{*}$Corresponding author: a.arik@hw.ac.uk}
\end{authgrp}

\begin{abstract}

 This paper investigates projection of two major causes of cancer mortality, breast cancer and lung cancer, by using a Bayesian modelling framework. We investigate patterns in 2001–2018 (as baseline) in cause-specific cancer mortality and project these by year of death and various risk factors: age, gender, regions of England, income deprivation quintile, average age-at-diagnosis, and non-smoker prevalence rates. We then assess excess cancer mortality during the COVID-19 pandemic years, and we examine the impact of diagnosis delays on lung cancer mortality across various scenarios. Our findings indicate that socio-economic disparities in lung cancer mortality will persist in the future. Additionally, we observe slight variations in breast cancer mortality across different regions up to 2036. Furthermore, marginal increases in excess deaths from lung and breast cancer are estimated in specific regions of England throughout the pandemic year (2020--2022), contrasting with the national trend. However, the excess lung cancer deaths markedly differ by age, region and deprivation as a result of delays in cancer diagnosis. Specifically, we find a notably higher number of excess deaths in the northern regions of England compared to the southern regions, as well as among individuals living in the most deprived areas compared to those in the least deprived areas.

\end{abstract}
\begin{keywords}
Bayesian modelling; cancer mortality; COVID-19 pandemic; inequalities; predictive modelling

\end{keywords}

\maketitle

\section{Introduction}

Cancer, the largest driving cause of avoidable mortality, is one of the major causes of mortality and morbidity in England, representing 27--28\% of all deaths per year \citep{ONSCovid2020, ONSAvoidableMortality}. 
Besides, cancer not only places strain on healthcare systems but also carries significant weight in the life insurance industry through, e.g. critical illness insurance (CII) contracts, paying a benefit on the occurrence of a serious illness \citep{Macdonaldetal2003}. 
In these contracts, cancer, apart from heart attack and stroke, accounts for one of the largest percentages of claims which was reported to be as high as 54\% in 2002 in the UK \citep{Kimball2002}.

Cancer has attracted more attention due to the global COVID-19 pandemic that was first identified in Wuhan, China in December 2019 
and then rapidly spread to other parts of world in 2020, 
by claiming more than 6.5 million lives worldwide as of November 2022 \citep{WHO2021}. 
As a response to the pandemic, the UK entered three national lockdowns, with the first being introduced on 23 March, 2020. 
These measures were followed by changes in health practices by leading to, for instance, a halt in cancer screening and considerable reductions in number of patients starting cancer treatment in the pandemic years \citep{CRUKCIT2021}. 
Re-occurring national lockdowns and the changes in health services resulted with worrying figures in cancer pathways, e.g. sharp declines in participants of cancer screening or fewer number of cancer patients starting a cancer treatment, in 2020 and 2021 \citep{CRUKCIT2022}. 
On one hand, the unprecedented changes in the cancer pathways have sparked the fear of a shift to later diagnosis for people having the disease but not diagnosed yet. 
This is considered to be a serious concern since 
a late cancer diagnosis would restrict the opportunities for feasible treatment and 
worsen cancer survival.
On the other hand, early empirical studies suggested that the COVID-19 pandemic has disproportionately affected certain groups, e.g. the elderly, people with comorbidities or people who are more deprived \citep{CRUK2018, Chenetal2020, Richardsonetal2020, Grassellietal2020, Zhouetal2020}. 
One potential implication of the indirect impact of the pandemic could be to exacerbate socio-economic inequalities in cancer risk, which has been a staggering issue in the last decades, 
mostly getting worse rather than better in several countries including the UK \citep{Ariketal2021, Brownetal2007, Mouwetal2008, Singhetal2011, Riazetal2011, Bennettetal2018}. 

Most of the recent published studies have focussed on identifying the effects of COVID-related health measures on cancer survival in England based on the National Health System (NHS) UK cancer registration and hospital administrative dataset. 
For example, \citet{Laietal2020} point out dramatic reductions in the demand for, and supply of, cancer services in response to the COVID-19 pandemic by showing that these reductions could largely contribute to excess mortality among cancer patients.
\citet{Sudetal2020} report a significant reduction in cancer survival as a result of treatment delay in England, whilst \citet{Maringeetal2020} note marked increases in avoidable cancer deaths as a result of diagnostic delays over a year on. 
Furthermore, \citet{Ariketal2021} report significant increases in cause-specific cancer mortality as a result of diagnostic delays based on a population-based study in England. 
\citet{Ariketal2023SAJ} further point out medium to large size increases in BC mortality from aged 65 and above 
based on a modelling study calibrated with respect to available population data of England and medical literature.

In this study our focus is on developing cancer projection models that can be used to address how socio-economic differences are expected to change in future years. 
We particularly choose to study on two cancer types representing the largest percentage of overall cancer deaths in the UK: malignant neoplasm of trachea, bronchus, and lung, lung cancer (LC) hereafter, and malignant neoplasm of BC \citep{ONSLeadingCOD}. Specifically, we have two main interests: (a) providing a deeper insight into future LC and BC mortality based on a detailed dataset; and (b) understanding the impact of diagnosis delays on future cancer mortality.  
Part of the contribution of this study is providing a modelling framework in order to project LC and BC mortality on regional and deprivation level, where appropriate, under future scenarios that can be linked to delays in cancer diagnosis. 

Our analysis is based on the population of England between 2001 and 2018, provided by the Office for National Statistics (ONS). 
We develop cause- and gender-specific Bayesian hierarchical models to project cancer mortality, together with 95\% credible intervals, where we use a Poisson distribution assumption for cancer deaths \citep{Ariketal2021, Wongetal2018, Czadoetal2005}. 
We investigate historical and future patterns in cause-specific cancer mortality by year of death and various risk factors: age, gender, regions of England, income deprivation quintile, average age-at-diagnosis (AAD), and non-smoker (NS) prevalence rates. 
Afterwards, we examine excess cancer deaths during the COVID-19 years, along with the impact of diagnosis delays on cancer mortality under separate scenarios.

We find that socio-economic differences persist in LC mortality in the future, up to 2036, whereas marginally significant regional differences remain relevant to BC mortality. 
We also show the value of considering other risk factors, such age-at-diagnosis and smoking status, in the context of cancer mortality modelling and projection.
Meanwhile, excess cancer deaths, that are the number of deaths above the expected number of deaths in a calendar year, are quantified during 2020--2022 in the regions of England. 
Importantly, future scenarios are developed under LC modelling based on the assumption of delays in cancer diagnosis through the AAD variable. 
As a result, excess LC deaths are calculated by age, gender, region, and deprivation levels of England by demonstrating marked variations across different population groups. 

This study is organised as follows. 
In Section~\ref{Sec:Data} we introduce the available data and important concepts used in different parts of this study. 
In Section~\ref{Sec:StatisticalAnalysis} we explain the modelling framework for LC and BC mortality rates.
In Section~\ref{Sec:NumericalResults} we present our main findings associated with LC mortality first, and then for BC mortality in England.
%the population data of England. 
In Section~\ref{Sec:Conclusion} we discuss the main implications of our findings and conclude.

\section{Data}\label{Sec:Data}

Data used in this study have been collected from the ONS.
The data are of the following types across England: (i) cause-specific cancer death counts; (ii) mid-year population estimates. 
Additionally, we have smoking prevalence rates based on Health Survey for England provided by the NHS Digital.

\subsection{Population and mortality data}

For each region of England, specifically the north east, the north west, Yorkshire and the Humber, the East Midlands, the West Midlands, the east, London, the south east, and the south west described by the Nomenclature of Territorial Units for Statistics \citep{Eurostat2007}, 
we have cause-specific cancer death counts, $D_{a,g, d,r,t}$, and mid-year population estimates $E_{a,g, d,r,t}$ from the ONS by five-year age-at-death group $a$, single calendar year $t$ from 2001 to 2018, gender $g$, region $r$, and deprivation deciles (1 to 10). 

Meanwhile, causes of death data at a lower granularity is accessible  up to 2022 through an ONS service, namely `NOMIS' \citep{NOMIS2022}. 
We use `NOMIS' as an additional source to obtain LC and BC deaths by gender and five-year age bands in the regions of England from 2019 to 2022.

Death rates for a given region $r$ and deprivation level $d$ are then 

%\vspace{-\baselineskip}
\begin{eqnarray*}
	\theta_{a,g, d,r,t} =  \frac{D_{a,g, d,r,t}}{E_{a,g, d,r,t}}, 
\end{eqnarray*}
%\vspace{-\baselineskip}

and at the regional level for region r
%\vspace*{-0.3cm}
%\vspace{-\baselineskip}
\begin{eqnarray*}
	\theta_{a,g,r,t} =  \frac{\sum_{d=1}^{10} {D_{a,g, d,r,t}} }{ \sum_{d=1}^{10}E_{a,g, d,r,t}}.
\end{eqnarray*}

\subsection{Smoking data}

We have three types of smoking prevalence rates, namely smokers, current smokers, and non-smokers, by 10-year age groups, sex and single year from 1993 to 2019 in England \citep{HealthEnglandSmoking2023}. 

\begin{figure}[htbp]
	\centering
	{\includegraphics[width=0.75\textwidth]{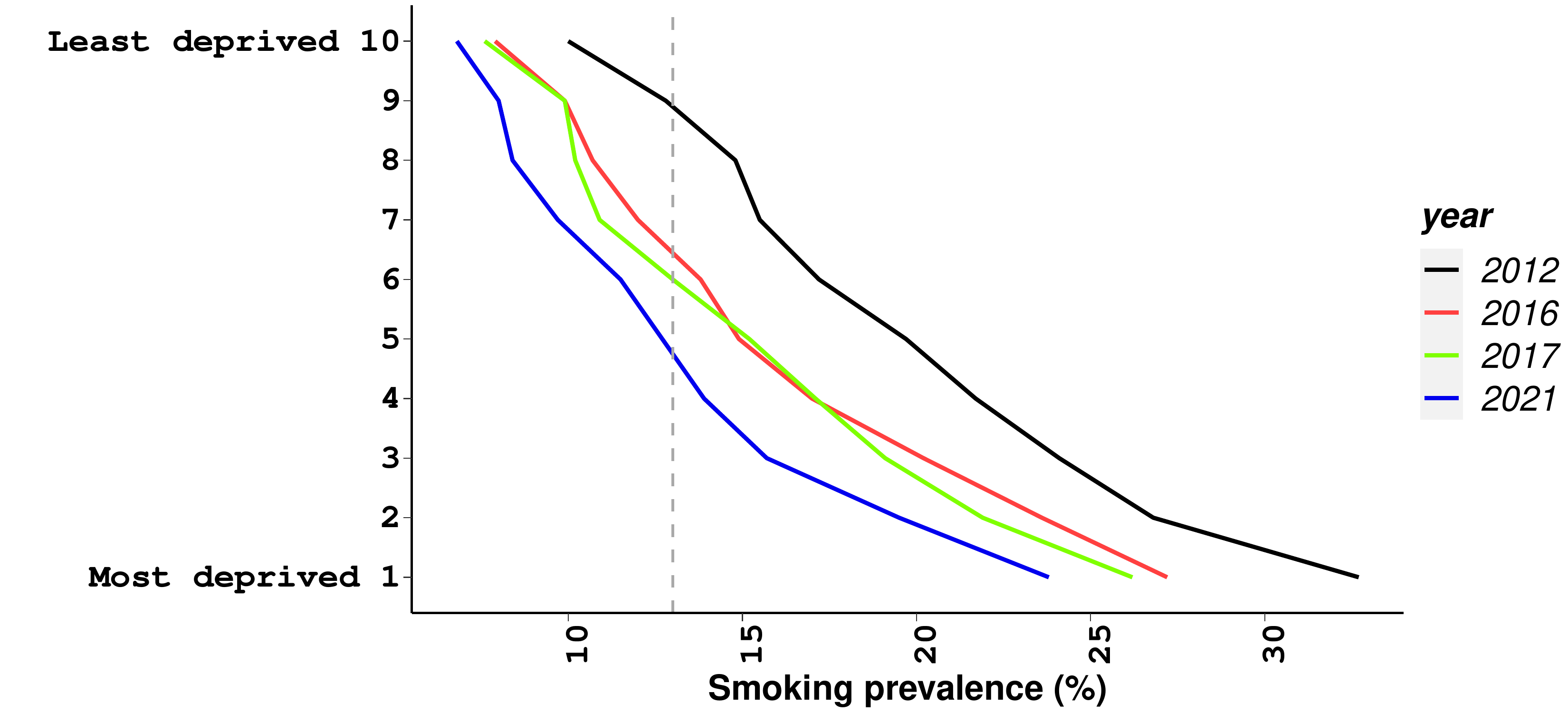}}
	\caption{Prevalence rates of current smokers by deprivation deciles, all persons aged 18 years and over, England, 2012 to 2021, where the dashed line is the national average, 13\%, in 2021. Source: Annual Population Survey from the Office for National Statistics.} 	
	\label{fig:NSPrevalenceEngland_byDeprivation}
\end{figure}

We note that, as part of the Annual Population Survey, the ONS started reporting smoking prevalence since 2012 by pointing out variation in smoking on regional and deprivation levels. 
\figref{fig:NSPrevalenceEngland_byDeprivation} displays that smoking prevalence has declined in each deprivation decile over the time with statistically higher smoking prevalence in the most deprived neighbourhoods as compared to the least deprived ones in England \citep{Archboldetal2023}. 

Cautious must be exercised regarding possible variations in health survey questions (even in the same survey) since these would impact the calculations of smoking estimates \citep{Windsoretal2018}. 
Differences in the survey questions are indicated to have a notable impact on smoking estimates, leading to modelling bias \citep{Ryanetal2012}. 
Thus, in this study, we consider NS prevalence from a single source, i.e. Health Survey for England, as a proxy for smoking in the implemented models. 
This is to: (a) avoid changes across different definitions of smoking information, and (b) have a simple and clear interpretation of smoking in the projection models. 

\begin{figure}[htbp]
	\centering
	\subfloat[ Males \label{fig:MalesNS}]{\includegraphics[width=0.5\textwidth]{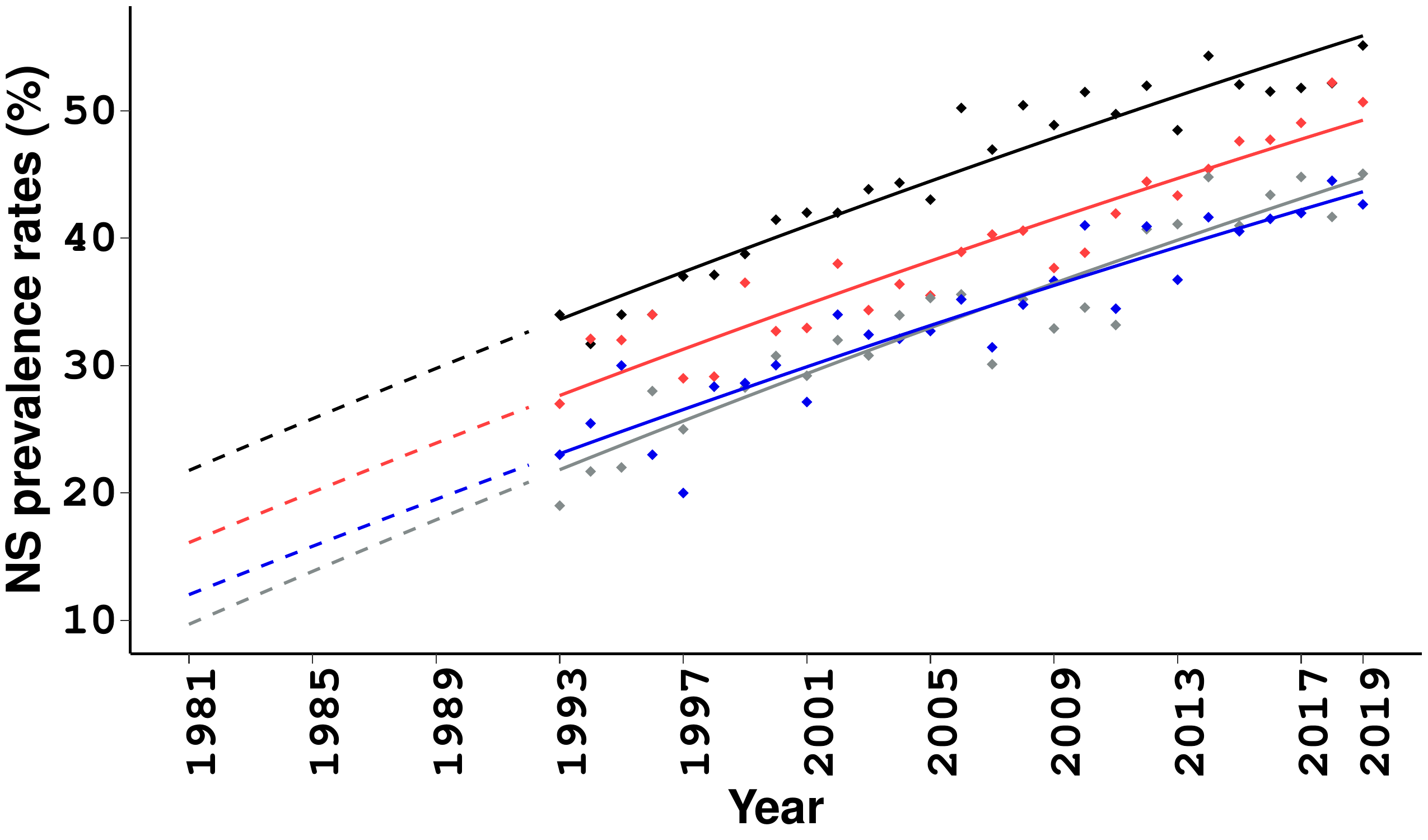}}
	\hfill
	\subfloat[Females \label{fig:FemalesNS}]{\includegraphics[width=0.5\textwidth]{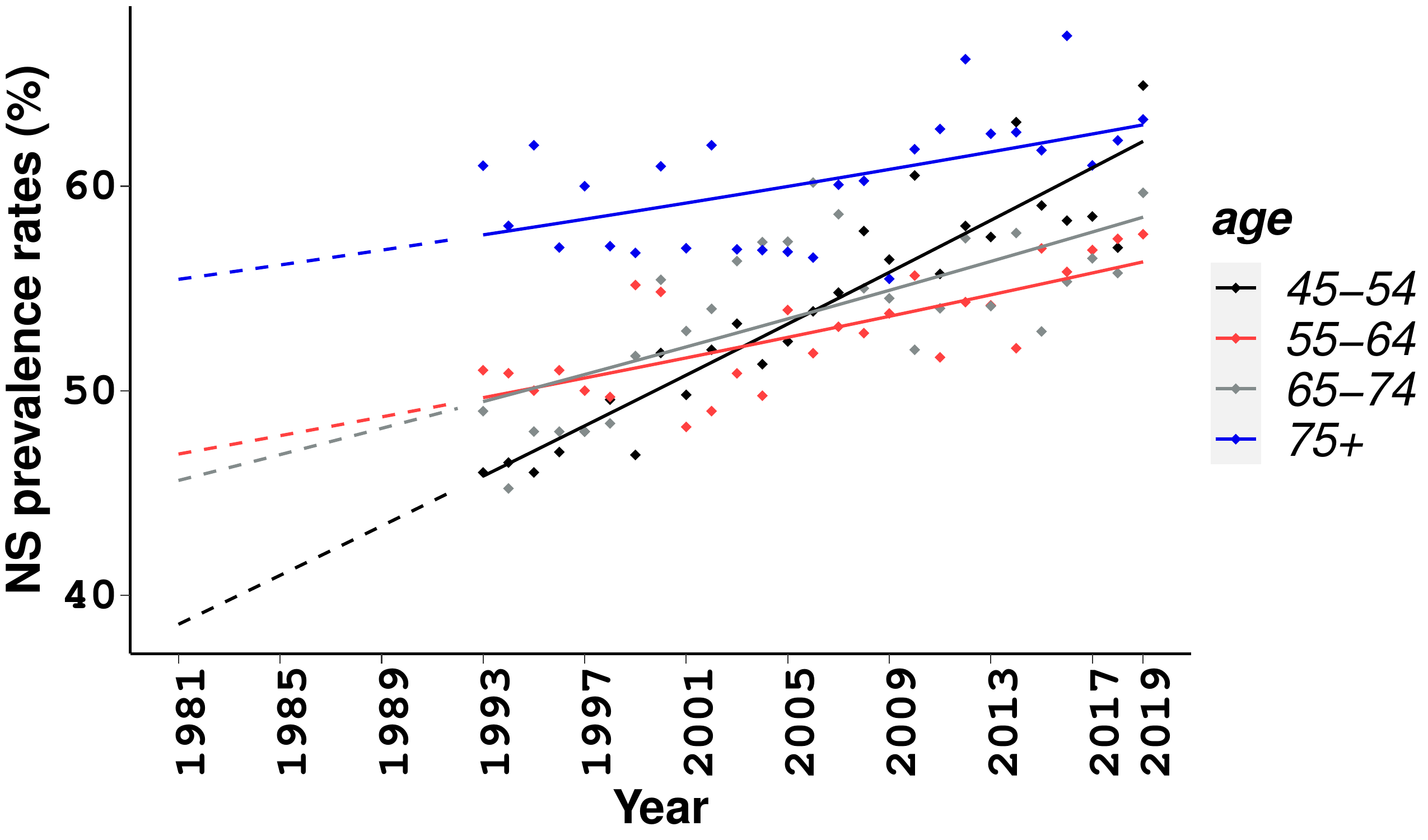}}
	\caption{Non-smoker prevalence rates at selected age groups between 1981 and 2019: observed rates (dots), fitted rates (solid lines), and re-constructed rates (dashed lines). } 	
	\label{fig:NSBackwardsBothGender}
\end{figure}

We assume a lag of 20 years while accounting for contribution of smoking to cancer mortality risk in our projection models \citep{Luoetal2022}. 
This means, for example, NS prevalence for men in 1981 must be used as an input to estimate, e.g., male LC mortality in 2001. 
Provided we don't have access to the related data before 1993, 
we relied on a simple modelling approach. 
%constructed based on the available dataset. 
Particularly, first, we have applied a gender-specific linear model to the available dataset between 1993 and 2019 as 

\begin{equation} \label{eq:NSLinearModel}
	{\text{NS}}_{a, t} = \beta_0 + \beta_{1,a} + \beta_{2} t  +  \beta_{3} t^2 + \beta_{4,a}t, 
\end{equation}
and then the parameter estimates in \eqref{eq:NSLinearModel} are employed to re-construct NS prevalence backwards to 1981 for each gender. The observed NS prevalence rates for men and women, along with estimated rates, are demonstrated from 1981 to 2019 in \figref{fig:NSBackwardsBothGender}. 
The figure reveals an increasing trend in NS prevalence for both genders, showing a more homogeneous and faster increase among men over the inspected period.

\section{Projection models for cancer mortality} \label{Sec:StatisticalAnalysis}

We assume that the number of cause-specific cancer deaths $D_{a,g,d,r,t}$ at age $a$ and year $t$ for gender $g$ in deprivation quintile $d$ of region $r$ in England follows a Poisson distribution. 
Although this is a common assumption in the literature since the study of \citet{Brouhnsetal2002}, there is an issue with this assumption. Specifically, the underlying assumption imposes mean-variance equality such that 

\[
\hat{\mathbb{E}}( D_{a,g, d,r,t} ) =\hat{\text{{var}}}( 	D_{a,g, d,r,t}  ) =  \hat{\theta}_{a,g, d,r,t} \,\, E_{a,g, d,r,t},
\]
where $\hat{\theta}_{a,g, d,r,t}$ shows the expected mean of cause-specific fitted mortality rates. 
This suggests that individuals born in the same year could have the same mortality experience despite several different factors, such as smoking, income, and education, impacting mortality \citep{Brown2003, LouresCairns2018, Ariketal2020}. 
This leads to an additional variation across individuals, also known to be `overdispersion'. 

Thus, in order to deal with overdispersion, or in other terms, to account for the heterogeneous structure in the sub-populations of England, 
we construct a baseline model for a given cause-specific cancer mortality using a Poisson-lognormal Bayesian hierarchical model. 
The general structure of our model is:

%\begin{subequations}
\begin{align}
	D_{a,g, d,r,t} &\sim \text{Poisson}( \theta_{a,g, d,r,t} \,\, E_{a,g, d,r,t} ) \nonumber \\
	\theta_{a,g, d,r,t} &\sim  \text{Lognormal}(\mu_{a,g, d,r,t}, \sigma^2) \nonumber \\
	\mu_{a,g, d,r,t} &= \bm{\beta} \bm{X} \nonumber\\
	{	\sigma^2 }&{ \sim  \text{Inv.Gamma}(1,0.1) } \nonumber \\ 
	\bm{\beta} &\sim   \text{Normal}(0, 10^4), 
	\label{eq:BaselineModelStructureLungCOD}
\end{align}
%\end{subequations}
where 
\begin{itemize}
	\item $\theta_{a,g, d,r,t}$ is the mortality rates at age-at-death $a$ in year $t$ for gender $g$ in deprivation quintile $d$ of region $r$, whenever applicable;
	\item $\mu_{a,g, d,r,t}$ is the location parameter of lognormal distribution for a given cancer type, which is defined based on different covariates, denoted by $\bm{X}$, and associated model parameters $\bm{\beta}$;
	\item non-informative prior distributions are assumed for model parameters $ \bm{\beta}$ and $\sigma^2$, apart from the ones linked to the period effect.
\end{itemize}

The structure of the location parameter, $\mu_{a,g, d,r,t}$, differs for each cause-specific cancer. 
It depends on main variables, namely age-at-death, year, deprivation quintile, region, AAD, and NS prevalence rates, along with two-way interaction terms between main variables, where appropriate. 
We note that the AAD variable is introduced and explained in the study of \citet{Ariketal2021}. 
This variable, which is based on related cancer morbidity, 
is to construct model-driven age-at-diagnosis for cancer mortality modelling (see Appendix~\ref{sec:AAD_desc}). 

Meanwhile, the structure of the location parameter, $\mu_{a,g, d,r,t}$, is determined through a Bayesian variable selection procedure in the R-INLA software \citep{LindgrenRue2015} based on two criteria: Deviance Information Criterion (DIC),  see \citet{Spiegelhalteretal2002}, and Bayes factors, see \citet{KassandRaftery1995}.
The DIC is calculated based on the effective number of parameters and marginal log-likelihood as

\[
\mbox{DIC} = -4 E_{\bm{\beta} \vert D} ( \log f(D \vert \bm{\beta} ) ) + 2 \log f( D \vert \hat{\bm{\beta}} ),
\]	
where $\hat{\bm{\beta}}$ is the posterior mean, mode or median of parameter vector ${\bm{\beta}}$, $f$ is the likelihood function, and $D$ is the related data.
The Bayes factor is defined as a ratio of posterior odds of models $H_j$ and $H_k$ based on data $D$, provided that the same prior distributions are assumed for both models, see \citet{LindgrenRue2015}, such that  
\[
B_{jk} =  \frac{P(D \vert H_j)}{ P(D \vert H_k)}; \,\, j \neq k.
\]

In our models, we consider age-at-death, deprivation quintile, and region as categorical variables, 
whereas NS prevalence rates and AAD variable are assumed to be numerical variables, {standardised to have zero mean and unit variance to facilitate the calculations}. 
Meanwhile, for model identifiability and interpretability, sum-to-zero (STZ) constraints are imposed to all categorical variables, apart from the period effect.
We note that STZ constraint allows us to make comparisons between a given level of a categorical variable with respect to the corresponding average effect, that is zero, as the reference level.

Furthermore, a random walk with drift is implemented to describe the period-related effects, denoted by $\bm{\kappa}_t = (\kappa_{1,t}, \kappa_{2,t})^T$, as follows:

\begin{align}\label{eq:kapparwd}
	\bm{\kappa} _t& = \bm{\psi}+ \bm{\kappa}_{t-1} + \bm{\epsilon}_{t} \nonumber \\ 
	\bm{\psi} &{ \sim  \text{Normal}( \bm{\psi}_0 , \bm{\sigma^2}_ {\bm{\psi}} )}\nonumber \\
	\bm{\epsilon}_{t}&\sim \text{Normal}( 	\bm{\epsilon}_{0} , \bm{\sigma_{\bm{\kappa}}^2} )\nonumber \\
	\bm{\sigma^{2}_\bm{\kappa} } &\sim \text{Inv.Gamma}(1, 0.001),
\end{align}
where $\bm{\psi} = (\psi_{\kappa_1}, \psi_{\kappa_2})^T$; $ \bm{\epsilon}_{t} = ({\epsilon}_{\kappa_1,t}, {\epsilon}_{\kappa_2, t})^T$; $\bm{\sigma^2}_ {\bm{\psi}} = ({\sigma^2}_ {{\psi}_{\kappa_1}},  {\sigma^2}_ {{\psi}_{\kappa_2}})^T$; $\bm{\sigma^{2}_\bm{\kappa} } = (\sigma^2_{\kappa_1}, \sigma^2_{\kappa_2})^T$; $\bm{\psi}_0 = (0, 0)^T$; $\bm{\epsilon}_{0}= (0,0)^T$; 
$\bm{\hat{\sigma}^2}_ {\bm{\psi}} =  \frac{1}{2018-2001}\,\bm{\hat{\sigma}^{2}_\bm{\kappa} }$.
We adopt corner constraint for the period effect $\bm{\kappa} _t$, $\bm{\kappa} _1 = (0, 0)^T$.
This changes the interpretation of the period-related coefficients by setting the first year as the baseline year. 
To be specific, the values related to the following years would be estimated with respect to $\bm{\kappa} _1$ and thus should be interpreted accordingly \citep{Wongetal2018}.

In order to project a given cause- and gender-specific cancer mortality beyond the observed calender period, we assume that the age-at-death-, region-, and deprivation-related effects remain unchanged over time, where appropriate. 
Accordingly, the future mortality rates for a given cancer type can be derived as 

\begin{align*}
	\theta_{a,g,d,r,t}^{*}&\sim  \text{Lognormal}(\mu_{a,g, d,r,t}^{*}, \sigma^2),
\end{align*}
where a new location parameter, $\mu_{a,g, d,r,t}^{*}$, is defined considering changes only in time-related terms. 
To be precise, the period-related effects would be extrapolated from 2019 to 2036 by setting the baseline year as the last year of the observed calendar year such that $\bm{\kappa}^* _1 = (\hat{\kappa}_{1, 18}, \hat{\kappa}_{2, 18})^T $ in \eqref{eq:kapparwd}. 
%BC data is available for a longer period, up to 2022. 
%%, that matches with the model described in \eqref{eq:FemaleBreastLocationPrmtr}. 
%However, the last four calendar years, 2019 to 2022, are used for model validation of BC mortality. 
%Hereby, the extrapolation in BC mortality goes until 2036 by choosing 2018 as the baseline year. 

Model validation is carried out both for LC and BC by quantifying Pearson residuals and checking with possible patterns across different ages and years for a given region and deprivation quintile. 
The residuals are obtained as follows:

\begin{subequations}
	\begin{align*}
		r_{a,d,r,t} = \frac{ D_{a,d,r,t}  - \hat{\mathbb{E}}( D_{a,d,r,t}  ) } { \sqrt{\hat{\text{var}}( D_{a,d,r,t}   ) } },
	\end{align*}
\end{subequations}
where  $\hat{ \mathbb{E}}( D_{a,d,r,t}  ) =  \hat{\theta}_{a,d,r,t} E_{a,d,r,t} $ and \\ $\hat{\text{var}}( D_{a,d,r,t} )  = \hat{ \mathbb{E}}( D_{a,d,r,t}  ) \times ( 1 +\hat{ \mathbb{E}}( D_{a,d,r,t}  )\, \exp( \sigma^2 - 1 ) )$ \citep{Wongetal2018}.
The corresponding fitted mortality rate, $\hat{\theta}_{a,d,r,t}  $, is derived using the mean of lognormal distribution as $ \hat{\theta}_{a,d,r,t}  = \exp( \mu_{a,d,r,t}  + \sigma^2/2 )$.

\subsection{Female lung cancer mortality}

From a modelling perspective, female LC mortality is more complicated than male LC mortality. 
For instance, the female LC mortality points out a change in time trend in the recent years, e.g., slowdown in mortality improvement at different age groups, as a result of various factors including changes in smoking patters \citep{ONS2015}.

We have established a model, where the location parameter of lognormal distribution in \eqref{eq:BaselineModelStructureLungCOD} is defined as 

%considering the main variables in the available data as follows:
\begin{align}
	\mu_{a,d,r,t}^{\text{lung}} &=  \beta_0 + \beta_{1,a} + \beta_{2,r} + \beta_{3,d} + \beta_{4} \text{AAD}_{r,d}^{ \text{morbidity}} + \beta_{5, d, a} + \beta_{6, r, a} + \kappa_{1,t}  +    \nonumber  \\  
	& \big( \kappa_{2,t} + \beta_{7, r} \big) \text{AAD}_{r,d}^{ \text{morbidity}} + \beta_8  \text{NS}^{\text{women}}_{a, t-20}.
	\label{eq:FemaleLungCODLocationPrmtr2}
\end{align}
Note that \eqref{eq:FemaleLungCODLocationPrmtr2} has been determined with respect to a forward variable selection procedure, where the details can be found in Appendix~\ref{Sec:VariableSelectionLungCODFemales}. 
Here, $\beta_{1,a}$ is the age coefficient for age group $a$ with levels $a= 1, 2, \ldots, 8$, where $a$ maps to $\{ 45-54, 55-59, 60-64, \ldots, 85-89\}$; $\kappa_{1,t}$ is the coefficient associated with period $t$ with levels $t= 1, 2, \ldots, 18$, where $t$ maps to $\{2001, 2002, \ldots, 2018\}$; $\beta_{2,r}$ is the region coefficient for region $r$ with levels $r = 1, 2, \ldots, 9$, where $r$ maps to \{North East, North West, Yorkshire and the Humber, East Midlands, West Midlands, East, London, South East and South West\}; $\beta_{3,d}$ is the deprivation coefficient for quintile $d$ with levels $d = 1, 2, \ldots, 5$, respectively; $\kappa_{2,t}$ is the coefficient of interaction between period effect and AAD component; $\beta_{5, d, a}$ is the coefficient of interaction between age-at-death and deprivation quintile; and $\beta_{6, r, a}$, for the interaction between age-at-death and region; $\beta_{7, r}$ is the coefficient of interaction between region effect and AAD component, and 
$\beta_8$ is the coefficient for the NS prevalence rates. %(\tabref{tab:FemaleLungCODVariableSelecNoSmokingv2}). 

\subsection{Male lung cancer mortality}

We have constructed the male LC mortality model as follows:

\begin{align}
	\mu_{a,d,r,t}^{\text{lung}} &=  \beta_0 + \beta_{1,a} + \beta_{2,r} + \beta_{3,d} + \beta_{4} \text{AAD}_{r,d}^{ \text{morbidity}} +\beta_{5, d, a}  + 	\kappa_{1,t}  +   \nonumber  \\ 
	& \big( \kappa_{2,t}  + \beta_{6,r} \big) \text{AAD}_{r,d}^{ \text{morbidity}}  + \beta_{7} \text{NS}^{\text{men}}_{a, t-20}, 
	\label{eq:MaleLungCODLocationPrmtr2}
\end{align}
where the main difference between the models in \eqref{eq:FemaleLungCODLocationPrmtr2} and \eqref{eq:MaleLungCODLocationPrmtr2} causes due to the additional interaction term between age and region in the former model.

We note that the LC models in \eqref{eq:FemaleLungCODLocationPrmtr2} and \eqref{eq:MaleLungCODLocationPrmtr2}, for women and men, respectively, have been used to establish future scenarios associated with cancer diagnosis. 
Specifically, we have introduced diagnosis delays with the aid of AAD component and estimated related increases in LC mortality accordingly.

We also note that several model specifications have been implemented before determining the final modelling structure(s). 
This is because different best fitted models can be identified by changing the description of null model in the variable selection process, as demonstrated in, e.g. Appendix~\ref{Sec:VariableSelectionLungCODMales}. 
The overall decision has been made with the aim of finding a compromise between model complexity, data fitting, and potential correlations across different variables.

\subsection{Female breast cancer mortality} \label{sec:BCOProjectionModel}
We focus on female BC mortality as there are few records regarding male BC. 
It is important to note that, in the existence of other variables, e.g. age and region, income deprivation is not found to be a significant risk factor to explain differences in BC mortality in England (see \tabref{tab:BreastCODVariableSelect} and \citet{Ariketal2021}). 
Hereby, BC mortality projection is considered on regional level. 
Furthermore, the AAD variable by the regions of England has not been found to be statistically important to explain BC mortality either.
This can be attributed to the empirical evidence, suggesting more `equality' in BC mortality as compared to a lifestyle-related cancer, e.g. LC mortality. 
This results with comparable AAD estimates across different regions of England.

Despite the fact that income deprivation and AAD variables were not significant for modelling purposes, female NS prevalence rates have been found to be an important risk factor that contributes to explain BC mortality.
The association between BC risk and smoking has been extensively studied by considering the amount of cigarette consumption \citep{DJHankinsonetal1997}, duration of smoking \citep{Reynoldsetal2004}, and smoking initiation at different ages \citep{Delaimyetal2004}, sometimes leading to conflicting results.
However, there is more evidence suggesting a potential causality between smoking and BC, 
especially in the case of long-term heavy smoking and smoking initiation at a young age \citep{Xueetal2011, Reynolds2013}.

Following the variable selection procedure, see \tabref{tab:BreastCODVariableSelect_wlSmoking}, we have come up with a much simpler projection model as opposed to LC models such that 

\begin{align}
	\mu_{a,r,t}^{\text{breast}} =  \beta_0 + \beta_{1,a} + \beta_{2,r} + \beta_{3} \text{NS}^{\text{women}}_{a,t-20} + \kappa_{1,t}.
	\label{eq:FemaleBreastLocationPrmtr}
\end{align}

Here, $\beta_{1,a}$ is the age coefficient for age group $a$ with levels $a= 1, 2, \ldots, 11$, where $a$ maps to $\{ 35-39, 40-44, 45-49, \ldots, 85-89\}$; $\kappa_{1,t}$ is the coefficient for the period component for period $t$ with levels $t= 1, 2, \ldots, 18$, where $t$ maps to $\{2001, 2002, \ldots, 2018\}$; $\beta_{2,r}$ is the coefficient of the region component; $\beta_3$ is the smoking coefficient. 

Provided regional-level BC mortality is available up to 2022, 
we have utilised observations from 2019 until 2022 in order to make comparisons between observed and projected BC mortality in those years.

\subsection{Measures of excess deaths and different future scenarios}\label{sec:PandemicScenarioModelAssumptions}

We can examine excess cancer deaths based on our modelling framework, following a similar approach to the ONS \citep{ONSExcessMortNewMeth2024}.
Additionally, we can investigate the impact of delays in cancer diagnosis on cancer mortality with the aid of AAD covariate. 
The latter analysis is motivated by the significant reductions in cancer registrations as a result of initial health disruptions caused by the COVID-19 pandemic, e.g. see \citet{CRUK2021}. 
We develop three scenarios by considering an increase in the AAD covariate. 
Specifically, Scenario 1 introduces a 1-month delay in AAD. 
Then, we assume a 3-month delay in AAD in Scenario 2, and a 6-month delay in Scenario 3. 
Provided that AAD has not been found important to explain differences in BC mortality, this part of the study
is only relevant to LC mortality. 

`Cumulative excess deaths' from a given cancer type during the pandemic years (2020--2022) for a given gender $g$ and region $r$ of England are calculated by using the observed and predicted number of deaths as 

\[
\text{CED}_{g,r} = \sum_{t=2020}^{2022} {\sum_{d} { \sum_{a} {{D}_{a,g, d,r,t} - \hat{\mathbb{E}}({D}^{\text{baseline}}_{a,g, d,r,t} )  } } 
} , 
\]
where $\hat{\mathbb{E}}( D^{\text{baseline}}_{a,g, d,r,t} )$ refers to the pre-pandemic estimates with no COVID-impact. 

Furthermore, `excess deaths' from a certain cancer at various age groups $a$ in deprivation level $d$ and region $r$ in the projection years, $\text{ED}_{a,g, d,r,t}$, 
are considered by subtracting the estimated number of deaths in the baseline calculations from those in a specific scenario as follows:
\[
\text{ED}_{a,g, d,r,t} = \hat{\mathbb{E}}({D}^{\text{scenario}}_{a,g, d,r,t} ) -  \hat{\mathbb{E}}({D}^{\text{baseline}}_{a,g, d,r,t} ).
\]

Meanwhile, `excess cause-specific cancer mortality' in the projection years is obtained by dividing excess cause-specific cancer deaths by the corresponding mid-year population estimates.  
Thus, in order to calculate age-specific excess cancer mortality for gender $g$ in a given projection year $t$, $\text{EAM}_{a,g,t}$, we use 

\[
\text{EAM}_{a,g,t}= \frac{ \hat{\mathbb{E}}( D^{\text{scenario}}_{a,g,t} ) - \hat{\mathbb{E}}( D^{\text{baseline}}_{a,g,t} )}{E_{a,g,t}}, 
\]
with, for instance, $\hat{\mathbb{E}}( D^{\text{scenario}}_{a,g,t} ) = \sum_{d}{  \sum_{r}{\hat{\mathbb{E}}( D^{\text{baseline}}_{a,g, d,r,t} )} }$. In a similar manner, region-specific excess cancer mortality in year $t$, $\text{ERM}_{g,r,t}$, is obtained as 

\[
\text{ERM}_{g,r, t} = \frac{ \hat{\mathbb{E}}( D^{\text{scenario}}_{g,r, t} ) - \hat{\mathbb{E}}( D^{\text{baseline}}_{g,r, t} )}{E_{g,r, t}}.
\]

\noindent Last, deprivation-specific cancer mortality in year $t$, $\text{EDM}^c_{g, d, t}$, is calculated as 

\[
\text{EDM}_{g,d,t} = \frac{ \hat{\mathbb{E}}( D^{\text{scenario}}_{g, d, t} ) - \hat{\mathbb{E}}( D^{\text{baseline}}_{g, d, t} )}{E_{g, d, t}}.
\]

\subsubsection{Assumption 1: cancer survival}

We take into account for net cancer survival to distribute an overall increase (1- to 6-month) in AAD over time. 
Any increase in AAD in a given year would lead to an increase in cause-specific cancer mortality under inspection in the same year. 
Hereby, the aim is to allow a gradual increase in the related cancer mortality in the future years. 
Particularly, we assume that a serious health disruption could cause a bigger increase in AAD in the first year of the projection by gradually declining later on. 

{LC survival} is reported to gradually decrease over time such that 40\% of people with LC would survive from this disease for one year or more, 15\% for 5 years or more, and 10\% for 10 years or more \citep{CRUK2021}. 
Hereby, we assume that a 60\% increase of a particular delay, e.g. 1-month, in the AAD variable would realise in the second year of the projection period, 2020. This would be followed by a 25\% increase up to 5 years, 10\% from 6 to 10 years, and 5\% in the rest of the projection period.

\subsubsection{Assumption 2: population estimates in future years}

Estimating cause-specific deaths in a particular year requires to know both relevant mortality rates and mid-year population estimates.
Our general modelling structure, \eqref{eq:BaselineModelStructureLungCOD}, provides the framework to obtain mortality rates in the projection years (2019--2036). Nevertheless, for the calculation of the related number of deaths, 
the corresponding mid-year population estimates must be provided as well.

We rely on the national population projections provided by the ONS in the future years. 
The population estimates, stratified by five-year age groups and gender in the regions of England, are available from 2019 to 2043 \citep{ONS2020SubnationalPopProj}. 
However, to facilitate our calculations, we specifically require the mid-year future population estimates by deprivation quintiles in a given region. 
Consequently, we make an additional assumption.
We accept that the distribution of population estimates across deprivation quintiles within a given region in the last observed calendar year, 2018, would remain unchanged throughout the projected years.

\section{Numerical results} \label{Sec:NumericalResults}

We present modelling results up to 2036, using the data from 2001 to 2018. 
We experimented with different levels of data granularity and selected two sets of data: age- and gender-specific LC mortality rates split by deprivation quintiles and regions of England, and age-specific BC mortality rates by region. 

In this part of the study, we mainly focus on age-standardised mortality rates. 
We have obtained age-standardised fitted and projected cause-specific cancer mortality to demonstrate mortality inequalities across different deprivation levels and regions of England. 
For this sake, the European Standard Population (ESP) 2013 is used as the reference population \citep{EurostatESP2013}. 

Note that we present parameter estimates for each cause-specific cancer mortality model in Appendix~\ref{sec:ParameterEstimates} and display age-specific estimates, together with Pearson residuals, 
in Appendix~\ref{sec:AgeSpecificFittedProjectedRates}.

\subsection{Regional and deprivation level lung cancer mortality}

\begin{figure}[htbp]
	\centering
	\includegraphics[width=0.8\textwidth, angle =0]{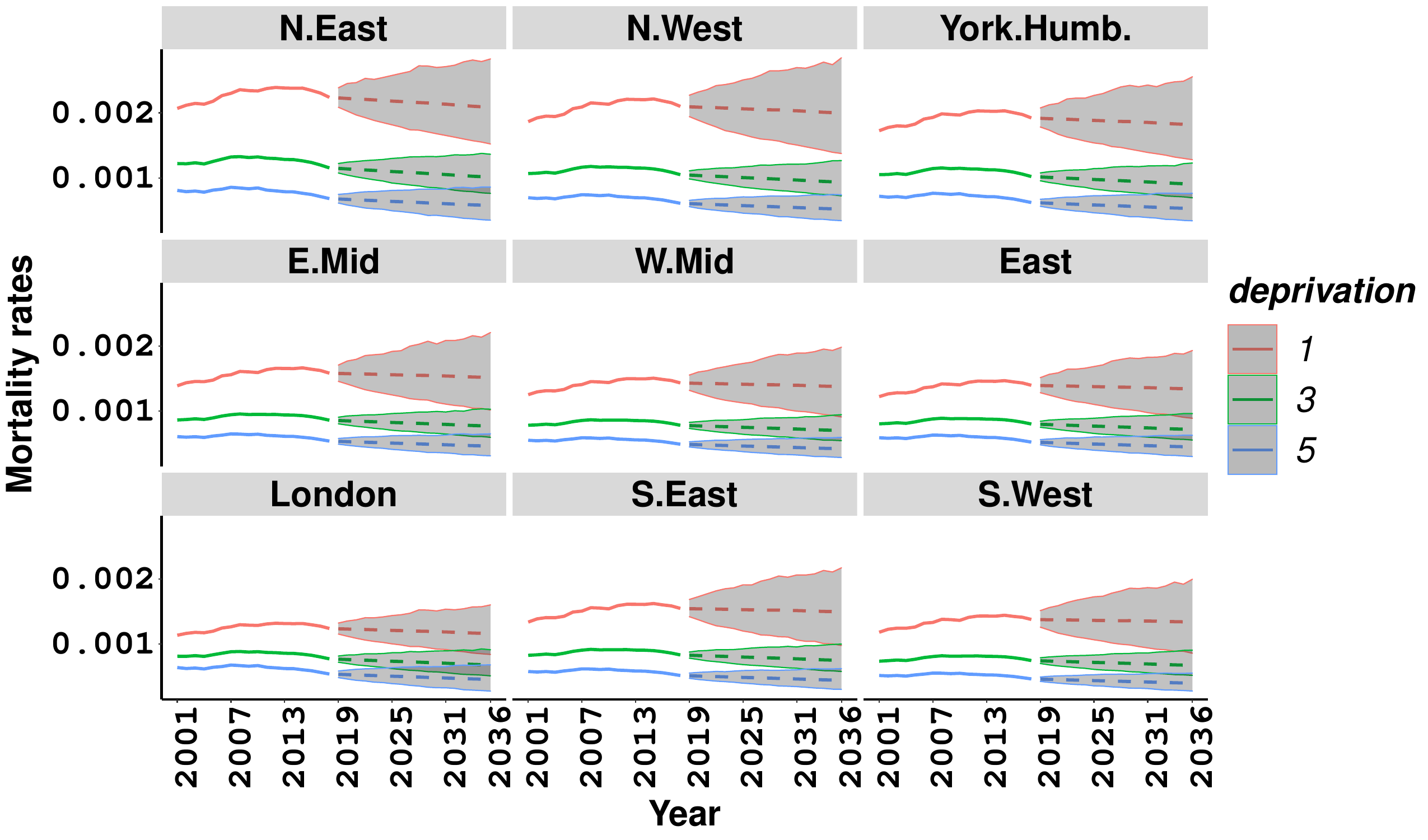}
	\caption{Age-standardised fitted (solid lines) and projected (dashed lines) lung cancer mortality, females, with 95\% credible intervals, in selected deprivation quintiles 1 (most deprived), 3, and 5 (least deprived) and regions of England based on \eqref{eq:FemaleLungCODLocationPrmtr2}.}
	\label{fig:ASRLungCOD_FeMales_v2}
\end{figure}

\begin{figure}[htbp]
	\centering
	\includegraphics[width=0.8\textwidth, angle =0]{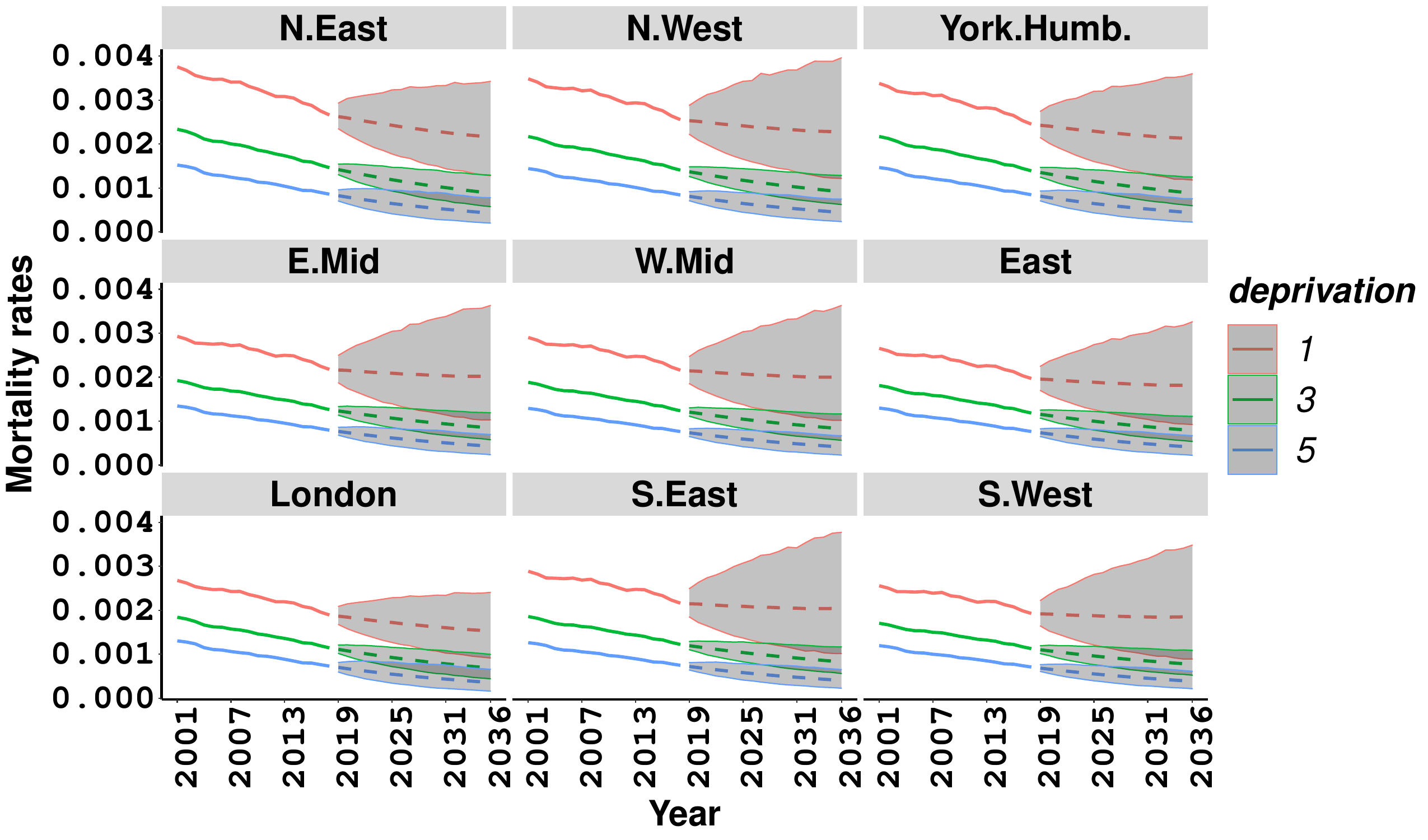}
	\caption{Age-standardised fitted (solid lines) and projected (dashed lines) lung cancer mortality, males, with 95\% credible intervals, in selected deprivation quintiles 1 (most deprived), 3, and 5 (least deprived) and regions of England based on \eqref{eq:MaleLungCODLocationPrmtr2}.}
	\label{fig:ASRLungCOD_Males_v2}
\end{figure}

Let us consider LC mortality in women, \figref{fig:ASRLungCOD_FeMales_v2}, and men, \figref{fig:ASRLungCOD_Males_v2}. There are some distinct differences and certain similarities in LC mortality by gender. 
First, both figures reveal markedly different time trends in the historical LC mortality by gender, 
where a clear mortality improvement has been estimated for men across varying deprivation and region levels but not for women. 
\figref{fig:ASRLungCOD_FeMales_v2} shows that 
women in the less deprived quintiles have actually entertained slight mortality improvements in the recent calendar years. However, this has not always been the case for the women in the most deprived quintiles of regions of England. 
Consequently, the projected rates for men are predicted to be more optimistic than the female counterparts, with a continuing mortality improvement at a varying degree in each deprivation quintile, e.g., slowing down in the most deprived quintiles as opposed to the less deprived levels. 
Meanwhile, the projected rates for women are estimated to be levelled up or to decline slightly over the time.

Second, both \figref{fig:ASRLungCOD_FeMales_v2} and \figref{fig:ASRLungCOD_Males_v2} point out substantial differences across deprivation quintiles in regions of England. 
Importantly, our findings show that socio-economic differences would persist in the future years. 
This is evidenced by the estimated mortality rates in the most (1) and least (5) deprived quintiles over the projection period, which have remained to be significantly different from each other in each region.

\subsection{Regional level breast cancer mortality}

We examine BC mortality in women. 
Provided that income deprivation and AAD were not found to be significant for modelling purposes, 
these variables have neither been used to estimate historical nor future BC mortality rates. 

\begin{figure}[htbp]
	\centering
	\includegraphics[width=0.8\textwidth, angle =0]{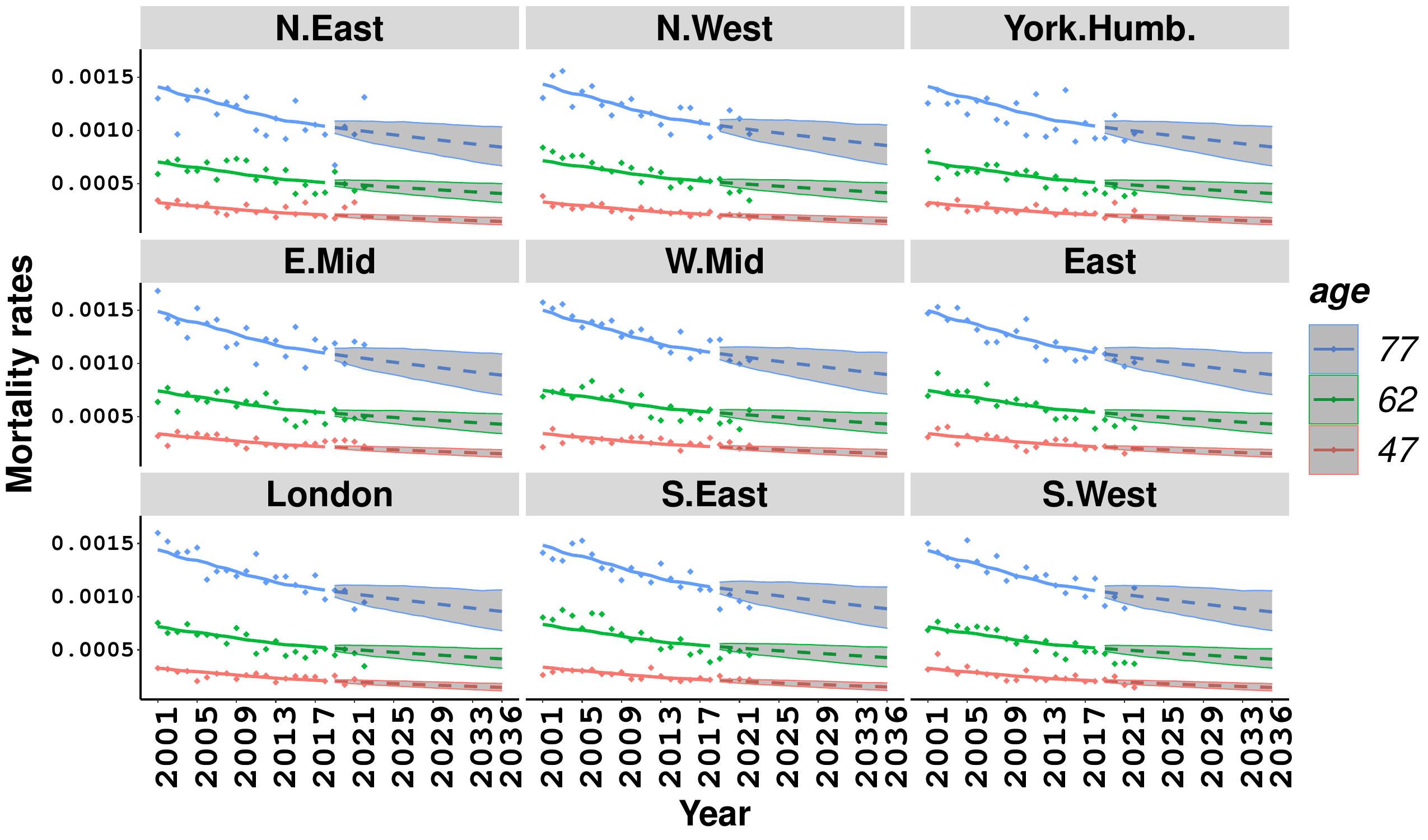}
	\caption{Breast cancer mortality, females, (screening) ages at death 47, 62, and 77, in regions of England based on \eqref{eq:FemaleBreastLocationPrmtr}: observed rates (dots), fitted rates (lines), projected rates (dashed lines) with 95\% credible intervals for the projected rates.}
	\label{fig:BreastCOD_ScreeningAges}
\end{figure}

\figref{fig:BreastCOD_ScreeningAges} present a declining trend in BC mortality among the screening age groups 45--49 to 75--79 in England.
Our model points out comparable observed and estimated mortality rates across the regions of England up to 2036. 
Nonetheless, the mortality rates between the youngest screening age group (47) and the oldest age group (77) remain significantly different from each other in the future years.

Furthermore, the crude BC rates between 2019 and 2022 are involved in this plot. 
This is to visually inspect if the projected rates in those years may fall into the related 95\% credible intervals. 
We note, once again, that the COVID-19 pandemic is considered to have an impact on cancer mortality from 2020 onwards due to, e.g. temporary halts in cancer treatment \citep{CRUKCIT2021}. 
It is possible to observe some increases in BC mortality at certain ages, e.g. 77, in those years in certain regions, e.g. the north east of England, as compared to other regions, such as the south east of England.

\begin{figure}[htbp]
	\centering
	\includegraphics[width=0.6\textwidth, angle =0]{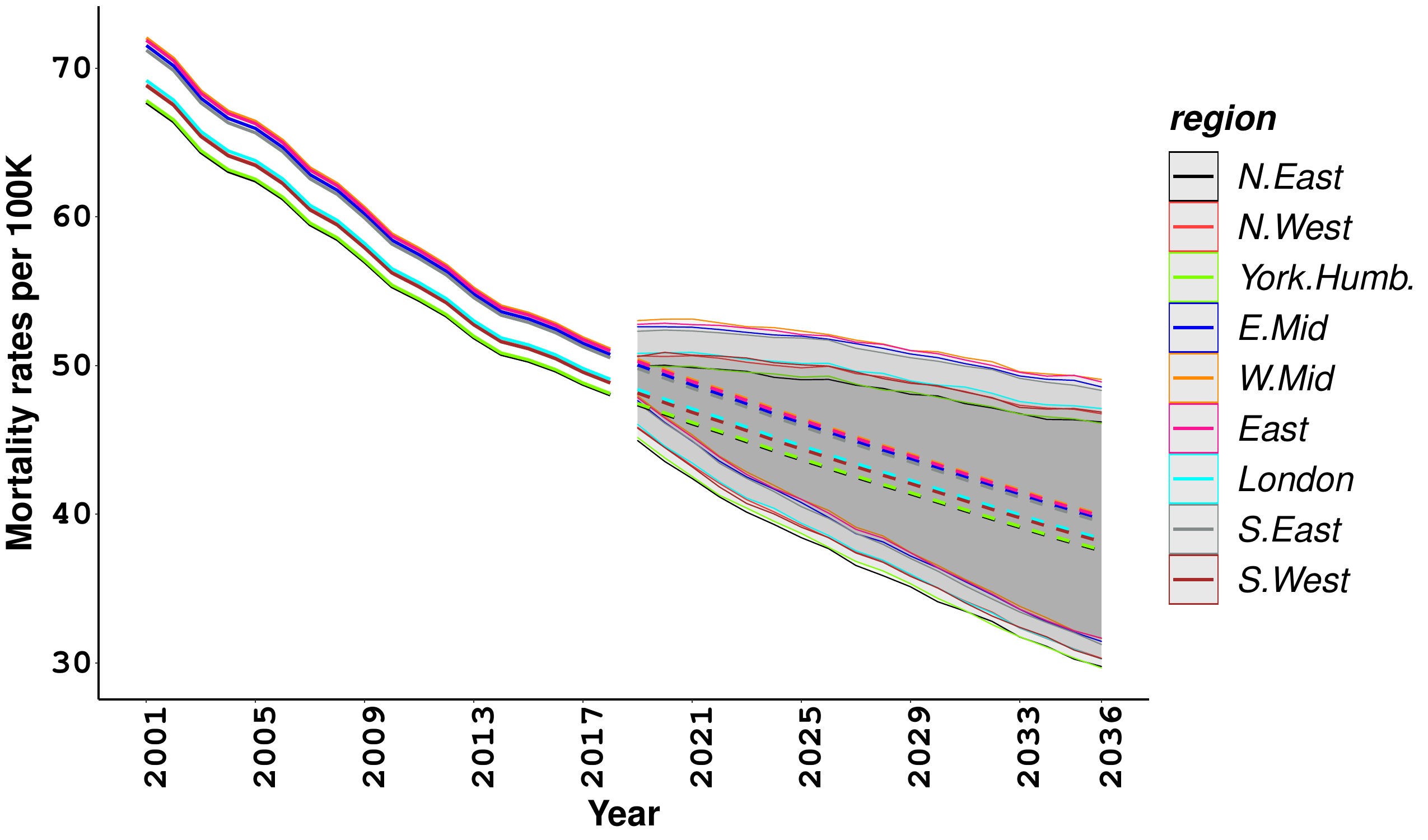}
	\caption{Age-standardised fitted and projected breast cancer mortality rates, females, in regions of England: fitted rates (solid lines), projected rates (dashed lines) with 95\% credible intervals for the projected rates.}
	\label{fig:BreastCOD_ASR}
\end{figure}

\figref{fig:BreastCOD_ASR} presents age-standardised fitted and projected BC mortality rates, with 95\% credible intervals, from 2001 to 2036. 
The figure conveys two main messages.
First, there is a declining trend in all regions over the calendar years, which is also expected to continue in the future years. 
Second, although region is a statistically significant risk factor to explain differences in BC mortality, 
only marginal differences are expected to occur across the regions of England in the projection period.

\subsection{Disparities in cancer mortality}

We use a national-level socio-economic variable, i.e. income deprivation, which allows us to make comparisons across different deprivation quintiles in a given region and also across different regions of England. 
The presence of varying socio-economic gap across different regions are evident, for example, in LC mortality for women. 
As remarked earlier, the risk of dying from LC in a given deprivation level, especially in the most and least deprivation levels, is not the same in different regions. 
For example, a woman in the lowest income bracket in the north east of England is twice as likely to die from LC as compared to another women in a similar socio-economic status in London (\figref{fig:ASRLungCOD_FeMales_v2}). 
Certainly, our findings associated with London are linked to a more complex underlying reason, which is not addressed as part of this study, including higher ethnic diversity \citep{Newtonetal2015}, healthy migrant effect, and bigger health investments in London compared to other English regions \citep{ONS2012}.

We consider a relative deprivation measure, $\text{RD}_{r, t}$, to quantify socio-economic variations in LC mortality for men and women across different regions as follows:
\[
	\text{RD}_{ r, t} = 	\frac{\hat{\text{ASR}}_{\text{quintile 1}, r, t} - \hat{\text{ASR}}_{\text{quintile 5}, r, t} }{\hat{\text{ASR}}_{\text{quintile 1}, r, t}},
\label{eq:RDgap}
\]
where fitted age-standardised mortality rates, $\hat{\text{ASR}}_{d, r, t}$, are used as an input. 
This measure examines the absolute gap across the most and least deprived quintiles in a given region and year, with respect to the most deprived quintile in the same region. 

\begin{figure}[htbp]
	\centering
	\subfloat[women \label{fig:LCFemaleDepGap}]{\includegraphics[width=0.5\textwidth]{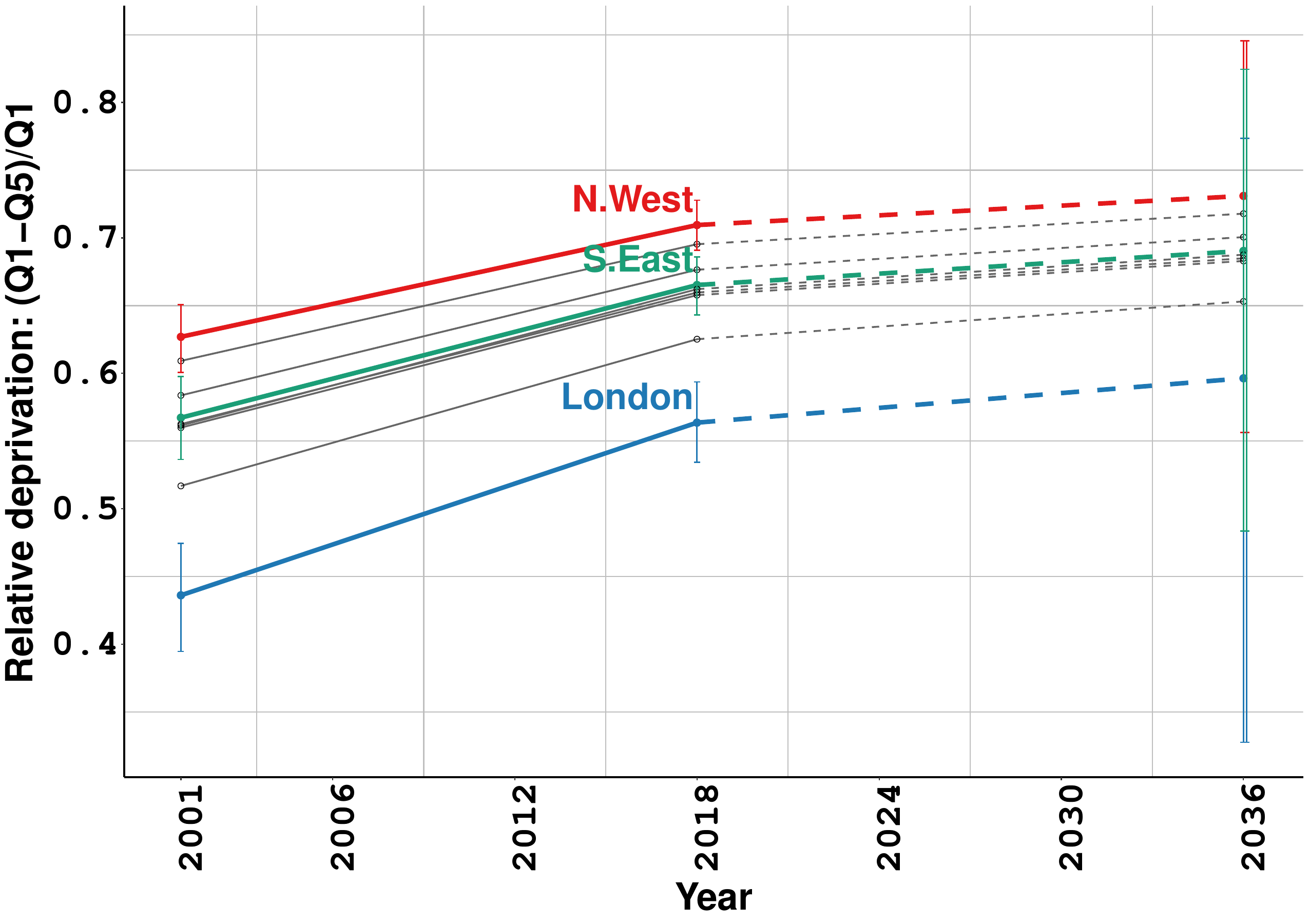}}
	\hfill
	\subfloat[men  \label{fig:LCMaleDepGap}]{\includegraphics[width=0.5\textwidth]{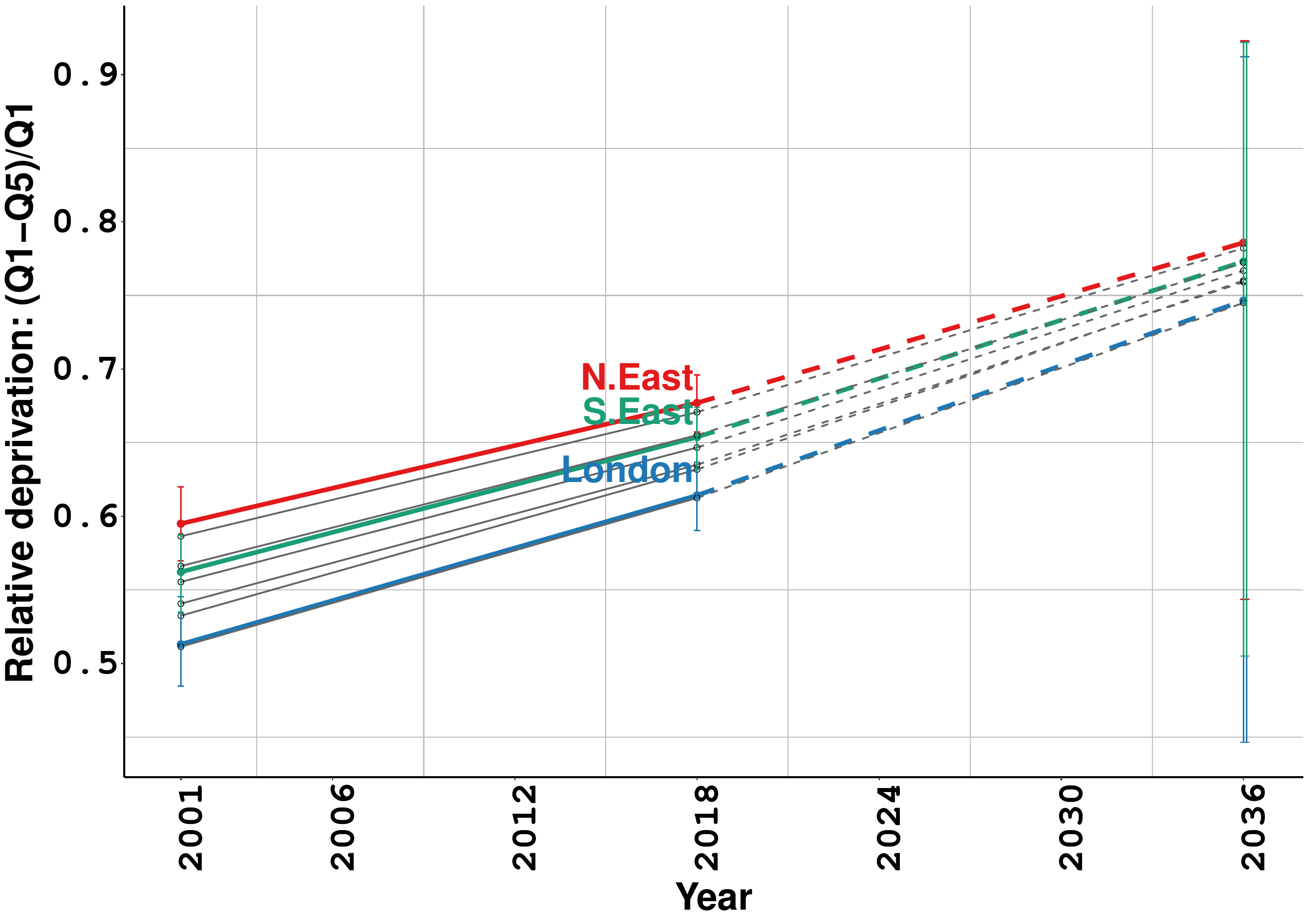}}
	\caption{Fitted (solid lines) and projected (dashed lines) relative deprivation estimates, $\text{RD}_{r, t}$, in comparison to the most deprived quintile, in lung cancer mortality by gender in 2001, 2018, and 2036 for the regions of England, with 95\% credible intervals.} 	
	\label{fig:DeprivationGapLC}
\end{figure}

The left-hand plot in \figref{fig:DeprivationGapLC} shows a wider deprivation gap in LC mortality among women compared to their male counterparts in the right-hand plot throughout the inspected period. 
Both plots display an increasing trend in the relative deprivation gap from 2001 to 2036. 
However, this trend seems to slow down for women after 2018 compared to the earlier estimates.
Notably, in the last observed calendar year, there is a significant discrepancy in the relative deprivation gap for women across the north west, the south east of England, and London, while other regions show comparable estimates. 
Meanwhile, for men, substantial differences are estimated between the north east of England and London, with similar outputs in other regions, in the same year. 
Last, our projection results point out comparable deprivation gap for both genders across different regions by 2036. 

%% To reduce spacing between figure and text
%\setlength\belowcaptionskip{-3ex}

\begin{figure}[htbp]
	\centering
{\includegraphics[width=0.6\textwidth]{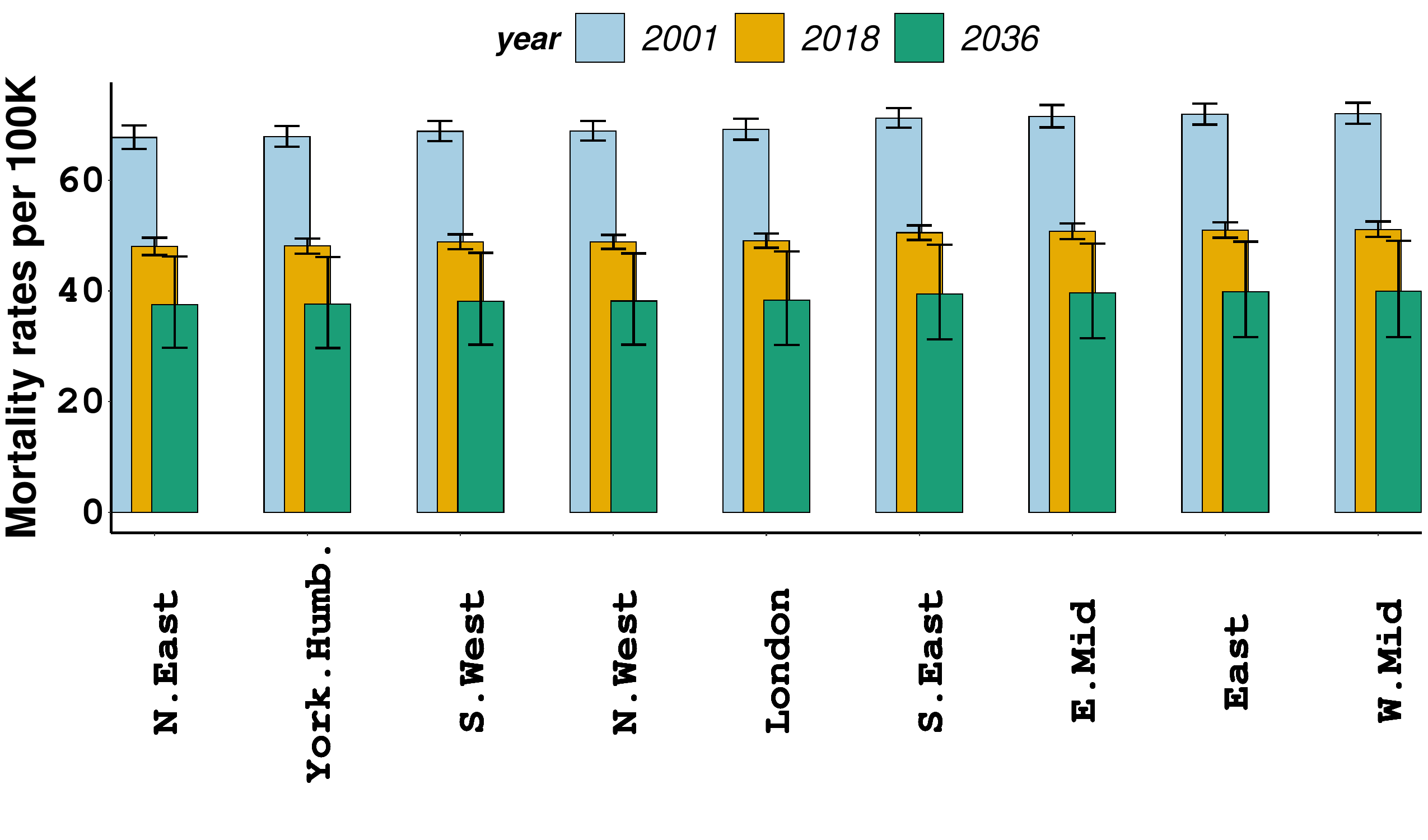}}
	\caption{Age-standardised breast cancer mortality estimates in 2001, 2018, and 2036 for the regions of England, with 95\% credible intervals.} 	
	\label{fig:RegionGapBC}
\end{figure}

\figref{fig:RegionGapBC} demonstrates that a notable improvement has been estimated in BC mortality from 2001 to 2018. 
A further significant improvement is predicted to happen up to 2036, with comparable results across all regions of England . 

\newpage
\subsection{Excess deaths}

\subsubsection{Excess deaths during the COVID-19 years}

We assess cumulative excess deaths from LC and BC by the regions of England in the most recent calendar years (2020--2022). 
Cumulative excess deaths are defined as the number of deaths exceeding the expected deaths in a calendar year, 
which are predicted using the pre-pandemic cancer deaths from 2001 to 2018 (Section~\ref{sec:PandemicScenarioModelAssumptions}). 
\tabref{tab:ExcessDeathsLCWomen} and \tabref{tab:ExcessDeathsLCMen} present registered and expected LC deaths in women and men, together with excess deaths, respectively. 
Likewise, \tabref{tab:ExcessDeathsBC} provides a similar summary for BC deaths. 

% Tue Mar 26 18:34:00 2024
\begin{table}[H]
	\centering
\caption{Summary of cumulative female lung cancer deaths in the regions of England from 2020 to 2022 }\label{tab:ExcessDeathsLCWomen}
\resizebox{0.75\columnwidth}{!}{%
\begin{tabular}{lrrrr}
  \hline
 &  \begin{tabular}[c]{@{}l@{}}Registered \\ deaths\end{tabular} &  \begin{tabular}[c]{@{}l@{}}Expected \\ deaths\end{tabular} &  \begin{tabular}[c]{@{}l@{}}Excess \\ cancer deaths\end{tabular}&  \begin{tabular}[c]{@{}l@{}}Ratio: \\ registered/expected \end{tabular} \\ 
  \hline
England & 35250 & 36511.51 & -1261.51 & 0.97 \\ 
North East& 2644  & 2909.99 & -265.99 & 0.91 \\ 
 North West & 6119  & 6334.92 & -215.92 & 0.97 \\ 
Yorkshire and the Humber & 4349 & 4417.48 & -68.48 & 0.98 \\ 
East Midlands & 3270 & 3136.31 & 133.69 & 1.04 \\ 
 West Midlands  & 3702  & 3633.07 & 68.93 & 1.02 \\ 
  East of England & 3564  & 3648.12 & -84.12 & 0.98 \\ 
  London & 3057  & 3840.38 & -783.38 & 0.80 \\ 
 South East  & 5050 & 5175.52 & -125.52 & 0.98 \\ 
  South West & 3495 & 3415.71 & 79.29 & 1.02 \\ 
   \hline
\end{tabular}
}
\end{table}

Both \tabref{tab:ExcessDeathsLCWomen} and  \tabref{tab:ExcessDeathsLCMen} show comparable results, with higher LC deaths in men in each region of England. 
We further estimate  a decrease of 3\% for women and 6\% for men in LC deaths in England between 2020 and 2022 as compared to the observed number of deaths in the same period. 
This is considered to be related to the COVID-19 pandemic as \citet{BakerandMansfield2023} state that, in 2020 and 2021, `a larger number of people died than usual (due to the Covid-19 pandemic)\ldots' (p. 12).
Nonetheless, unlike men, there is a 2--4\% higher rate of LC deaths in women than expected in three regions: the East and West Midlands, and the south west of England. 

\begin{table}[H]
	\centering
\caption{Summary of cumulative male lung cancer deaths in the regions of England from 2020 to 2022 }\label{tab:ExcessDeathsLCMen}
\resizebox{0.75\columnwidth}{!}{%
	\begin{tabular}{lrrrr}
  \hline
 &  \begin{tabular}[c]{@{}l@{}}Registered \\ deaths\end{tabular} &  \begin{tabular}[c]{@{}l@{}}Expected \\ deaths\end{tabular} &  \begin{tabular}[c]{@{}l@{}}Excess \\ cancer deaths\end{tabular}&  \begin{tabular}[c]{@{}l@{}}Ratio: \\ registered/expected \end{tabular} \\ 
  \hline
England  & 39652 & 42235.37 & -2583.37 & 0.94 \\ 
  North East & 2663 & 2908.62 & -245.62 & 0.92 \\ 
  North West & 6345 & 6783.03 & -438.03 & 0.94 \\ 
Yorkshire and the Humber& 4552  & 4809.02 & -257.02 & 0.95 \\ 
  East Mid & 3732  & 3810.11 & -78.11 & 0.98 \\ 
  West Mid & 4270 & 4681.48 & -411.48 & 0.91 \\ 
  East of England & 4201 & 4388.18 & -187.18 & 0.96 \\ 
  London & 3808 & 4529.62 & -721.62 & 0.84 \\ 
  South East & 5893 & 6094.42 & -201.42 & 0.97 \\ 
  South West & 4188 & 4230.89 & -42.89 & 0.99 \\ 
   \hline
\end{tabular}
}
\end{table}

\tabref{tab:ExcessDeathsBC} presents 2\% decrease in BC deaths in England compared to the expected numbers during 2020--2022. 
A closer examination reveals that three regions of England, which are the north east of England, the East and West Midlands, have witnessed an increase in BC deaths (1--4\%) compared to what was expected.

\begin{table}[H]
	\centering
	\caption{Summary of cumulative breast cancer deaths in the regions of England from 2020 to 2022 }\label{tab:ExcessDeathsBC}
\resizebox{0.75\columnwidth}{!}{%
\begin{tabular}{lrrrr}
  \hline
 &  \begin{tabular}[c]{@{}l@{}}Registered \\ deaths\end{tabular} &  \begin{tabular}[c]{@{}l@{}}Expected \\ deaths\end{tabular} &  \begin{tabular}[c]{@{}l@{}}Excess \\ cancer deaths\end{tabular}&  \begin{tabular}[c]{@{}l@{}}Ratio: \\ registered/expected \end{tabular} \\ 
  \hline
England & 24330 & 24861.25 & -531.25 & 0.98  \\ 
North East & 1265 & 1216.00 & 49.00 & 1.04 \\ 
North West & 3083 & 3213.86 & -130.86 & 0.96  \\ 
Yorkshire and the Humber & 2365 & 2378.59 & -13.59 & 0.99   \\ 
 East Midlands & 2297 &  2263.78 & 33.22 & 1.01 \\ 
 West Midlands & 2694 & 2690.02 & 3.98 & 1.00 \\ 
  East of England& 2835 & 2990.98 & -155.98 & 0.95\\ 
  London & 2795 & 2951.61 & -156.61 & 0.95\\ 
  South East & 4262 & 4344.02 & -82.02 & 0.98  \\ 
  South West & 2734 & 2812.37 & -78.37 & 0.97 \\ 
   \hline
\end{tabular}
}
\end{table}

\subsubsection{Impact of diagnosis delays: lung cancer}

We further investigate excess LC deaths among both women and men resulting from 
delays in diagnosis ranging from 1 to 6 months, which may occur due to significant disruptions in health services (e.g. see \citet{CRUK2022}).

First, we have predicted 2,340 (1,743; 2,869) and 10,180 (7,944; 12,340) cumulative excess LC deaths for women in England due to 1-month and 6-month diagnosis delays, respectively, over 17 years, from 2020 until 2036 based on \eqref{eq:FemaleLungCODLocationPrmtr2}. 
Meanwhile, in the same period, 1-month delay in diagnosis is estimated to cause 5,164 (4,353 to 6,066) cumulative excess LC deaths for men under \eqref{eq:MaleLungCODLocationPrmtr2}, with 28,660 (23,040 to 35,090) deaths due to 6-month delay in diagnosis. 

Second, cumulative excess deaths are plotted in \figref{fig:ExcessDeathsFemSmokRegionDep} for females.
Result for males are similar (see Appendix~\ref{sec:ExcessLCMen}). 
The two plots show the cumulative excess deaths in women for the regions and deprivation quintiles of England. 
The left-hand plot in \figref{fig:ExcessDeathsFemSmokRegionDep} shows considerable differences in excess deaths  (a) as a result of 1- to 6-month diagnosis delays in a given region; and (b) by region, such as the south east vs. the north east of England. 
The right-hand plot additionally shows marked differences between the most and least deprived quintiles as a result of any delay in diagnosis.

\begin{figure}[htbp]
	\centering
	\subfloat[ Excess deaths by region \label{fig:ExcessDeathsFemSmokRegion}]{\includegraphics[width=0.5\textwidth]{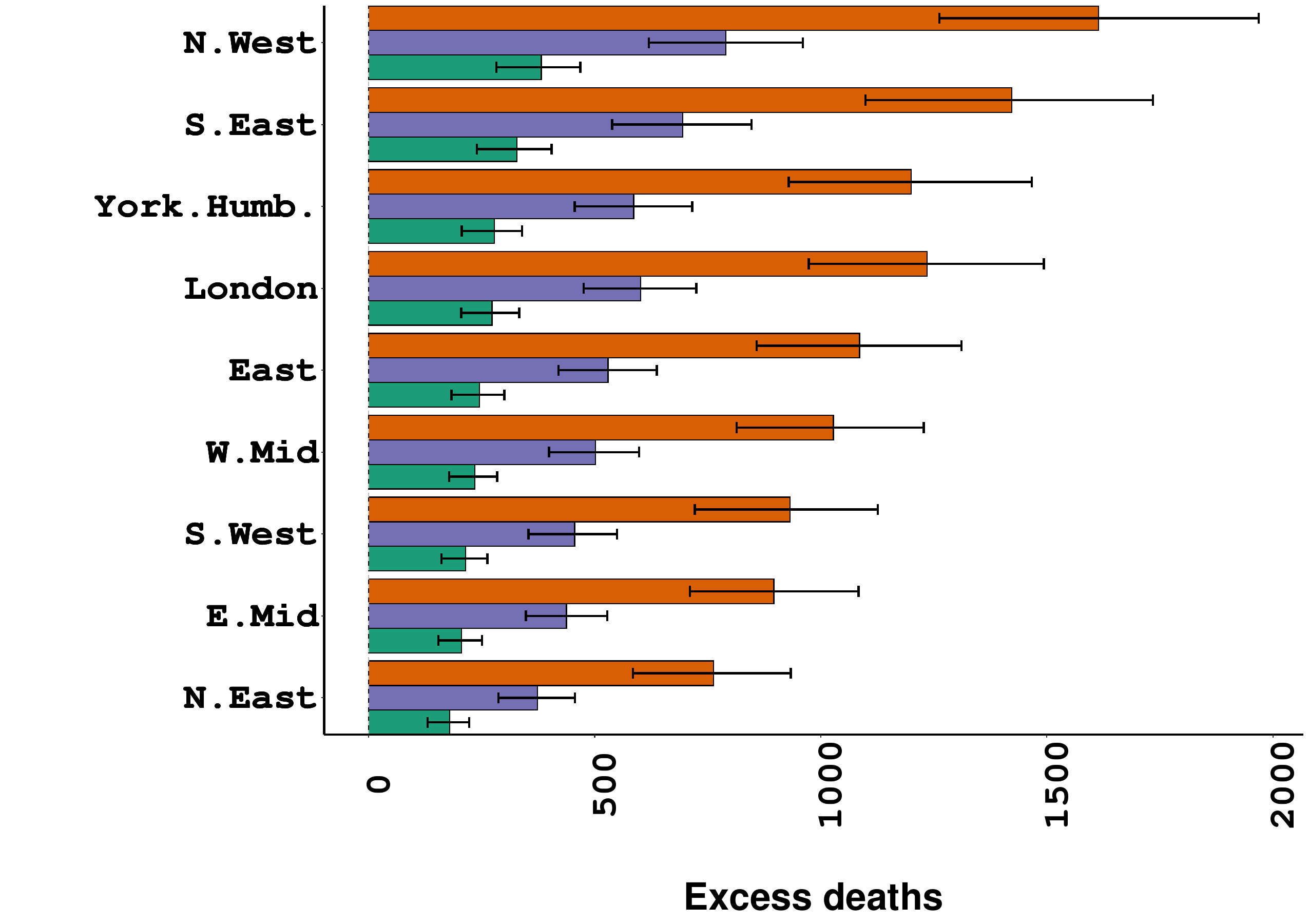}}
	\hfill
	\subfloat[ Excess deaths by deprivation quintiles \label{fig:ExcessDeathsFemSmokDepriv}]{\includegraphics[width=0.5\textwidth]{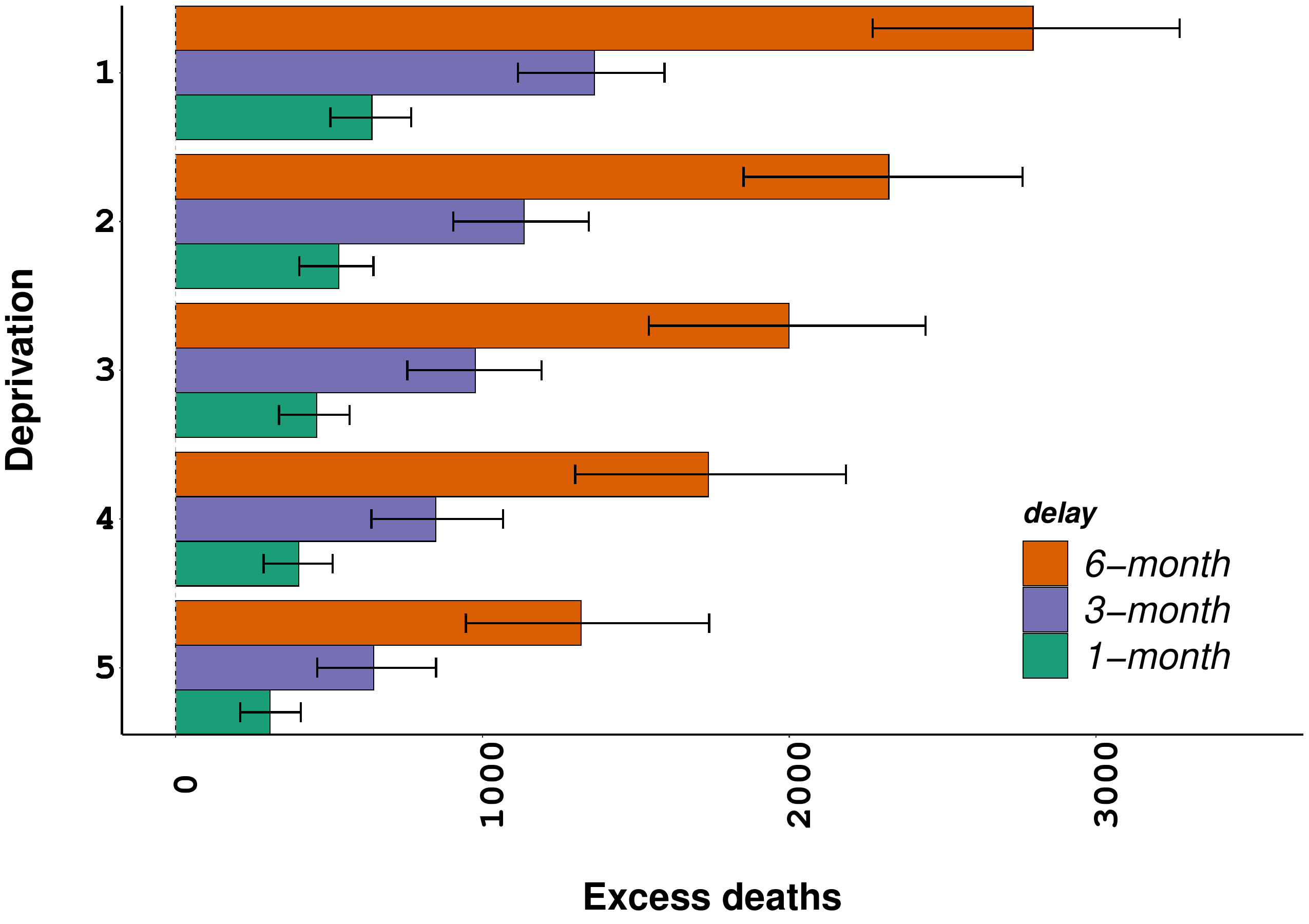}}
	\caption{Total lung cancer excess deaths in women, $\text{ED}^{\text{lung}}_{\text{women},d}$ and $\text{ED}^{\text{lung}}_{\text{women},r}$, respectively, in different deprivation quintiles and regions of England from 2020 to 2036, over 17 years, with 95\% credible intervals, based on \eqref{eq:FemaleLungCODLocationPrmtr2}.} 	
	\label{fig:ExcessDeathsFemSmokRegionDep}
\end{figure}

\begin{figure}[htbp]
	\centering
	\subfloat[1-month delay \label{fig:ExcessRate1monthFemSmok}]{\includegraphics[width=0.5\textwidth]{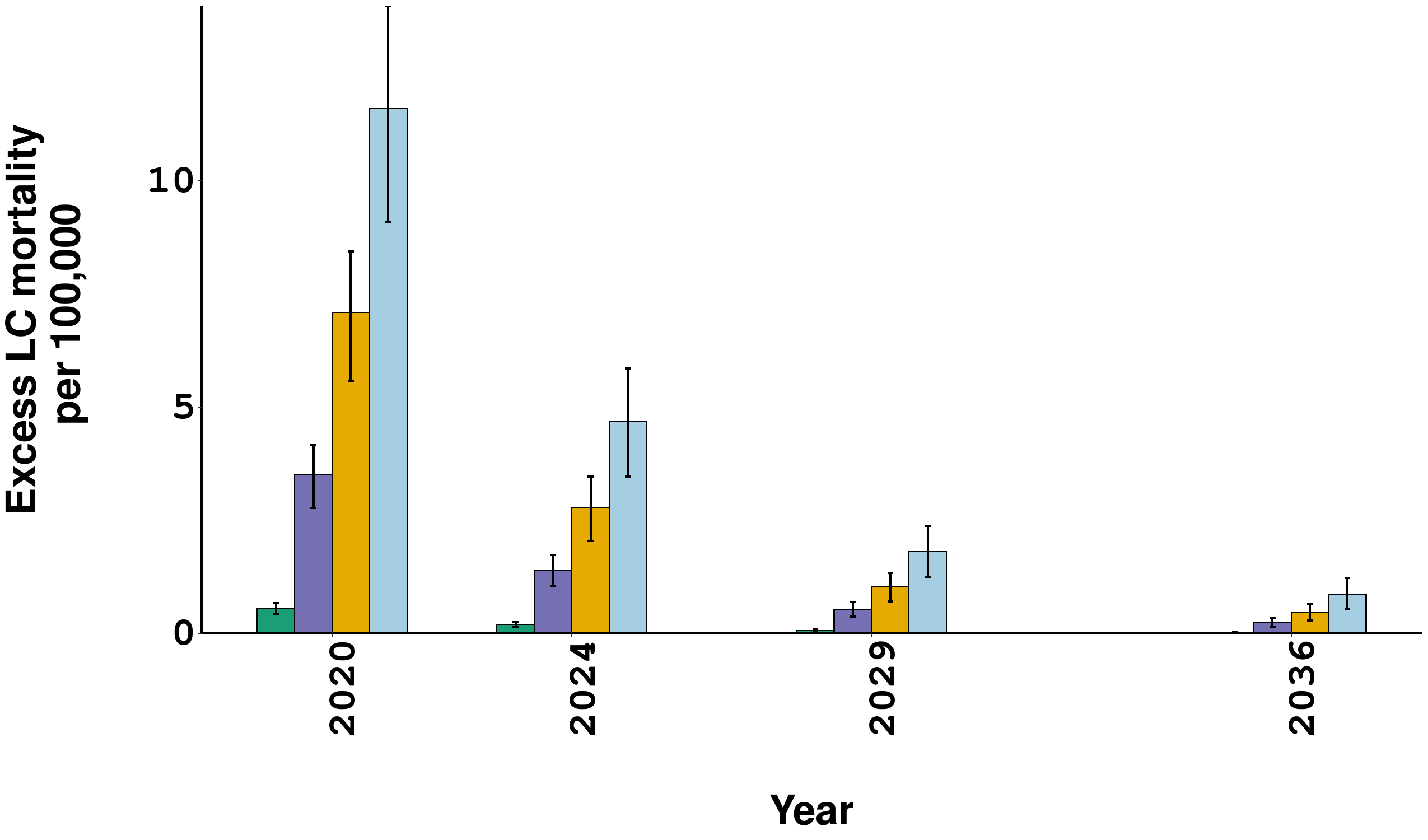}}
	\hfill
	\subfloat[ 6-month delay \label{fig:ExcessRate1monthFemSmokv2}]{\includegraphics[width=0.5\textwidth]{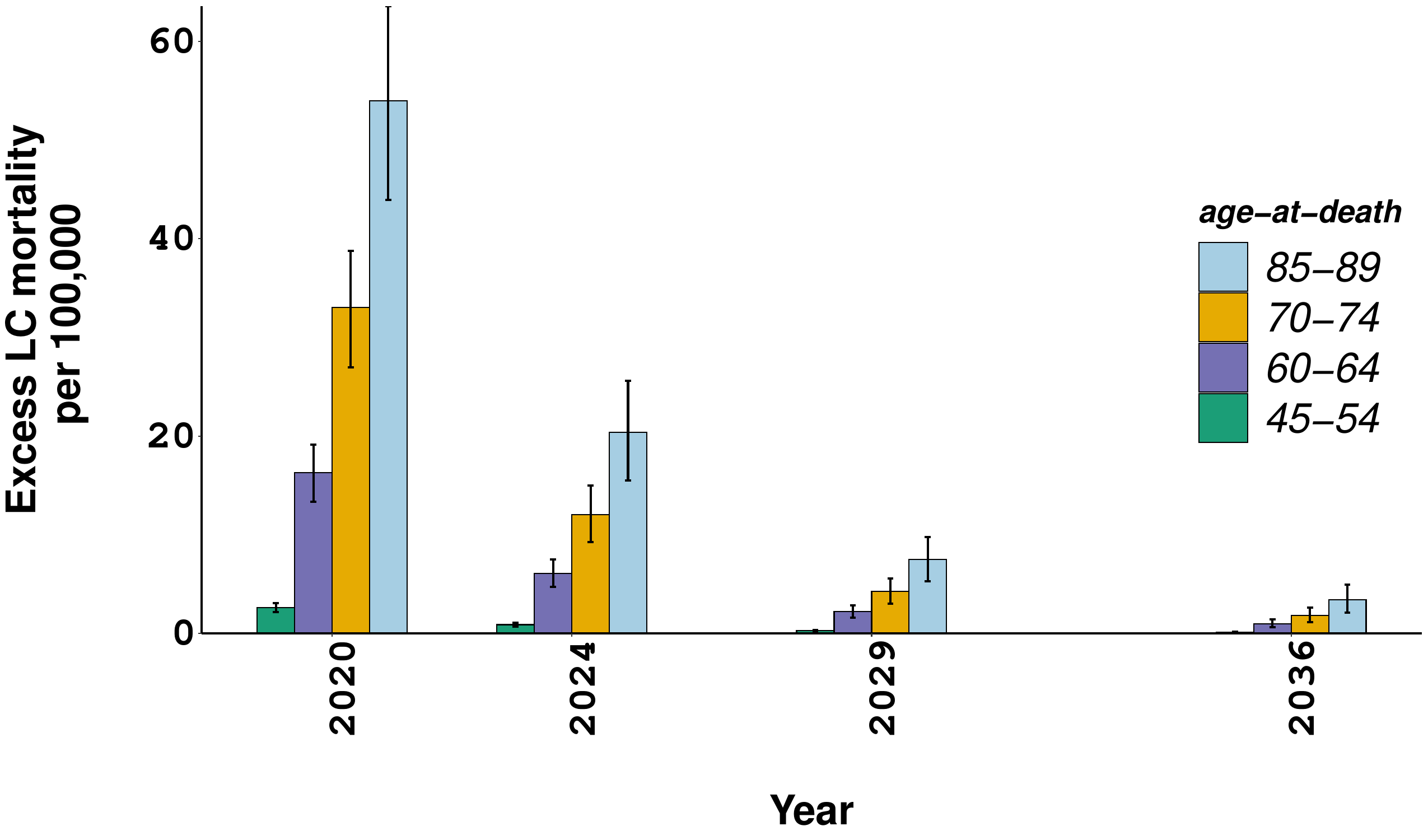}}
	\caption{Lung cancer excess mortality, per 100,000 women, $\text{EAM}^{\text{lung}}_{a, \text{women},t}$, by age-at-death in England from 2020 to 2036, with 95\% credible intervals, based on \eqref{eq:FemaleLungCODLocationPrmtr2}. Note that differences in lung cancer excess mortality at other ages in a given year, and differences in intermediate years, are negligible.} 	
	\label{fig:ExcessRateFemAgeSmok}
\end{figure}

\begin{figure}[htbp]
	\centering
	\subfloat[1-month delay \label{fig:ExcessRate1monthFemRegSmok}]{\includegraphics[width=0.5\textwidth]{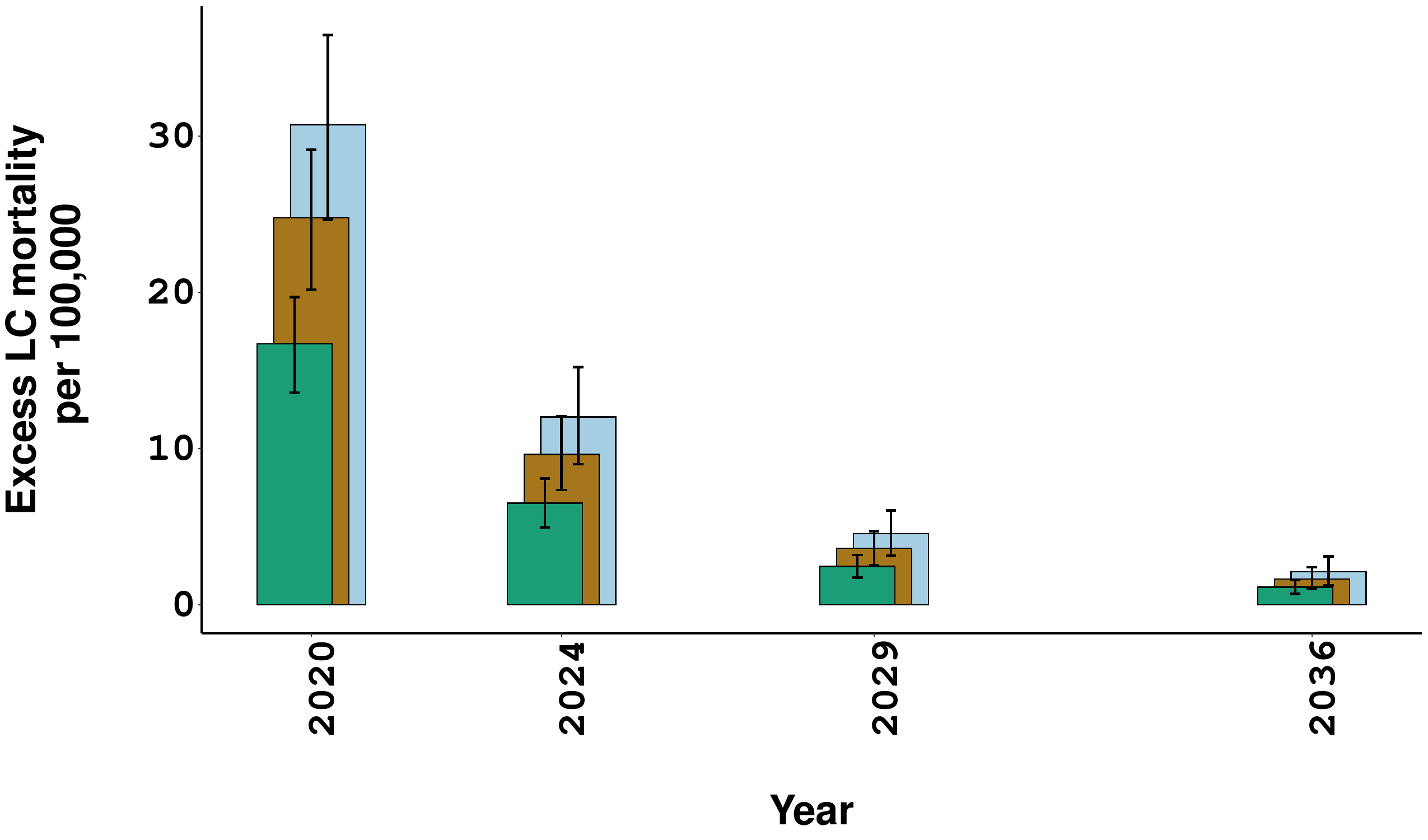}}
	\hfill
	\subfloat[ 6-month delay \label{fig:ExcessRate6monthFemReg2Smok}]{\includegraphics[width=0.5\textwidth]{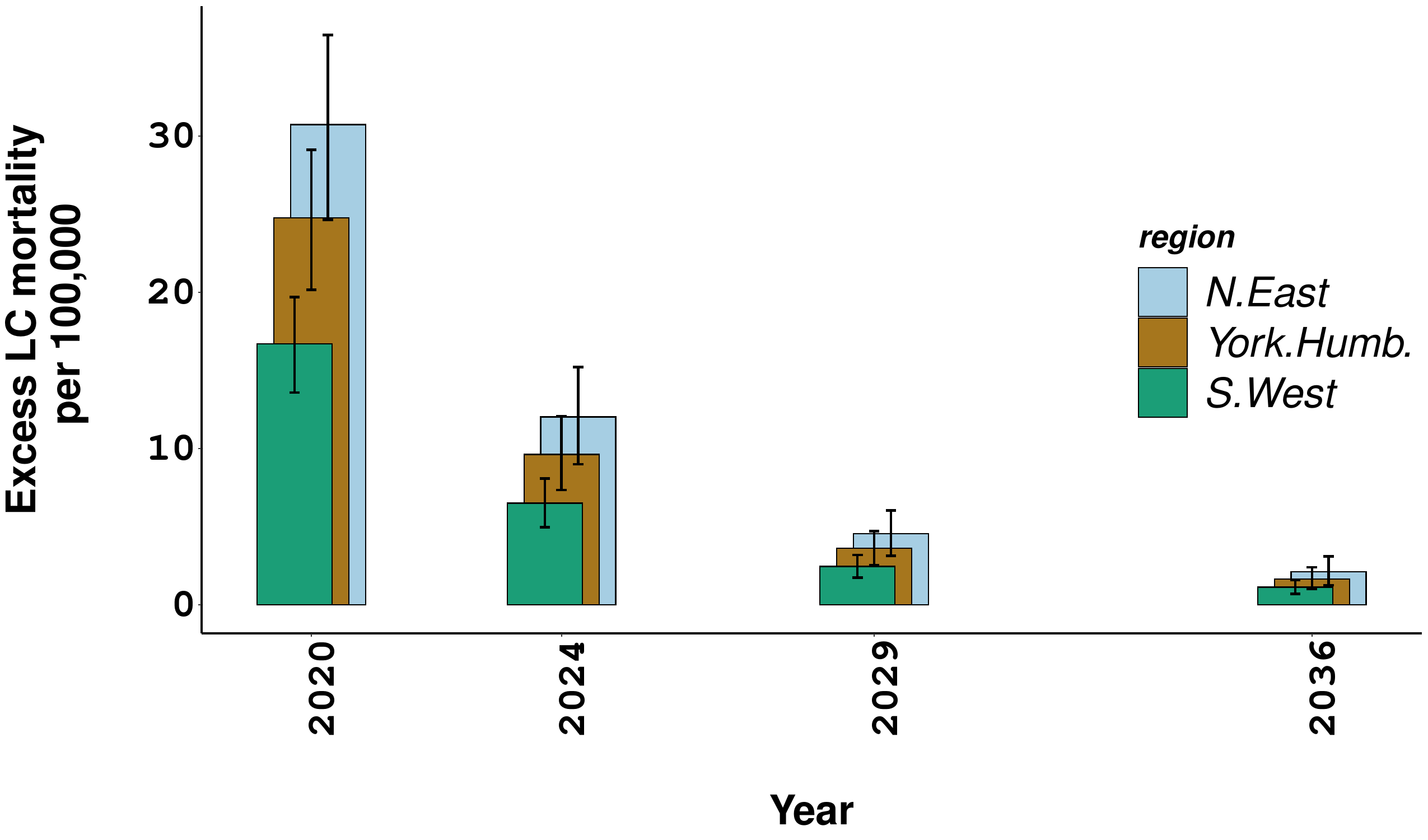}}
	\caption{Lung cancer excess mortality, per 100,000 women, $\text{ERM}^{\text{lung}}_{\text{women},r, t}$, by selected regions of England from 2020 to 2036, with 95\% credible intervals, based on \eqref{eq:FemaleLungCODLocationPrmtr2}. Note that differences in lung cancer excess mortality in other regions in a given year, and differences in intermediate years, are negligible. } 	
	\label{fig:ExcessRateFemRegionSmok}
\end{figure}

\begin{figure}[htbp]
	\centering
	\subfloat[1-month delay \label{fig:ExcessRate1monthFemDepSmok}]{\includegraphics[width=0.5\textwidth]{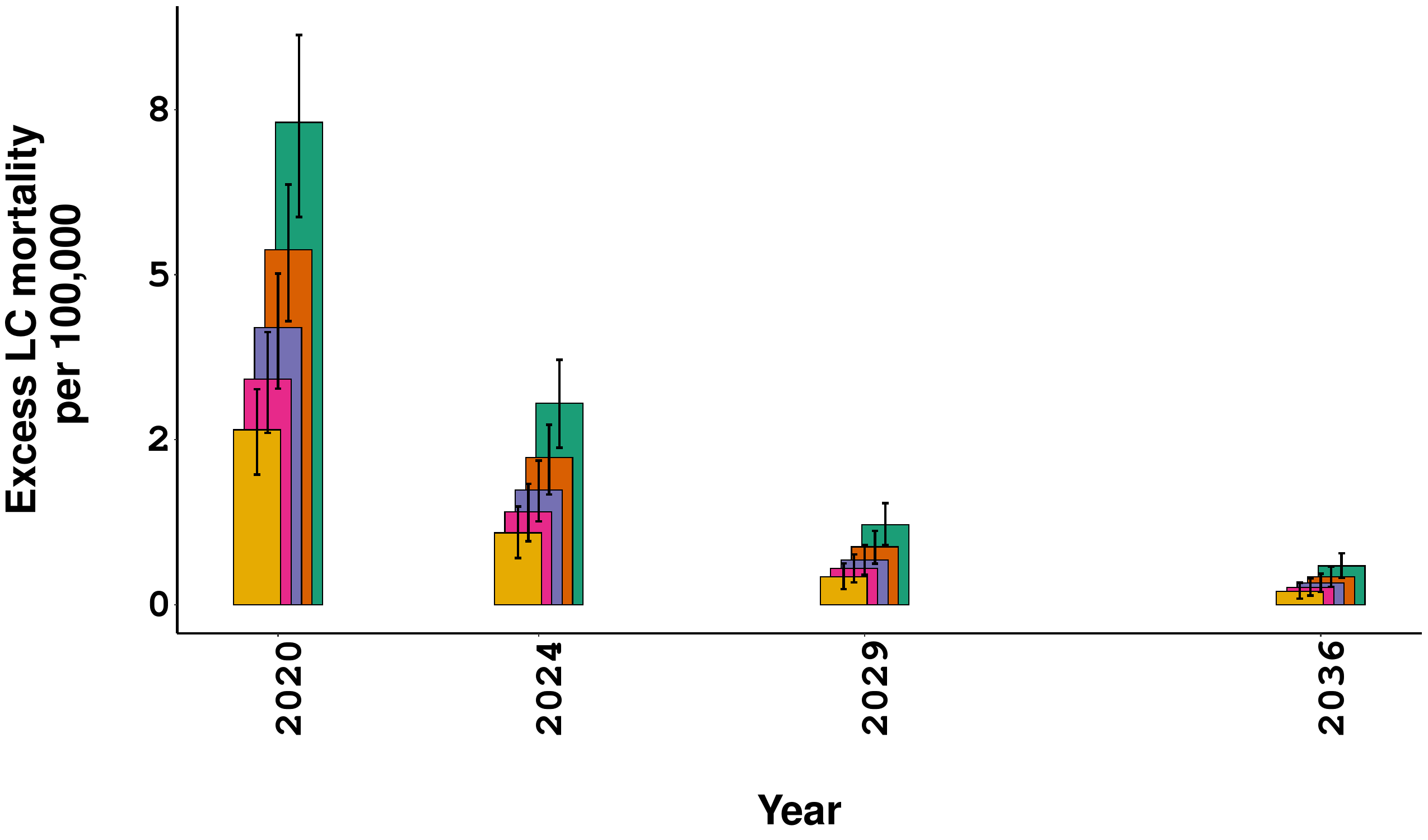}}
	\hfill
	\subfloat[ 6-month delay \label{fig:ExcessRate6monthFemDep2Smok}]{\includegraphics[width=0.5\textwidth]{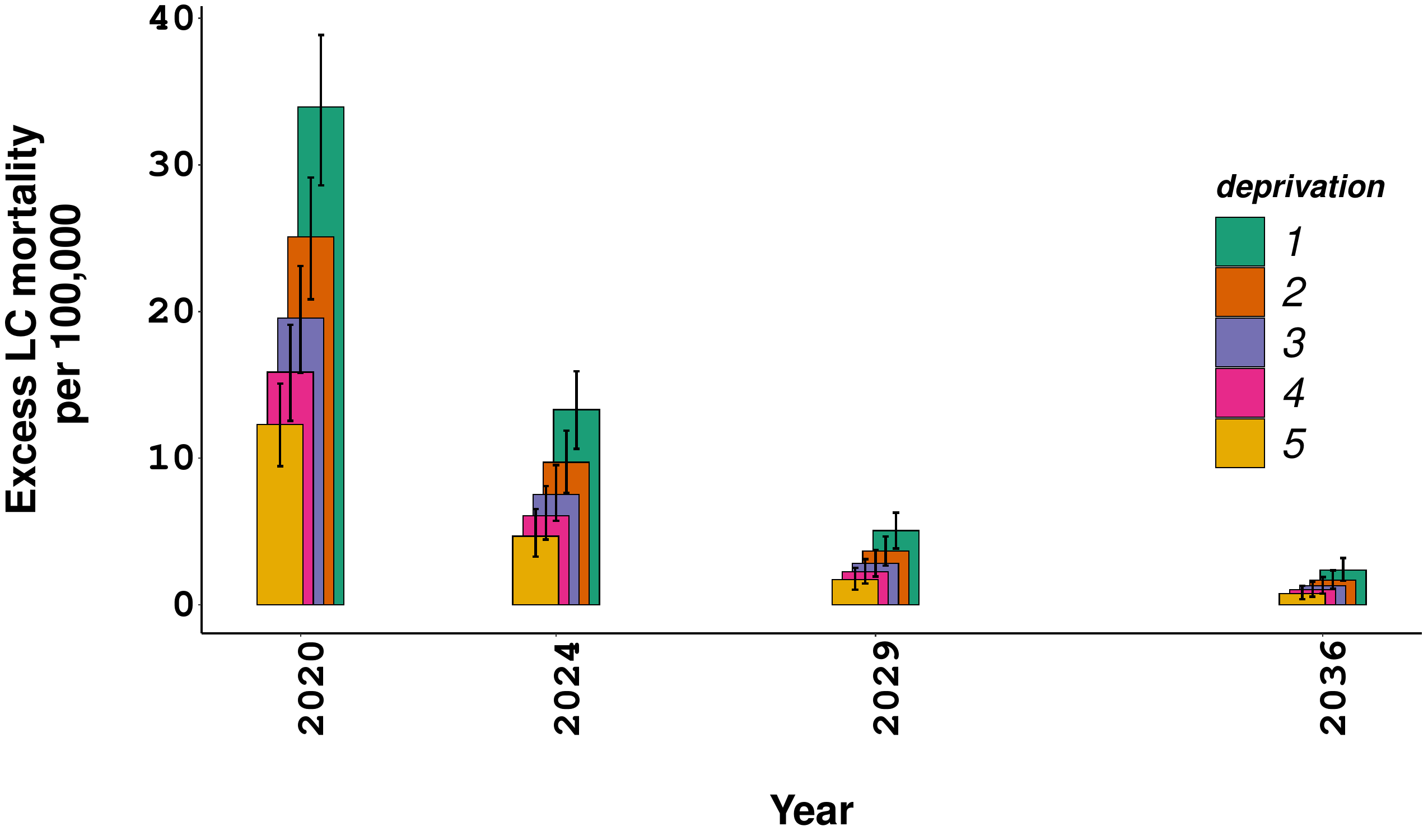}}
	\caption{Lung cancer excess mortality, per 100,000 women, $\text{EDM}^{\text{lung}}_{\text{women},d, t}$, by deprivation quintiles in England from 2020 to 2036, with 95\% credible intervals, based on \eqref{eq:FemaleLungCODLocationPrmtr2}. Note that differences in lung cancer excess mortality by deprivation in intermediate years are negligible.} 	
	\label{fig:ExcessRateFemDepSmok}
\end{figure}

Third, LC excess mortality, per 100,000 women, are presented by age (\figref{fig:ExcessRateFemAgeSmok}), region (\figref{fig:ExcessRateFemRegionSmok}) or deprivation quintiles (\figref{fig:ExcessRateFemDepSmok}) in single projection years. 
In each figure, the left-hand plot shows the relevant results for a 1-month diagnosis delay whereas the right-hand plot demonstrates the results associated with a 6-month diagnosis delay. 
Although excess mortality rates for men are significantly higher than women, 
the distribution of these rates by age, region, or deprivation is comparable to the female counterparts (Figures~\ref{fig:ExcessRateMaleAgeSmok}--\ref{fig:ExcessRateMenDepSmok}).

\figref{fig:ExcessRateFemAgeSmok} points out markedly different excess mortality, per 100,000 women, across the middle, e.g. 45--54, and old age groups, e.g. 70--74. 

\figref{fig:ExcessRateFemRegionSmok} shows the lowest excess mortality in the south west of England in 2020, with significantly higher excess mortality in the north east.  

Last, \figref{fig:ExcessRateFemDepSmok} points out substantial differences, up to a factor of 4, across the most and least deprived quintiles as a result of a given diagnosis delay.

\newpage
\section{Conclusion} \label{Sec:Conclusion}

Identifying future trends in cancer mortality is an important topic to inform 

\begin{itemize}
	\item policy makers for targeted health initiatives by outlining expected socio-economic and regional variations in cancer risk in the future years; and
	\item life insurers by providing a better understanding of this risk for modelling and pricing of CII-type contracts. 
\end{itemize}

This paper has focussed on two primary cancer types: LC and BC mortality. 
We have utilised the population data of England at regional and deprivation levels between 2001 and 2018.
Accordingly, we have carried out a detailed analysis to understand how inequalities in cancer mortality are expected to change in the future. 
Later, we have quantified excess cancer deaths based on the difference between the observed and expected number deaths by region between 2020 and 2022. 
Additionally, we have explored scenarios concerning LC modelling that pertain to delays in cancer diagnosis, aiming to assess its effect on LC mortality. 
We have used a Bayesian hierarchical modelling which is well suited to large heterogeneous datasets and can handle parameter uncertainty and inclusion of several predictive variables, involving non-linear relationships between these variables. 

Our findings confirm that both LC and BC mortality vary by age, gender, and region over the time. 
Also, socio-economic differences are confirmed to be relevant to LC mortality based on a neighbourhood-level variable (i.e. income deprivation) (Figures~\ref{fig:ASRLungCOD_FeMales_v2}--\ref{fig:BreastCOD_ASR}). 
We found that NS prevalence, used as a proxy for smoking, is significant to explain differences in both cancer types. 
Meanwhile, AAD variable, used as a proxy for age-at-diagnosis, is found to be significant for LC mortality by allowing us to establish future scenarios relevant to an extreme event, e.g. diagnosis delays due to unprecedented health disruptions.

We found evidence suggesting socio-economic differences in LC mortality for both genders have increased, with notable differences across the regions of England in 2018, and are expected to continue similar increasing trend up to 2036 (\figref{fig:DeprivationGapLC}). 
We also found that delays in cancer diagnosis lead to significantly higher deaths from LC, varying markedly from delays of 1 to 6 months.  
Specifically,  for a 1-month diagnosis delay, our models have estimated 2,340 (95\% CI 1,743 to 2,869) cumulative excess deaths for women and 5,164 (4,353 to 6,066) for men from 2020 to 2036. When a 6-month delay is considered, our models suggest 10,180 (7,944 to 12,340) and 28,660 (23,040 to 35,090) excess deaths for women and men respectively.
Importantly, the related LC excess deaths were obtained to be substantially 
\begin{itemize}
	\item higher at older age groups including 60--64 years old; 
	
	\item higher in the northern regions of England compared to the southern regions; and
	
	\item higher for those living in the most deprived quintiles compared to those in the least deprived quintiles.
\end{itemize}

The modelling results are broadly consistent with the existing literature (e.g. \citet{Luoetal2022, NHSDigital2023Cancer}). 
For example, recent data from NHS Digital reveals that the age-standardised cancer mortality in England was highest for those living in the most deprived areas in 2020, in addition to an increasing deprivation gap compared to 2019 \citep{NHSDigital2023Cancer}. 
Specifically, the age-standardised all-cause cancer mortality in the most deprived quintiles was reported to be 53\% higher for men and 55\% higher for women, in contrast to those in the least deprived areas. 
This is noted to mark an increase from the 2019 cancer mortality, which showed a 49\% disparity for both genders.
Besides, LC is reported to be impacted the most in 2020, with age-standardised mortality of 103, per 100,000 people, for men in the most deprived areas, compared to 37 for those in the least deprived areas, 
and 78 for women as opposed to 26 for the ones in the least deprived areas. 
Aligned with these observations, 
our modelling approach has provided estimates of excess deaths from LC with 95\% credible intervals, pointing out highest excess mortality in the most deprived quintiles or in the northern regions of England 
under cancer diagnosis-related scenarios.

Furthermore, we conducted a comparison between the observed and projected BC mortality rates on regional level between 2019 and 2022 in order to evaluate the performance of our models to some extent. 
Despite the disruptions caused by the COVID-19 pandemic between 2020 and 2022, 
the observed age-specific BC mortality rates were largely consistent with 
the corresponding 95\% credible intervals established by our projections. 
Besides, our predictions suggested a decline in LC and BC deaths on national level during the pandemic years perhaps due to larger deaths from or with the COVID-19 pandemic \citep{BakerandMansfield2023}. 

Our results can be of use in informing decision makers to implement evidence-based health interventions so that they can tackle with inequalities in cancer risk. 
Our findings can also help life insurers to understand and assess the implications of late diagnosis on cancer mortality and survival rates. 
This is relevant to long-term insurance policies, e.g. CII insurance products, since understanding trends and changes in cancer mortality would be an important underlying assumption for price guarantees. 
This study can further add value while considering insurance pricing and valuation assumptions related to different sub-populations. 
For example, examining variations in cancer mortality among the most and least deprived population groups can provide valuable insights into potential differences between insured and general populations.

We note challenges in accessing further cancer data at both deprivation and regional levels in the most recent calendar years. 
Besides, smoking data has not been available in the same granularity as for cancer data. 
Similar challenges appeared when mid-year population estimates have been required by region and deprivation quintiles for obtaining excess deaths. 
This limited our ability to make data-driven inferences. 
Nonetheless, suitable adjustments were used when data, e.g. mid-year population estimates, are not provided to suitable resolution. 
Thus, our analysis is mostly based on data and estimation.

% \paragraph{Acknowledgement.} 
\mbox{}

\paragraph{\textbf{Funding Statement.}}
\mbox{}

GS received financial support by the Society of Actuaries under a research project on `Predictive modelling for medical morbidity trends related to insurance'. 
The funders did not play any role in the study design, data collection and analysis, decision to publish, or preparation of the manuscript. 
All the views presented in this paper are of the authors only.

\paragraph{\textbf{Competing Interests.}} 
\mbox{}

None.

%----------------------------------------------------------------------------------------
%	BIBLIOGRAPHY
%----------------------------------------------------------------------------------------

\bibliographystyle{aaslike.bst}
\bibliography{Bibliography.bib}

\begin{thebibliography}{}

\bibitem[Al-Delaimy et~al., 2004]{Delaimyetal2004}
{\bf Al-Delaimy, W.}, {\bf Cho, E.}, {\bf Chen, W.}, {\bf Colditz, G.},
  \textbf{and} {\bf Willet, W.} (2004).
\newblock A prospective study of smoking and risk of breast cancer in young
  adult women.
\newblock {\em Cancer Epidemiol Biomarkers Prev.}, 13(3):398--404.

\bibitem[Archbold et~al., 2023]{Archboldetal2023}
{\bf Archbold, M.}, {\bf Davies, B.}, \textbf{and} {\bf Mais, D.} ({2023}).
\newblock Deprivation and the impact on smoking prevalence, {E}ngland and
  {W}ales: 2017 to 2021.
\newblock Technical report, Office for National Statistics.

\bibitem[Ar{\i}k et~al., 2023]{Ariketal2023SAJ}
{\bf Ar{\i}k, A.}, {\bf Cairns, A.}, {\bf Dodd, E.}, {\bf Macdonald, A.},
  \textbf{and} {\bf Streftaris, G.} (2023).
\newblock The effect of the covid-19 health disruptions on breast cancer
  mortality for older women: a semi-markov modelling approach.

\bibitem[Ar{\i}k et~al., 2021]{Ariketal2021}
{\bf Ar{\i}k, A.}, {\bf Dodd, E.}, {\bf Cairns, A.}, \textbf{and} {\bf
  Streftaris, G.} ({2021}).
\newblock Socioeconomic disparities in cancer incidence and mortality in
  {E}ngland and the impact of age-at-diagnosis on cancer mortality.
\newblock {\em PLoS One}, 16(7).

\bibitem[Ar{\i}k et~al., 2020]{Ariketal2020}
{\bf Ar{\i}k, A.}, {\bf Dodd, E.}, \textbf{and} {\bf Streftaris, G.} ({2020}).
\newblock Cancer morbidity trends and regional differences in {E}ngland - a
  bayesian analysis.
\newblock {\em PLoS One}, 15(5).

\bibitem[Baker and Mansfield, 2023]{BakerandMansfield2023}
{\bf Baker, C.} \textbf{and} {\bf Mansfield, Z.} (2023).
\newblock Cancer statistics for {E}ngland.

\bibitem[Bennett et~al., 2018]{Bennettetal2018}
{\bf Bennett, J.}, {\bf Pearson-Stuttard, J.}, {\bf Kontis, V.}, {\bf Capewell,
  S.}, {\bf Wolfe, I.}, \textbf{and} {\bf Ezzati, M.} ({2018}).
\newblock Contributions of diseases and injuries to widening life expectancy
  inequalities in {E}ngland from 2001 to 2016: a population-based analysis of
  vital registration data.
\newblock {\em LANCET}, 3:e586--97.

\bibitem[Brouhns et~al., 2002]{Brouhnsetal2002}
{\bf Brouhns, N.}, {\bf Denuit, M.}, \textbf{and} {\bf Vermunt, J.} ({2002}).
\newblock A poisson log-bilinear regression approach to the construction of
  projected life tables.
\newblock {\em Insurance: Mathematics and Economics}, 31:373--393.

\bibitem[Brown, 2003]{Brown2003}
{\bf Brown, J.} ({2003}).
\newblock Redistribution and insurance: Mandatory annuitization with mortality
  heterogenity.
\newblock {\em Journal of Risk and Insurance}, 70(1):17--41.

\bibitem[Brown et~al., 2007]{Brownetal2007}
{\bf Brown, S.}, {\bf Hole, D.}, \textbf{and} {\bf Cooke, T.} ({2007}).
\newblock Breast cancer incidence trends in deprived and affluent {S}cottish
  women.
\newblock {\em Breast Cancer Res Treat}, 103:233--238.

\bibitem[Chen et~al., 2020]{Chenetal2020}
{\bf Chen, N.}, {\bf Zhou, M.}, {\bf Dong, X.}, \textbf{and} {\bf et~al.}
  ({2020}).
\newblock Epidemiological and clinical characteristics of 99 cases of 2019
  coronavirus pneumonia in {W}uhan, {C}hina: a descriptive study.
\newblock {\em Lancet}, 395(10223):507--13.

\bibitem[CRUK, 2020]{CRUK2018}
{\bf CRUK} ({2020}).
\newblock Cancer in the {UK} 2020: Socio-economic deprivation.
\newblock Technical report, Cancer Research UK.

\bibitem[CRUK, 2021a]{CRUKCIT2021}
{\bf CRUK} ({2021})a.
\newblock Evidence of the impact of covid-19 across the cancer pathway: Key
  stats.
\newblock Technical report, Cancer Research UK.

\bibitem[CRUK, 2021b]{CRUK2021}
{\bf CRUK} (2021)b.
\newblock Survival for all stages of lung cancer.

\bibitem[CRUK, 2022a]{CRUKCIT2022}
{\bf CRUK} ({2022})a.
\newblock Performance measures across the cancer pathway: Key stats.
\newblock Technical report, Cancer Research UK.

\bibitem[CRUK, 2022b]{CRUK2022}
{\bf CRUK} ({2022})b.
\newblock Performance measures across the cancer pathway: Key stats.
\newblock Technical report, Cancer Research UK.

\bibitem[Czado et~al., 2005]{Czadoetal2005}
{\bf Czado, C.}, {\bf Delwarde, A.}, \textbf{and} {\bf Denuit, M.} ({2005}).
\newblock Bayesian {P}oisson log-bilinear mortality projections.
\newblock {\em Insurance: Mathematics and Economics}, 36:260--284.

\bibitem[Digital, 2020]{HealthEnglandSmoking2023}
{\bf Digital, N.} (2020).
\newblock Health survey for {E}ngland.

\bibitem[Digital, 2023]{NHSDigital2023Cancer}
{\bf Digital, N.} (2023).
\newblock Deaths from cancer increased with deprivation.

\bibitem[Eurostat, 2007]{Eurostat2007}
{\bf Eurostat} ({2007}).
\newblock Regions in the {E}uropean {U}nion: {N}omenclature of territorial
  units for statistics.
\newblock {\em Eurostat:Methodologies and Working Papers}.

\bibitem[Eurostat, 2013]{EurostatESP2013}
{\bf Eurostat} (2013).
\newblock Revision of the european standard population - report of eurostat's
  task force.

\bibitem[Grasselli et~al., 2020]{Grassellietal2020}
{\bf Grasselli, G.}, {\bf Zangrillo, A.}, {\bf Zanella, A.}, \textbf{and} {\bf
  et~al.} ({2020}).
\newblock Baseline characteristics and outcomes of 1591 patients infected with
  {SARS-CoV-2} admitted to {ICU}s of the {L}ombardy {R}egion, {I}taly.
\newblock {\em JAMA}, 323(16):1574--81.

\bibitem[Hunter et~al., 1997]{DJHankinsonetal1997}
{\bf Hunter, D.}, {\bf Hankinson, S.}, {\bf Hough, H.}, {\bf Gertig, D.},
  \textbf{and} {\bf et~al.} (1997).
\newblock A prospective study of nat2 acetylation genotype, cigarette smoking,
  and risk of breast cancer.
\newblock {\em Carcinogenesis}, 18(11):2127--32.

\bibitem[Kass and Raftery, 1995]{KassandRaftery1995}
{\bf Kass, R.} \textbf{and} {\bf Raftery, A.} ({1995}).
\newblock Bayes factors.
\newblock {\em J. Am. Statist. Ass.}, 90:773--795.

\bibitem[Kimball, 2002]{Kimball2002}
{\bf Kimball, S.} ({2002}).
\newblock Product matters.
\newblock Technical report, Society of Actuaries.

\bibitem[Lai et~al., 2020]{Laietal2020}
{\bf Lai, A.}, {\bf Pasea, L.}, {\bf Banerjee, A.}, \textbf{and} {\bf et~al.}
  ({2020}).
\newblock Estimated impact of the covid-19 pandemic on cancer services and
  excess 1-year mortality in people with cancer and multimorbidity: near
  real-time data on cancer care, cancer deaths and a population-based cohort
  study.
\newblock {\em BMJ Open}.

\bibitem[Lindgren and Rue, 2015]{LindgrenRue2015}
{\bf Lindgren, F.} \textbf{and} {\bf Rue, H.} ({2015}).
\newblock Bayesian spatial modelling with r-inla.
\newblock {\em Journal of Statistical Software}, 63(19).

\bibitem[Luo et~al., 2022]{Luoetal2022}
{\bf Luo, Q.}, {\bf O'Connell, D.}, {\bf Yu, X.}, {\bf Kahn, C.}, {\bf Caruana,
  M.}, \textbf{and} {\bf et~al.} ({2022}).
\newblock Cancer incidence and mortality in {A}ustralia from 2020 to 2044 and
  an exploratory analysis of the potential effect of treatment delays during
  the {COVID}-19 pandemic: a statistical modelling study.
\newblock {\em Lancet Public Health}, 7:537--48.

\bibitem[Macdonald et~al., 2003]{Macdonaldetal2003}
{\bf Macdonald, A.}, {\bf Waters, H.}, \textbf{and} {\bf Wekwete, C.} ({2003}).
\newblock The genetics of breast and ovarian cancer ii: A model of critical
  illness insurance.
\newblock {\em Scandinavian Actuarial Journal}, 1:28--50.

\bibitem[Maringe et~al., 2020]{Maringeetal2020}
{\bf Maringe, C.}, {\bf Spicer, J.}, {\bf Morris, M.}, {\bf Purushotham, A.},
  {\bf Nolte, E.}, \textbf{and} {\bf Sullivan, R. e.~a.} ({2020}).
\newblock The impact of the {COVID}-19 pandemic on cancer deaths due to delays
  in diagnosis in {E}ngland, {UK}: a national, population-based, modelling
  study.
\newblock {\em The LANCET Oncology}, 21(8):1023--1034.

\bibitem[Mouw et~al., 2008]{Mouwetal2008}
{\bf Mouw, T.}, {\bf Koster, A.}, {\bf Wright, M.}, {\bf Blank, M.}, {\bf
  Moore, S.}, {\bf Hollenbeck, A.}, \textbf{and} {\bf Schatzkin, A.} ({2008}).
\newblock Education and risk of cancer in a large cohort of men and women in
  the {U}nited {S}tates.
\newblock {\em PLOS ONE}.

\bibitem[Nash, 2020]{ONS2020SubnationalPopProj}
{\bf Nash, A.} ({2020}).
\newblock Subnational population projections for england: 2018-based.
\newblock Technical report, Office for National Statistics.

\bibitem[Newton et~al., 2015]{Newtonetal2015}
{\bf Newton, J.}, {\bf Briggs, A.}, {\bf Murray, C.}, {\bf Dicker, D.}, {\bf
  Foreman, K.}, {\bf Wang, H.}, \textbf{and} {\bf et~al.} ({2015}).
\newblock Changes in health in {E}ngland, with analysis by {E}nglish regions
  and areas of deprivation, 1990-2013: a systematic analysis for the {G}lobal
  {B}urden of {D}isease {S}tudy 2013.
\newblock {\em Lancet}, 386.

\bibitem[ONS, 2012]{ONS2012}
{\bf ONS} ({2012}).
\newblock Ethnicity and national identity in {E}ngland and {W}ales 2011.
\newblock Technical report, Office for National Statistics.

\bibitem[ONS, 2017]{ONS2015}
{\bf ONS} ({2017}).
\newblock Cancer registration statistics, {E}ngland, 2015.
\newblock Technical report, Office for National Statistics.

\bibitem[ONS, 2020a]{ONSCovid2020}
{\bf ONS} (2020)a.
\newblock Coronavirus {(COVID-19}): 2020 in charts.

\bibitem[ONS, 2020b]{ONSLeadingCOD}
{\bf ONS} (2020)b.
\newblock Leading causes of death, {UK}: 2001 to 2018.

\bibitem[ONS, 2022a]{ONSAvoidableMortality}
{\bf ONS} (2022)a.
\newblock Avoidable mortality in {G}reat {B}ritain: 2020.

\bibitem[ONS, 2022b]{NOMIS2022}
{\bf ONS} (2022)b.
\newblock Nomis - offical census and labour statistics.

\bibitem[ONS, 2024]{ONSExcessMortNewMeth2024}
{\bf ONS} (2024).
\newblock Estimating excess deaths in the {UK}, methodology changes: {F}ebruary
  2024.

\bibitem[Redondo~Loures and Cairns, 2019]{LouresCairns2018}
{\bf Redondo~Loures, C.} \textbf{and} {\bf Cairns, A.} ({2019}).
\newblock Mortality in the {US} by education level.
\newblock {\em Annals of Actuarial Science}, 14(2):384--419.

\bibitem[Reynolds, 2013]{Reynolds2013}
{\bf Reynolds, P.} ({2013}).
\newblock Smoking and breast cancer.
\newblock {\em Journal of Mammary Gland Biology and Neoplasia}, 18:15--23.

\bibitem[Reynolds et~al., 2004]{Reynoldsetal2004}
{\bf Reynolds, P.}, {\bf Hurley, S.}, {\bf Goldberg, D.}, {\bf Anton-Culver,
  H.}, \textbf{and} {\bf et~al.} ({2004}).
\newblock Active smoking, household passive smoking, and breast cancer:
  evidence from the {C}alifornia {T}eachers {S}tudy.
\newblock {\em Journal of the National Cancer Institute}, 96(1):29--37.

\bibitem[Riaz et~al., 2011]{Riazetal2011}
{\bf Riaz, S.}, {\bf Horton, M.}, {\bf Kang, J.}, {\bf Mak, V.}, {\bf
  Luchtenborg, M.}, \textbf{and} {\bf M{\o}ller, H.} ({2011}).
\newblock Lung cancer incidence and survival in {E}ngland: an analysis by
  socioeconomic deprivation and urbanization.
\newblock {\em Journal of Thoracic Oncology}, 6(12).

\bibitem[Richardson et~al., 2020]{Richardsonetal2020}
{\bf Richardson, S.}, {\bf Hirsch, J.}, {\bf Narasimhan, M.}, \textbf{and} {\bf
  et~al.} ({2020}).
\newblock Presenting characteristics, comorbidities, and outcomes among 5700
  patients hospitalized with {COVID-19} in the {N}ew {Y}ork city area.
\newblock {\em JAMA}, 323(20):2052--9.

\bibitem[Ryan et~al., 2012]{Ryanetal2012}
{\bf Ryan, H.}, {\bf Trosclair, A.}, \textbf{and} {\bf Gfroerer, J.} ({2012}).
\newblock Adult current smoking: Differences in definitions and prevalence
  estimates -- {NHIS} and {NSDUH}, 2008.
\newblock {\em Journal of Environmental and Public Health}, 2012:918368.

\bibitem[Singh et~al., 2011]{Singhetal2011}
{\bf Singh, G.}, {\bf Williams, S.}, {\bf Siahpush, M.}, \textbf{and} {\bf
  Mulhollen, A.} ({2011}).
\newblock Socioeconomic, rural-urban, and racial inequalities in {US} cancer
  mortality: {P}art i-all cancers and lung cancer and {P}art ii-colorectal,
  prostate, breast, and cervical cancers.
\newblock {\em Journal of Cancer Epidemiology}, 2011.

\bibitem[Spiegelhalter et~al., 2002]{Spiegelhalteretal2002}
{\bf Spiegelhalter, D.}, {\bf Best, N.}, {\bf Carlin, B.}, \textbf{and} {\bf
  Van Der~Linde, A.} ({2002}).
\newblock Bayesian measures of model complexity and fit.
\newblock {\em Journal of the Royal Statistical Society}, B(64):583--640.

\bibitem[Sud et~al., 2020]{Sudetal2020}
{\bf Sud, A.}, {\bf Torr, B.}, {\bf Jones, M.}, \textbf{and} {\bf et~al.}
  ({2020}).
\newblock Effect of delays in the 2-week-wait cancer referral pathway during
  the {COVID}-19 pandemic on cancer survival in the {UK}: a modelling study.
\newblock {\em The LANCET: Oncology}.

\bibitem[WHO, 2022]{WHO2021}
{\bf WHO} (2022).
\newblock {WHO} coronavirus ({COVID}-19) dashboard.

\bibitem[Windsor-Shellard et~al., 2018]{Windsoretal2018}
{\bf Windsor-Shellard, B.}, {\bf Pullin, L.}, \textbf{and} {\bf Horton, M.}
  ({2018}).
\newblock Adult smoking habits in the {UK}:2017.
\newblock Technical report, Office for National Statistics.

\bibitem[Wong et~al., 2018]{Wongetal2018}
{\bf Wong, J.}, {\bf Forster, J.}, \textbf{and} {\bf Smith, P.} ({2018}).
\newblock Bayesian mortality forecasting with overdispersion.
\newblock {\em Insurance: Mathematics and Economics}, 83:206--221.

\bibitem[Xue et~al., 2011]{Xueetal2011}
{\bf Xue, F.}, {\bf Willett, W.}, {\bf Rosner, B.}, \textbf{and} {\bf et~al.}
  ({2011}).
\newblock Cigarette smoking and the incidence of breast cancer.
\newblock {\em JAMA Internal Medicine}, 171(2):125--133.

\bibitem[Zhou et~al., 2020]{Zhouetal2020}
{\bf Zhou, F.}, {\bf Yu, T.}, {\bf Du, R.}, \textbf{and} {\bf et~al.} ({2020}).
\newblock Clinical course and risk factors for mortality of adult inpatients
  with {COVID-19} in {W}uhan, {C}hina: a retrospective cohort study.
\newblock {\em Lancet}, 395(10229):1054--62.

\end{thebibliography}

%\endctxt{Jose A., Macdonald A.S., Tzougas G. and Streftaris G. Interpretable zero-inflated neural network models for predicting admission counts. {\it Annals of Actuarial Science}, 1--8. \href{10.1017/xxxxx}{10.1017/xxxxx}}

%----------------------------------------------------------------------------------------
%	APPENDIX
%----------------------------------------------------------------------------------------

\appendix

\section{A useful variable: average age-at-diagnosis}\label{sec:AAD_desc}

Age-at-diagnosis is known to be a crucial factor for cancer survival. 
In the study of \citet{Ariketal2021}, a link has been established between cancer morbidity and mortality through a variable, namely average age-at-diagnosis (AAD). 

AAD for gender $g$ in deprivation quintile $d$ of region $r$ at year at diagnosis $t$, denoted by $\mbox{AAD}_{g, d,r,t}$, is estimated as follows: 

\begin{align}
	\mbox{AAD}_{g, d,r,t} = \frac{\sum_a{a \hat{\lambda}_{a, g, d,r,t} E_a^{\text{std}}} }{\sum_a { \hat{\lambda}_{a, g, d,r,t} E_a^{\text{std}} } },
	\label{eq:AADformula1}
\end{align}
where $E_a^{\text{std}}$ shows population numbers at age-at-diagnosis $a$ according to the ESP 2013, and $\hat{\lambda}_{a, g, d,r,t}$ is the relevant cause-specific fitted incidence rate
obtained based on the best fitted models (\citep{Ariketal2021}).
For modelling purposes, AAD is then weighted over years as described below:

\begin{align}
	\mbox{AAD}_{g, d,r} = \frac{\sum_{t} {\mbox{AAD}^{\text{morbidity}}_{g, d,r,t}  E_{g, d,r,t}} }{\sum_{t}{ E_{g, d,r,t} }},
\end{align}
by using the relevant mid-year population estimates $E_{g, d,r,t}$ in deprivation quintile $d$ of region $r$.
Note that, if deprivation is not a significant variable in the model under inspection, AAD could also be averaged over deprivation quintiles so that it would be by region only.

\begin{figure}[htbp]
	\centering
	\includegraphics[width=0.75\textwidth, angle =0]{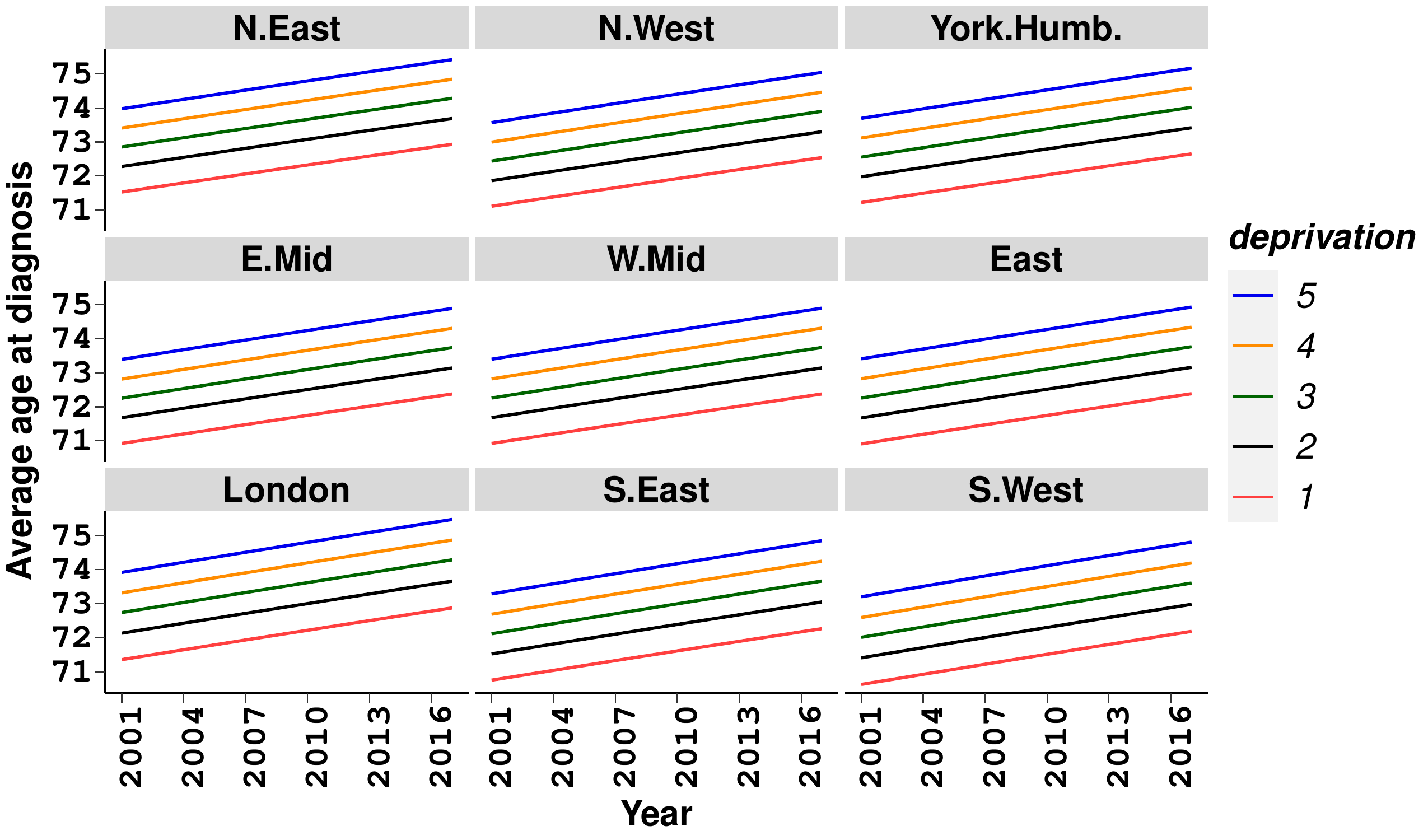}
	\caption{Average age-at-diagnosis in lung cancer mortality, females, in deprivation quintiles 1 (most deprived) to 5 (least deprived), of regions of England between 2001 and 2017.}
	\label{fig:LungCOD_AAD_f}
\end{figure}

\figref{fig:LungCOD_AAD_f} shows estimated AAD values based on LC morbidity in women from 2001 to 2017 (the latest available calendar year). 
An increasing trend in AAD values is estimated over the calendar years, with comparable results in the regions of England. 
%However, the estimates across deprivation quintiles in a given region are notably different, where lower AAD values are obtained in more deprived quintiles (as opposed to higher mortality rates in the same quintiles, see \figref{fig:ASRLungCOD_Female_CrudeRates}). 

\begin{figure}[htbp]
	\centering
	\includegraphics[width=0.75\textwidth, angle =0]{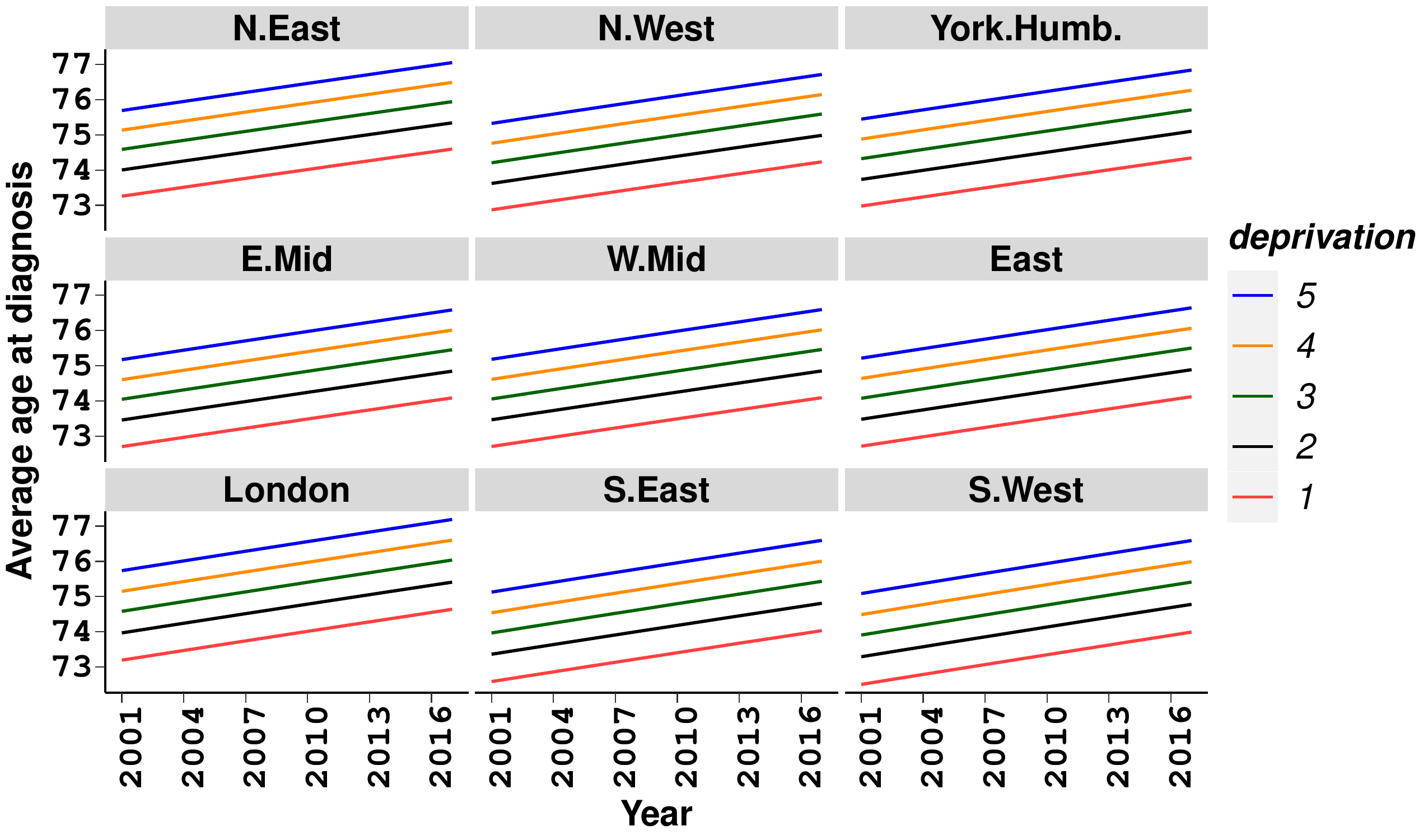}
	\caption{Average age-at-diagnosis in lung cancer mortality, males, in deprivation quintiles 1 (most deprived) to 5 (least deprived), of regions of England between 2001 and 2017.}
	\label{fig:LungCOD_AAD_m}
\end{figure}

\figref{fig:LungCOD_AAD_m} demonstrates AAD estimates in LC morbidity for men between 2001 and 2017. 
Similar to the female counterparts, there is an increasing trend in calculated AAD values over the time, 
with lower AAD values estimated in more deprived quintiles of a given region. 
We note higher AAD estimates for men as opposed to women. 

\section{Forward variable selection for cancer mortality models} \label{sec:VariableSelection}

\subsection{Variable selection for female lung cancer mortality}\label{Sec:VariableSelectionLungCODFemales}	

\begin{table}[H]
	\centering
	\caption{Variable selection procedure in the R-INLA software to determine the best fitted model for female lung cancer mortality, without considering `smoking proxy'.}\label{tab:FemaleLungCODVariableSelecNoSmoking}
	\resizebox{0.75\columnwidth}{!}{%
		\begin{tabular}{lllll}
			\hline
			variable added & Bayes factor & marginal likelihood & diff. in marginal likelihood & DIC \\ 
			\hline
			null &  & -30356.43 &  & 45717.28 \\ 
			age & Inf & -25067.03 & 5289.39 & 44475.35 \\ 
			income & Inf & -22314.66 & 2752.36 & 42788.21 \\ 
			region & Inf & -21187.68 & 1126.98 & 41748.87 \\ 
			year & 126951622561447328 & -21148.30 & 39.38 & 41652.67 \\ 
			AAD & 28463406452108.4 & -21117.31 & 30.97 & 41614.40 \\ 
			income:age & 2.26320928808427e+125 & -20828.67 & 288.63 & 41114.45 \\ 
			region:age & 312056207.62 & -20809.11 & 19.55 & 40943.17 \\
			\hline
	\end{tabular}}
	{\parbox{5.5in}{
			\footnotesize Note: The null model only includes the related offset variable, and age and year are defined to be categorical variables.
		}
	}
\end{table}

\begin{table}[H]
	\centering
	\caption{Variable selection procedure in the R-INLA software to determine the best fitted model for female lung cancer mortality, with `smoking proxy'. 
		%		Note that null model includes ALL variables, except age and year interaction term, in \tabref{tab:FemaleLungCODVariableSelecNoSmoking}, and age and year are defined to be categorical variables. \emph{Note that if the interaction term between age and year is included in the null model, smoking has not become a significant variable in the variable selection.
			%	} 
	}\label{tab:FemaleLungCODVariableSelecNoSmokingv2}
	\resizebox{0.75\columnwidth}{!}{%
		\begin{tabular}{lllll}
			\hline
			variable added & Bayes factor & marginal likelihood & diff. in marginal likelihood & DIC \\ 
			\hline
			null &  & -20809.12 &  & 40943.37  \\ 
			smoking & 1.53e+43 & -20709.68 & 99.43 & 40757.45 \\ 
			AAD:region & 4.00e+20 & -20662.25 & 47.44 & 40644.90 \\ 
			AAD:year & 36.24 & -20658.66 & 3.59 & 40564.79 \\ 
			\hline
		\end{tabular}
	}
	{\parbox{5.5in}{
			\footnotesize Note: The null model includes ALL variables in \tabref{tab:FemaleLungCODVariableSelecNoSmoking}; age and year are defined to be categorical variables; and non-smoker prevalence rates are used with 20-year time lag. 
		}
	}
\end{table}

\begin{table}[H]
	\centering
	\caption{An alternative variable selection procedure in the R-INLA software to determine the best fitted model for female lung cancer mortality, with `smoking proxy'.}\label{tab:FemaleLungCODVariableSelecSmoking}
	\resizebox{0.75\columnwidth}{!}{%
		\begin{tabular}{lllll}
			\hline
			variable added & Bayes factor & marginal likelihood & diff. in marginal likelihood & DIC \\ 
			\hline
			null &  & -30356.45 &  & 45718.20 \\ 
			age & Inf & -25069.12 & 5287.33 & 44475.38 \\ 
			income & Inf & -22318.36 & 2750.76 & 42788.29 \\ 
			region & Inf & -21193.58 & 1124.77 & 41748.84  \\ 
			year & 129115643505238544 & -21154.18 & 39.39 & 41652.72  \\ 
			smoking & 5.59e+41 & -21058.05 & 96.12 & 41491.94 \\ 
			AAD & 52164756508979.5 & -21026.46 & 31.58  & 41452.27 \\ 
			age:income & 1.05e+117 & -20757.01 & 269.45 & 40937.04 \\ 
			age:smoking & 3.80e+73 & -20587.58 & 169.42 & 40649.96 \\ 
			region:AAD & 8.93e+19 & -20541.64 & 45.93 & 40535.07 \\ 
			income:smoking & 3500548.57 & -20526.58 & 15.06  & 40490.82 \\ 
			AAD:smoking & 985210.84 & -20512.78 & 13.80 & 40462.12 \\ 
			year:AAD & 9706822596297934848 & -20469.06 & 43.71 & 40310.30 \\  
			\hline
		\end{tabular}
	}
	{\parbox{5.5in}{
			\footnotesize Note: This is a two-stage variable selection procedure. 
			The first stage involves all main variables; the second stage involves all two-way interaction terms between them; and non-smoker prevalence rates are used with 20-year time lag.
		}
	}
\end{table}

\subsection{Variable selection for male lung cancer mortality}\label{Sec:VariableSelectionLungCODMales}	

\begin{table}[H]
	\centering
	\caption{Variable selection procedure in the R-INLA software to determine the best fitted model for male lung cancer mortality, without considering `smoking proxy'. }\label{tab:MaleLungCODVariableSelecNoSmoking}
	\resizebox{0.75\columnwidth}{!}{%
		\begin{tabular}{lllll}
			\hline
			variable added & Bayes factor & marginal likelihood & diff. in marginal likelihood & DIC \\ 
			\hline
			null &  & -33057.97 &  & 47607.99 \\ 
			age & Inf & -26665.01 & 6392.96 & 46598.55\\ 
			income & Inf & -23252.59 & 3412.41 & 44710.56 \\ 
			year & Inf & -22404.99 & 847.59 & 43789.62 \\ 
			region & 1.62e+286 & -21745.97 & 659.02& 43003.53 \\ 
			AAD & 8634573.64 & -21730.00& 15.97 & 42981.00 \\ 
			income:age & 6.67e+228 & -21203.12 & 526.88 & 42012.54 \\ 
			AAD:year & 60480674313.05 & -21178.29 & 24.82& 41888.20 \\ 
			AAD:region & 35.60 & -21174.72 & 3.57& 41843.68 \\ 
			\hline
	\end{tabular}}
	{\parbox{5.5in}{
			\footnotesize Note: The null model only includes the related offset variable, and age and year are defined to be categorical variables.
		}
	}
\end{table}

\begin{table}[H]
	\centering
	\caption{Variable selection procedure in the R-INLA software to determine the best fitted model for male lung cancer mortality, with `smoking proxy'.}\label{tab:MaleLungCODVariableSelecNoSmokingv2}
	\resizebox{0.75\columnwidth}{!}{%
		\begin{tabular}{lllll}
			\hline
			variable added & Bayes factor & marginal likelihood & diff. in marginal likelihood & DIC \\ 
			\hline
			null &  & -21174.72 &  & 41843.63 \\ 
			smoking & 1399391.13 & -21160.57 & 14.15 & 41814.23 \\ 
			\hline
		\end{tabular}
	}
	{\parbox{5.5in}{
			\footnotesize Note: The null model includes ALL variables in \tabref{tab:MaleLungCODVariableSelecNoSmoking}; age and year are defined to be categorical variables; and non-smoker prevalence rates are used with 20-year time lag.
		}
	}
\end{table}

\begin{table}[H]
	\centering
	\caption{An alternative variable selection procedure in the R-INLA software to determine the best fitted model for male lung cancer mortality, with `smoking proxy'.}
	\resizebox{0.75\columnwidth}{!}{%
		\begin{tabular}{lllll}
			\hline
			variable added & Bayes factor & marginal likelihood & diff. in marginal likelihood & DIC \\ 
			\hline
			null &  & -33057.98 &  & 47608.01 \\ 
			age & Inf & -26665.02 & 6392.96  & 46598.59 \\ 
			income & Inf & -23252.59 & 3412.42 & 44710.58  \\ 
			smoking & Inf & -22361.73 & 890.86 & 43784.52 \\ 
			region & 4.63e+285 & -21703.96 & 657.76 & 43002.49 \\ 
			AAD & 8245920.92 & -21688.03 & 15.92  & 42980.08 \\ 
			income:age & 6.26e+226 & -21165.81 & 522.21 & 42019.89 \\ 
			AAD:smoking & 2.73e+38 & -21077.30 & 88.50 & 41855.82 \\ 
			age:smoking & 2.97e+42 & -20979.50  & 97.79 & 41638.24  \\ 
			income:smoking & 322067.75 & -20966.82 & 12.68 & 41593.15 \\ 
			AAD:region & 28.06 & -20963.49 & 3.33 & 41545.50  \\ 
			\hline		
		\end{tabular}
	}
	{\parbox{5.5in}{
			\footnotesize Note: This is a two-stage variable selection procedure. 
			The first stage involves all main variables; the second stage involves all two-way interaction terms between them; and non-smoker prevalence rates are used with 20-year time lag.
		}
	}
\end{table}

\subsection{Variable selection for breast cancer mortality}\label{Sec:VariableSelectionBreastCOD}	
		
		\begin{table}[H]
			\centering
			\caption{Variable selection procedure in the R-INLA software to determine the best fitted model for breast cancer mortality, without considering `smoking proxy'. 
				%	, and there is no time lag in smoking variable.
			}\label{tab:BreastCODVariableSelect}
			\resizebox{0.75\columnwidth}{!}{%
				\begin{tabular}{lllll}
					\hline
					variable added & Bayes factor & marginal likelihood & diff. in marginal likelihood & DIC \\ 
					\hline
					null &  & -10133.46 &  & 14460.45 \\ 
					age & Inf & -7120.84  & 3012.62 & 13686.88 \\ 
					year & 3.31e+204 & -6649.91 & 470.92 & 13093.15 \\ 
					region & 13.28 & -6647.33 & 2.58 & 13053.13 \\ 			
					\hline
				\end{tabular}
			}
			{\parbox{5.5in}{
					\footnotesize Note: The null model only includes the related offset variable; 
					and age and year are defined to be categorical variables.
				}
			}
		\end{table}
		
		\begin{table}[H]
			\centering
			\caption{Variable selection procedure in the R-INLA software to determine the best fitted model for breast cancer mortality, with `smoking proxy'.}\label{tab:BreastCODVariableSelect_wlSmoking}
			\resizebox{0.75\columnwidth}{!}{%
				\begin{tabular}{lllll}
					\hline
					variable added & Bayes factor & marginal likelihood & diff. in marginal likelihood & DIC \\ 
					\hline
					null &  & -10133.46  &  & 14460.46 \\ 
					age & Inf & -7120.84 & 3012.62 & 13686.90 \\ 
					year & 3.32e+204 & -6649.915 & 470.92 & 13093.17  \\ 
					smoking & 1822.77 & -6642.40 & 7.50 & 13075.32 \\ 
					region & 25.68 & -6639.16 & 3.24 & 13039.60 \\ 
					%age:smoking & 1.16e+32 & -6565.31 & 73.83 & 12897.37 \\ 
					\hline
				\end{tabular}
			}
			{\parbox{5.5in}{
					\footnotesize Note: The null model only includes the related offset variable; age and year are defined to be categorical variables; and non-smoker prevalence rates are used with 20-year time lag.
				}
			}
		\end{table}

\newpage
\section{Parameter estimates for cancer mortality models} \label{sec:ParameterEstimates}

%----------------------------------------------------------------------------------------
% PARAMETER ESTIMATES - LUNG CANCER, FEMALES 
%----------------------------------------------------------------------------------------	

% latex table generated in R 4.1.2 by xtable 1.8-4 package
% Thu Nov 16 22:24:32 2023
\begin{table}[H]
	\centering
	%\captionsetup{labelformat=empty}
	\caption{{{Estimated coefficients for lung cancer mortality in women based on \eqref{eq:FemaleLungCODLocationPrmtr2}. }}}
	\addtolength{\tabcolsep}{-4pt}
	\resizebox{0.8\columnwidth}{!}{
		\begin{tabular}{llrrrr|llrrrr}
			\hline
			Covariate& Parameter & Mean & SD & \%2.5 & \%97.5 &Covariate& Parameter & Mean & SD & \%2.5 & \%97.5 \\ 
			\hline
			& {$\beta_{0}$} & -7.2760 & 0.0264 & -7.3240 & -7.2260 & {     } & {$\beta_{5,\text{deprivation}_{2},\text{age}_{7}}$} & 0.0859 & 0.0142 & 0.0573 & 0.1146 \\ 
			& {$\beta_{1,\text{age}_{1}}$} & -2.4300 & 0.0339 & -2.4960 & -2.3670 & {     } & {$\beta_{5,\text{deprivation}_{3},\text{age}_{7}}$} & -0.1328 & 0.0239 & -0.1826 & -0.0895 \\ 
			& {$\beta_{1,\text{age}_{2}}$} & -1.0340 & 0.0132 & -1.0590 & -1.0080 & {     } & {$\beta_{5,\text{deprivation}_{4},\text{age}_{7}}$} & -0.0941 & 0.0185 & -0.1306 & -0.0564 \\ 
			& {$\beta_{1,\text{age}_{3}}$} & -0.4921 & 0.0121 & -0.5149 & -0.4687 & {     } & {$\beta_{5,\text{deprivation}_{5},\text{age}_{7}}$} & -0.0649 & 0.0165 & -0.0955 & -0.0302 \\ 
			& {$\beta_{1,\text{age}_{4}}$} & -0.0717 & 0.0132 & -0.0973 & -0.0458 & {     } & {$\beta_{5,\text{deprivation}_{1},\text{age}_{8}}$} & -0.0852 & 0.0145 & -0.1137 & -0.0549 \\ 
			& {$\beta_{1,\text{age}_{5}}$} & 0.3056 & 0.0133 & 0.2810 & 0.3316 & {     } & {$\beta_{5,\text{deprivation}_{2},\text{age}_{8}}$} & -0.0311 & 0.0137 & -0.0575 & -0.0029 \\ 
			& {$\beta_{1,\text{age}_{6}}$} & 1.1480 & 0.0253 & 1.1010 & 1.1970 & {     } & {$\beta_{5,\text{deprivation}_{3},\text{age}_{8}}$} & 0.0351 & 0.0140 & 0.0089 & 0.0631 \\ 
			& {$\beta_{1,\text{age}_{7}}$} & 1.2830 & 0.0257 & 1.2340 & 1.3330 & {     } & {$\beta_{5,\text{deprivation}_{4},\text{age}_{8}}$} & 0.1340 & 0.0141 & 0.1071 & 0.1613 \\ 
			& {$\beta_{1,\text{age}_{8}}$} & 1.2920 & 0.0265 & 1.2410 & 1.3430 & {     } & {$\beta_{5,\text{deprivation}_{5},\text{age}_{8}}$} & 0.2390 & 0.0160 & 0.2081 & 0.2696 \\ 
			& {$\beta_{2,\text{region}_{1}}$} & -0.1256 & 0.0386 & -0.1850 & -0.0552 & {     } & {$\beta_{6,\text{region}_{1},\text{age}_{1}}$} & -0.1074 & 0.0322 & -0.1710 & -0.0481 \\ 
			& {$\beta_{2,\text{region}_{2}}$} & 0.1451 & 0.0073 & 0.1310 & 0.1592 & {     } & {$\beta_{6,\text{region}_{2},\text{age}_{1}}$} & -0.0668 & 0.0279 & -0.1201 & -0.0084 \\ 
			& {$\beta_{2,\text{region}_{3}}$} & 0.0152 & 0.0145 & -0.0096 & 0.0437 & {     } & {$\beta_{6,\text{region}_{3},\text{age}_{1}}$} & -0.0325 & 0.0243 & -0.0784 & 0.0160 \\ 
			& {$\beta_{2,\text{region}_{4}}$} & 0.1079 & 0.0154 & 0.0780 & 0.1358 & {     } & {$\beta_{6,\text{region}_{4},\text{age}_{1}}$} & -0.0331 & 0.0206 & -0.0747 & 0.0063 \\ 
			& {$\beta_{2,\text{region}_{5}}$} & 0.0113 & 0.0153 & -0.0175 & 0.0378 & {     } & {$\beta_{6,\text{region}_{5},\text{age}_{1}}$} & 0.0299 & 0.0192 & -0.0067 & 0.0664 \\ 
			& {$\beta_{2,\text{region}_{6}}$} & 0.0305 & 0.0136 & 0.0020 & 0.0534 & {     } & {$\beta_{6,\text{region}_{6},\text{age}_{1}}$} & 0.0750 & 0.0182 & 0.0382 & 0.1086 \\ 
			& {$\beta_{2,\text{region}_{7}}$} & -0.5302 & 0.0344 & -0.5821 & -0.4674 & {     } & {$\beta_{6,\text{region}_{7},\text{age}_{1}}$} & 0.0618 & 0.0192 & 0.0264 & 0.0992 \\ 
			& {$\beta_{2,\text{region}_{8}}$} & 0.1846 & 0.0228 & 0.1423 & 0.2207 & {     } & {$\beta_{6,\text{region}_{8},\text{age}_{1}}$} & 0.0731 & 0.0239 & 0.0264 & 0.1169 \\ 
			& {$\beta_{2,\text{region}_{9}}$} & 0.1611 & 0.0293 & 0.1088 & 0.2087 & {     } & {$\beta_{6,\text{region}_{9},\text{age}_{1}}$} & -0.0718 & 0.0229 & -0.1184 & -0.0293 \\ 
			& {$\beta_{3,\text{deprivation}_{1}}$} & 1.8050 & 0.1168 & 1.5980 & 1.9890 & {     } & {$\beta_{6,\text{region}_{1},\text{age}_{2}}$} & -0.0535 & 0.0200 & -0.0918 & -0.0136 \\ 
			& {$\beta_{3,\text{deprivation}_{2}}$} & 0.7603 & 0.0488 & 0.6740 & 0.8345 & {     } & {$\beta_{6,\text{region}_{2},\text{age}_{2}}$} & 0.0328 & 0.0175 & -0.0020 & 0.0657 \\ 
			& {$\beta_{3,\text{deprivation}_{3}}$} & -0.0806 & 0.0070 & -0.0937 & -0.0668 & {     } & {$\beta_{6,\text{region}_{3},\text{age}_{2}}$} & 0.0255 & 0.0165 & -0.0068 & 0.0573 \\ 
			& {$\beta_{3,\text{deprivation}_{4}}$} & -0.8448 & 0.0547 & -0.9253 & -0.7480 & {     } & {$\beta_{6,\text{region}_{4},\text{age}_{2}}$} & 0.0339 & 0.0147 & 0.0049 & 0.0623 \\ 
			& {$\beta_{3,\text{deprivation}_{5}}$} & -1.6400 & 0.1067 & -1.8030 & -1.4500 & {     } & {$\beta_{6,\text{region}_{5},\text{age}_{2}}$} & 0.0355 & 0.0146 & 0.0063 & 0.0641 \\ 
			& {$\beta_{4}$} & 0.9437 & 0.0803 & 0.8022 & 1.0700 & {     } & {$\beta_{6,\text{region}_{6},\text{age}_{2}}$} & 0.0102 & 0.0152 & -0.0185 & 0.0399 \\ 
			& {$\beta_{5,\text{deprivation}_{1},\text{age}_{1}}$} & 0.1318 & 0.0167 & 0.1005 & 0.1634 & {     } & {$\beta_{6,\text{region}_{7},\text{age}_{2}}$} & -0.0127 & 0.0173 & -0.0457 & 0.0228 \\ 
			& {$\beta_{5,\text{deprivation}_{2},\text{age}_{1}}$} & 0.1150 & 0.0150 & 0.0869 & 0.1455 & {     } & {$\beta_{6,\text{region}_{8},\text{age}_{2}}$} & -0.0956 & 0.0264 & -0.1475 & -0.0420 \\ 
			& {$\beta_{5,\text{deprivation}_{3},\text{age}_{1}}$} & 0.0810 & 0.0132 & 0.0551 & 0.1069 & {     } & {$\beta_{6,\text{region}_{9},\text{age}_{2}}$} & 0.0341 & 0.0216 & -0.0098 & 0.0775 \\ 
			& {$\beta_{5,\text{deprivation}_{4},\text{age}_{1}}$} & 0.0797 & 0.0118 & 0.0563 & 0.1028 & {     } & {$\beta_{6,\text{region}_{1},\text{age}_{3}}$} & 0.0281 & 0.0194 & -0.0122 & 0.0655 \\ 
			& {$\beta_{5,\text{deprivation}_{5},\text{age}_{1}}$} & 0.0109 & 0.0116 & -0.0109 & 0.0347 & {     } & {$\beta_{6,\text{region}_{2},\text{age}_{3}}$} & 0.0046 & 0.0174 & -0.0298 & 0.0394 \\ 
			& {$\beta_{5,\text{deprivation}_{1},\text{age}_{2}}$} & -0.0592 & 0.0111 & -0.0810 & -0.0365 & {     } & {$\beta_{6,\text{region}_{3},\text{age}_{3}}$} & -0.0002 & 0.0160 & -0.0323 & 0.0312 \\ 
			& {$\beta_{5,\text{deprivation}_{2},\text{age}_{2}}$} & -0.1391 & 0.0114 & -0.1616 & -0.1171 & {     } & {$\beta_{6,\text{region}_{4},\text{age}_{3}}$} & 0.0144 & 0.0153 & -0.0162 & 0.0441 \\ 
			& {$\beta_{5,\text{deprivation}_{3},\text{age}_{2}}$} & -0.2201 & 0.0135 & -0.2458 & -0.1924 & {     } & {$\beta_{6,\text{region}_{5},\text{age}_{3}}$} & -0.0042 & 0.0164 & -0.0378 & 0.0269 \\ 
			& {$\beta_{5,\text{deprivation}_{4},\text{age}_{2}}$} & 0.0635 & 0.0177 & 0.0296 & 0.0973 & {     } & {$\beta_{6,\text{region}_{6},\text{age}_{3}}$} & 0.0186 & 0.0188 & -0.0167 & 0.0575 \\ 
			& {$\beta_{5,\text{deprivation}_{5},\text{age}_{2}}$} & 0.0298 & 0.0159 & -0.0020 & 0.0608 & {     } & {$\beta_{6,\text{region}_{7},\text{age}_{3}}$} & 0.0476 & 0.0299 & -0.0083 & 0.1060 \\ 
			& {$\beta_{5,\text{deprivation}_{1},\text{age}_{3}}$} & 0.0403 & 0.0140 & 0.0127 & 0.0674 & {     } & {$\beta_{6,\text{region}_{8},\text{age}_{3}}$} & 0.0372 & 0.0281 & -0.0215 & 0.0929 \\ 
			& {$\beta_{5,\text{deprivation}_{2},\text{age}_{3}}$} & 0.0371 & 0.0119 & 0.0135 & 0.0591 & {     } & {$\beta_{6,\text{region}_{9},\text{age}_{3}}$} & 0.0095 & 0.0221 & -0.0307 & 0.0539 \\ 
			& {$\beta_{5,\text{deprivation}_{3},\text{age}_{3}}$} & 0.0263 & 0.0118 & 0.0026 & 0.0499 & {     } & {$\beta_{6,\text{region}_{1},\text{age}_{4}}$} & 0.0305 & 0.0200 & -0.0092 & 0.0698 \\ 
			& {$\beta_{5,\text{deprivation}_{4},\text{age}_{3}}$} & -0.0126 & 0.0115 & -0.0353 & 0.0081 & {     } & {$\beta_{6,\text{region}_{2},\text{age}_{4}}$} & 0.0261 & 0.0180 & -0.0077 & 0.0588 \\ 
			& {$\beta_{5,\text{deprivation}_{5},\text{age}_{3}}$} & -0.0684 & 0.0121 & -0.0930 & -0.0440 & {     } & {$\beta_{6,\text{region}_{3},\text{age}_{4}}$} & -0.0153 & 0.0169 & -0.0481 & 0.0172 \\ 
			& {$\beta_{5,\text{deprivation}_{1},\text{age}_{4}}$} & -0.1159 & 0.0135 & -0.1423 & -0.0889 & {     } & {$\beta_{6,\text{region}_{4},\text{age}_{4}}$} & -0.0403 & 0.0190 & -0.0764 & -0.0021 \\ 
			& {$\beta_{5,\text{deprivation}_{2},\text{age}_{4}}$} & -0.0237 & 0.0199 & -0.0624 & 0.0146 & {     } & {$\beta_{6,\text{region}_{5},\text{age}_{4}}$} & -0.0953 & 0.0221 & -0.1360 & -0.0514 \\ 
			& {$\beta_{5,\text{deprivation}_{3},\text{age}_{4}}$} & 0.0030 & 0.0180 & -0.0340 & 0.0378 & {     } & {$\beta_{6,\text{region}_{6},\text{age}_{4}}$} & 0.0052 & 0.0272 & -0.0475 & 0.0546 \\ 
			& {$\beta_{5,\text{deprivation}_{4},\text{age}_{4}}$} & 0.0151 & 0.0147 & -0.0141 & 0.0434 & {     } & {$\beta_{6,\text{region}_{7},\text{age}_{4}}$} & 0.0301 & 0.0285 & -0.0256 & 0.0828 \\ 
			& {$\beta_{5,\text{deprivation}_{5},\text{age}_{4}}$} & -0.0107 & 0.0137 & -0.0367 & 0.0181 & {     } & {$\beta_{6,\text{region}_{8},\text{age}_{4}}$} & -0.0267 & 0.0217 & -0.0687 & 0.0176 \\ 
			& {$\beta_{5,\text{deprivation}_{1},\text{age}_{5}}$} & -0.0143 & 0.0125 & -0.0393 & 0.0098 & {     } & {$\beta_{6,\text{region}_{9},\text{age}_{4}}$} & 0.0088 & 0.0192 & -0.0282 & 0.0471 \\ 
			& {$\beta_{5,\text{deprivation}_{2},\text{age}_{5}}$} & 0.0087 & 0.0116 & -0.0139 & 0.0308 & {     } & {$\beta_{6,\text{region}_{1},\text{age}_{5}}$} & -0.0006 & 0.0168 & -0.0320 & 0.0316 \\ 
			& {$\beta_{5,\text{deprivation}_{3},\text{age}_{5}}$} & 0.0108 & 0.0124 & -0.0123 & 0.0351 & {     } & {$\beta_{6,\text{region}_{2},\text{age}_{5}}$} & 0.0109 & 0.0170 & -0.0204 & 0.0443 \\ 
			& {$\beta_{5,\text{deprivation}_{4},\text{age}_{5}}$} & 0.0111 & 0.0139 & -0.0155 & 0.0368 & {     } & {$\beta_{6,\text{region}_{3},\text{age}_{5}}$} & 0.0088 & 0.0179 & -0.0261 & 0.0433 \\ 
			& {$\beta_{5,\text{deprivation}_{5},\text{age}_{5}}$} & -0.0388 & 0.0207 & -0.0820 & 0.0012 & {     } & {$\beta_{6,\text{region}_{4},\text{age}_{5}}$} & -0.0365 & 0.0215 & -0.0781 & 0.0056 \\ 
			& {$\beta_{5,\text{deprivation}_{1},\text{age}_{6}}$} & -0.0537 & 0.0183 & -0.0893 & -0.0169 & {     } & {$\beta_{6,\text{region}_{5},\text{age}_{5}}$} & 0.0282 & 0.0304 & -0.0336 & 0.0866 \\ 
			& {$\beta_{5,\text{deprivation}_{2},\text{age}_{6}}$} & -0.0714 & 0.0151 & -0.1018 & -0.0433 & {     } & {$\beta_{6,\text{region}_{6},\text{age}_{5}}$} & 0.0347 & 0.0250 & -0.0139 & 0.0845 \\ 
			& {$\beta_{5,\text{deprivation}_{3},\text{age}_{6}}$} & -0.0209 & 0.0133 & -0.0459 & 0.0046 & {     } & {$\beta_{6,\text{region}_{7},\text{age}_{5}}$} & 0.0055 & 0.0211 & -0.0382 & 0.0485 \\ 
			& {$\beta_{5,\text{deprivation}_{4},\text{age}_{6}}$} & 0.0082 & 0.0130 & -0.0176 & 0.0325 & {     } & {$\beta_{6,\text{region}_{8},\text{age}_{5}}$} & 0.0028 & 0.0187 & -0.0339 & 0.0393 \\ 
			& {$\beta_{5,\text{deprivation}_{5},\text{age}_{6}}$} & 0.0280 & 0.0125 & 0.0026 & 0.0522 & {     } & {$\beta_{6,\text{region}_{9},\text{age}_{5}}$} & -0.0058 & 0.0183 & -0.0423 & 0.0295 \\ 
			& {$\beta_{5,\text{deprivation}_{1},\text{age}_{7}}$} & 0.0626 & 0.0122 & 0.0392 & 0.0874 & {     } & {$\beta_{6,\text{region}_{1},\text{age}_{6}}$} & -0.0362 & 0.0174 & -0.0700 & -0.0036 \\ 
			\hline
		\end{tabular}
	}
\end{table}

% latex table generated in R 4.1.2 by xtable 1.8-4 package
% Thu Nov 16 22:24:32 2023
\begin{table}[H]
	\centering
	%\captionsetup{labelformat=empty}
	%	\caption{{{Estimated coefficients for lung cancer mortality in women (continuing) based on \eqref{eq:FemaleLungCODLocationPrmtr2}. }}}
	\addtolength{\tabcolsep}{-4pt}
	\resizebox{0.8\columnwidth}{!}{
		\begin{tabular}{llrrrr|llrrrr}
			\hline
			Covariate& Parameter & Mean & SD & \%2.5 & \%97.5 &Covariate& Parameter & Mean & SD & \%2.5 & \%97.5 \\ 
			\hline
			& {$\beta_{6,\text{region}_{2},\text{age}_{6}}$} & -0.0016 & 0.0167 & -0.0339 & 0.0316 & {     } & {$\kappa^{*}_{1,\text{year}_{3}}$} & 0.3224 & 0.0532 & 0.2238 & 0.4296 \\ 
			& {$\beta_{6,\text{region}_{3},\text{age}_{6}}$} & -0.0274 & 0.0212 & -0.0721 & 0.0120 & {     } & {$\kappa^{*}_{1,\text{year}_{4}}$} & 0.3353 & 0.0606 & 0.2211 & 0.4599 \\ 
			& {$\beta_{6,\text{region}_{4},\text{age}_{6}}$} & -0.0904 & 0.0253 & -0.1400 & -0.0406 & {     } & {$\kappa^{*}_{1,\text{year}_{5}}$} & 0.3472 & 0.0655 & 0.2217 & 0.4758 \\ 
			& {$\beta_{6,\text{region}_{5},\text{age}_{6}}$} & -0.1121 & 0.0240 & -0.1575 & -0.0633 & {     } & {$\kappa^{*}_{1,\text{year}_{6}}$} & 0.3596 & 0.0721 & 0.2189 & 0.5008 \\ 
			& {$\beta_{6,\text{region}_{6},\text{age}_{6}}$} & -0.0584 & 0.0221 & -0.1020 & -0.0153 & {     } & {$\kappa^{*}_{1,\text{year}_{7}}$} & 0.3695 & 0.0796 & 0.2139 & 0.5235 \\ 
			& {$\beta_{6,\text{region}_{7},\text{age}_{6}}$} & -0.0204 & 0.0191 & -0.0569 & 0.0158 & {     } & {$\kappa^{*}_{1,\text{year}_{8}}$} & 0.3822 & 0.0854 & 0.2199 & 0.5491 \\ 
			& {$\beta_{6,\text{region}_{8},\text{age}_{6}}$} & -0.0135 & 0.0167 & -0.0482 & 0.0185 & {     } & {$\kappa^{*}_{1,\text{year}_{9}}$} & 0.3936 & 0.0915 & 0.2164 & 0.5855 \\ 
			& {$\beta_{6,\text{region}_{9},\text{age}_{6}}$} & 0.0310 & 0.0164 & -0.0001 & 0.0641 & {     } & {$\kappa^{*}_{1,\text{year}_{10}}$} & 0.4060 & 0.0970 & 0.2287 & 0.6055 \\ 
			& {$\beta_{6,\text{region}_{1},\text{age}_{7}}$} & 0.1050 & 0.0175 & 0.0692 & 0.1369 & {     } & {$\kappa^{*}_{1,\text{year}_{11}}$} & 0.4174 & 0.1028 & 0.2307 & 0.6234 \\ 
			& {$\beta_{6,\text{region}_{2},\text{age}_{7}}$} & 0.1588 & 0.0204 & 0.1195 & 0.1991 & {     } & {$\kappa^{*}_{1,\text{year}_{12}}$} & 0.4292 & 0.1059 & 0.2389 & 0.6426 \\ 
			& {$\beta_{6,\text{region}_{3},\text{age}_{7}}$} & 0.0845 & 0.0238 & 0.0360 & 0.1307 & {     } & {$\kappa^{*}_{1,\text{year}_{13}}$} & 0.4405 & 0.1112 & 0.2390 & 0.6685 \\ 
			& {$\beta_{6,\text{region}_{4},\text{age}_{7}}$} & 0.0423 & 0.0221 & 0.0005 & 0.0867 & {     } & {$\kappa^{*}_{1,\text{year}_{14}}$} & 0.4509 & 0.1160 & 0.2306 & 0.6871 \\ 
			& {$\beta_{6,\text{region}_{5},\text{age}_{7}}$} & 0.0304 & 0.0188 & -0.0089 & 0.0665 & {     } & {$\kappa^{*}_{1,\text{year}_{15}}$} & 0.4623 & 0.1203 & 0.2380 & 0.7114 \\ 
			& {$\beta_{6,\text{region}_{6},\text{age}_{7}}$} & -0.0044 & 0.0165 & -0.0358 & 0.0283 & {     } & {$\kappa^{*}_{1,\text{year}_{16}}$} & 0.4728 & 0.1240 & 0.2419 & 0.7243 \\ 
			& {$\beta_{6,\text{region}_{7},\text{age}_{7}}$} & -0.0120 & 0.0162 & -0.0433 & 0.0199 & {     } & {$\kappa^{*}_{1,\text{year}_{17}}$} & 0.4856 & 0.1271 & 0.2570 & 0.7562 \\ 
			& {$\beta_{6,\text{region}_{8},\text{age}_{7}}$} & -0.0449 & 0.0157 & -0.0749 & -0.0154 & {     } & {$\kappa^{*}_{1,\text{year}_{18}}$} & 0.4956 & 0.1324 & 0.2544 & 0.7788 \\ 
			& {$\beta_{6,\text{region}_{9},\text{age}_{7}}$} & -0.0635 & 0.0158 & -0.0944 & -0.0326 & {     } & {$\kappa_{2,\text{year}_{2}}$} & -0.0178 & 0.0123 & -0.0431 & 0.0049 \\ 
			& {$\beta_{6,\text{region}_{1},\text{age}_{8}}$} & -0.0325 & 0.0175 & -0.0665 & 0.0015 & {     } & {$\kappa_{2,\text{year}_{3}}$} & -0.0189 & 0.0138 & -0.0481 & 0.0090 \\ 
			& {$\beta_{6,\text{region}_{2},\text{age}_{8}}$} & 0.1998 & 0.0278 & 0.1482 & 0.2558 & {     } & {$\kappa_{2,\text{year}_{4}}$} & -0.0251 & 0.0123 & -0.0512 & -0.0031 \\ 
			& {$\beta_{6,\text{region}_{3},\text{age}_{8}}$} & 0.0540 & 0.0254 & 0.0023 & 0.1022 & {     } & {$\kappa_{2,\text{year}_{5}}$} & -0.0185 & 0.0129 & -0.0425 & 0.0064 \\ 
			& {$\beta_{6,\text{region}_{4},\text{age}_{8}}$} & 0.0112 & 0.0220 & -0.0292 & 0.0559 & {     } & {$\kappa_{2,\text{year}_{6}}$} & -0.0268 & 0.0136 & -0.0527 & 0.0010 \\ 
			& {$\beta_{6,\text{region}_{5},\text{age}_{8}}$} & -0.0144 & 0.0193 & -0.0520 & 0.0243 & {     } & {$\kappa_{2,\text{year}_{7}}$} & -0.0200 & 0.0136 & -0.0466 & 0.0057 \\ 
			& {$\beta_{6,\text{region}_{6},\text{age}_{8}}$} & -0.0580 & 0.0181 & -0.0923 & -0.0219 & {     } & {$\kappa_{2,\text{year}_{8}}$} & -0.0339 & 0.0136 & -0.0593 & -0.0076 \\ 
			& {$\beta_{6,\text{region}_{7},\text{age}_{8}}$} & -0.0703 & 0.0182 & -0.1059 & -0.0338 & {     } & {$\kappa_{2,\text{year}_{9}}$} & -0.0373 & 0.0143 & -0.0672 & -0.0100 \\ 
			& {$\beta_{6,\text{region}_{8},\text{age}_{8}}$} & -0.0763 & 0.0191 & -0.1114 & -0.0373 & {     } & {$\kappa_{2,\text{year}_{10}}$} & -0.0313 & 0.0138 & -0.0589 & -0.0062 \\ 
			& {$\beta_{6,\text{region}_{9},\text{age}_{8}}$} & -0.0460 & 0.0210 & -0.0862 & -0.0023 & {     } & {$\kappa_{2,\text{year}_{11}}$} & -0.0521 & 0.0134 & -0.0801 & -0.0268 \\ 
			& {$\beta_{7,\text{region}_{1}}$} & -0.0294 & 0.0098 & -0.0480 & -0.0103 & {     } & {$\kappa_{2,\text{year}_{12}}$} & -0.0604 & 0.0131 & -0.0861 & -0.0358 \\ 
			& {$\beta_{7,\text{region}_{2}}$} & -0.0461 & 0.0068 & -0.0598 & -0.0326 & {     } & {$\kappa_{2,\text{year}_{13}}$} & -0.0655 & 0.0141 & -0.0946 & -0.0384 \\ 
			& {$\beta_{7,\text{region}_{3}}$} & -0.0128 & 0.0070 & -0.0267 & 0.0008 & {     } & {$\kappa_{2,\text{year}_{14}}$} & -0.0664 & 0.0141 & -0.0965 & -0.0394 \\ 
			& {$\beta_{7,\text{region}_{4}}$} & 0.0103 & 0.0085 & -0.0065 & 0.0271 & {     } & {$\kappa_{2,\text{year}_{15}}$} & -0.0774 & 0.0135 & -0.1044 & -0.0511 \\ 
			& {$\beta_{7,\text{region}_{5}}$} & 0.0091 & 0.0076 & -0.0058 & 0.0242 & {     } & {$\kappa_{2,\text{year}_{16}}$} & -0.0811 & 0.0137 & -0.1090 & -0.0542 \\ 
			& {$\beta_{7,\text{region}_{6}}$} & 0.0344 & 0.0085 & 0.0180 & 0.0518 & {     } & {$\kappa_{2,\text{year}_{17}}$} & -0.0912 & 0.0140 & -0.1199 & -0.0643 \\ 
			& {$\beta_{7,\text{region}_{7}}$} & 0.0741 & 0.0081 & 0.0584 & 0.0893 & {     } & {$\kappa_{2,\text{year}_{18}}$} & -0.0975 & 0.0144 & -0.1257 & -0.0708 \\ 
			& {$\beta_{7,\text{region}_{8}}$} & -0.0145 & 0.0076 & -0.0289 & -0.0000 & {     } & {$\kappa^{*}_{2,\text{year}_{1}}$} & -0.0997 & 0.0226 & -0.1451 & -0.0578 \\ 
			& {$\beta_{7,\text{region}_{9}}$} & -0.0252 & 0.0090 & -0.0427 & -0.0080 & {     } & {$\kappa^{*}_{2,\text{year}_{2}}$} & -0.1022 & 0.0293 & -0.1604 & -0.0465 \\ 
			& {$\beta_{8}$} & -0.3542 & 0.0223 & -0.3958 & -0.3101 & {     } & {$\kappa^{*}_{2,\text{year}_{3}}$} & -0.1053 & 0.0356 & -0.1818 & -0.0373 \\ 
			& {$\kappa_{1,\text{year}_{2}}$} & 0.0235 & 0.0146 & -0.0104 & 0.0471 & {     } & {$\kappa^{*}_{2,\text{year}_{4}}$} & -0.1077 & 0.0411 & -0.1975 & -0.0262 \\ 
			& {$\kappa_{1,\text{year}_{3}}$} & 0.0533 & 0.0162 & 0.0204 & 0.0830 & {     } & {$\kappa^{*}_{2,\text{year}_{5}}$} & -0.1099 & 0.0459 & -0.2049 & -0.0220 \\ 
			& {$\kappa_{1,\text{year}_{4}}$} & 0.0593 & 0.0159 & 0.0264 & 0.0871 & {     } & {$\kappa^{*}_{2,\text{year}_{6}}$} & -0.1126 & 0.0510 & -0.2145 & -0.0076 \\ 
			& {$\kappa_{1,\text{year}_{5}}$} & 0.1041 & 0.0163 & 0.0709 & 0.1347 & {     } & {$\kappa^{*}_{2,\text{year}_{7}}$} & -0.1155 & 0.0550 & -0.2309 & -0.0057 \\ 
			& {$\kappa_{1,\text{year}_{6}}$} & 0.1529 & 0.0147 & 0.1207 & 0.1791 & {     } & {$\kappa^{*}_{2,\text{year}_{8}}$} & -0.1175 & 0.0590 & -0.2329 & 0.0047 \\ 
			& {$\kappa_{1,\text{year}_{7}}$} & 0.1924 & 0.0160 & 0.1578 & 0.2218 & {     } & {$\kappa^{*}_{2,\text{year}_{9}}$} & -0.1198 & 0.0638 & -0.2485 & 0.0106 \\ 
			& {$\kappa_{1,\text{year}_{8}}$} & 0.2208 & 0.0165 & 0.1849 & 0.2482 & {     } & {$\kappa^{*}_{2,\text{year}_{10}}$} & -0.1224 & 0.0678 & -0.2559 & 0.0152 \\ 
			& {$\kappa_{1,\text{year}_{9}}$} & 0.2292 & 0.0178 & 0.1940 & 0.2606 & {     } & {$\kappa^{*}_{2,\text{year}_{11}}$} & -0.1263 & 0.0721 & -0.2719 & 0.0126 \\ 
			& {$\kappa_{1,\text{year}_{10}}$} & 0.2505 & 0.0181 & 0.2156 & 0.2848 & {     } & {$\kappa^{*}_{2,\text{year}_{12}}$} & -0.1278 & 0.0760 & -0.2891 & 0.0153 \\ 
			& {$\kappa_{1,\text{year}_{11}}$} & 0.2641 & 0.0193 & 0.2258 & 0.3002 & {     } & {$\kappa^{*}_{2,\text{year}_{13}}$} & -0.1312 & 0.0802 & -0.2892 & 0.0197 \\ 
			& {$\kappa_{1,\text{year}_{12}}$} & 0.2813 & 0.0200 & 0.2412 & 0.3181 & {     } & {$\kappa^{*}_{2,\text{year}_{14}}$} & -0.1341 & 0.0845 & -0.3026 & 0.0290 \\ 
			& {$\kappa_{1,\text{year}_{13}}$} & 0.2906 & 0.0202 & 0.2498 & 0.3309 & {     } & {$\kappa^{*}_{2,\text{year}_{15}}$} & -0.1363 & 0.0881 & -0.3160 & 0.0328 \\ 
			& {$\kappa_{1,\text{year}_{14}}$} & 0.3061 & 0.0208 & 0.2665 & 0.3448 & {     } & {$\kappa^{*}_{2,\text{year}_{16}}$} & -0.1388 & 0.0923 & -0.3199 & 0.0360 \\ 
			& {$\kappa_{1,\text{year}_{15}}$} & 0.3145 & 0.0221 & 0.2719 & 0.3557 & {     } & {$\kappa^{*}_{2,\text{year}_{17}}$} & -0.1407 & 0.0956 & -0.3304 & 0.0452 \\ 
			& {$\kappa_{1,\text{year}_{16}}$} & 0.3138 & 0.0234 & 0.2706 & 0.3566 & {     } & {$\kappa^{*}_{2,\text{year}_{18}}$} & -0.1437 & 0.0998 & -0.3472 & 0.0584 \\ 
			& {$\kappa_{1,\text{year}_{17}}$} & 0.3049 & 0.0238 & 0.2587 & 0.3473 & {     } & {$\psi_{\kappa_1}$} & 0.0112 & 0.0040 & 0.0048 & 0.0204 \\ 
			& {$\kappa_{1,\text{year}_{18}}$} & 0.2881 & 0.0240 & 0.2400 & 0.3329 & {     } & {$\psi_{\kappa_2}$} & -0.0026 & 0.0032 & -0.0095 & 0.0031 \\ 
			& {$\kappa^{*}_{1,\text{year}_{1}}$} & 0.2999 & 0.0349 & 0.2328 & 0.3711 & {     } & {$\sigma^2$} & 0.0068 & 0.0007 & 0.0056 & 0.0082 \\ 
			& {$\kappa^{*}_{1,\text{year}_{2}}$} & 0.3113 & 0.0455 & 0.2230 & 0.4041 & {     } & {$\sigma^2_{\psi_{\kappa_2}}$} & 0.0000 & 0.0000 & 0.0000 & 0.0000 \\ 
			\hline
		\end{tabular}
	}
\end{table}

%----------------------------------------------------------------------------------------
% PARAMETER ESTIMATES - LUNG CANCER, MEN 
%----------------------------------------------------------------------------------------	

% latex table generated in R 4.1.2 by xtable 1.8-4 package
% Fri Nov 17 14:05:16 2023
\begin{table}[H]
	\centering
	%\captionsetup{labelformat=empty}
	\caption{{{Estimated coefficients for lung cancer mortality in men based on \eqref{eq:MaleLungCODLocationPrmtr2}. }}}
	\addtolength{\tabcolsep}{-4pt}
	\resizebox{0.6\columnwidth}{!}{
		\begin{tabular}{llrrrr|llrrrr}
			\hline
			Covariate& Parameter & Mean & SD & \%2.5 & \%97.5 &Covariate& Parameter & Mean & SD & \%2.5 & \%97.5 \\ 
			\hline
			& {$\beta_{0}$} & -6.8200 & 0.0326 & -6.8960 & -6.7620 & {     } & {$\kappa_{1,\text{year}_{3}}$} & 0.0186 & 0.0105 & -0.0019 & 0.0380 \\ 
			& {$\beta_{1,\text{age}_{1}}$} & -1.9120 & 0.0185 & -1.9480 & -1.8740 & {     } & {$\kappa_{1,\text{year}_{4}}$} & 0.0168 & 0.0119 & -0.0041 & 0.0414 \\ 
			& {$\beta_{1,\text{age}_{2}}$} & -0.9831 & 0.0088 & -1.0010 & -0.9657 & {     } & {$\kappa_{1,\text{year}_{5}}$} & 0.0319 & 0.0122 & 0.0084 & 0.0568 \\ 
			& {$\beta_{1,\text{age}_{3}}$} & -0.3204 & 0.0074 & -0.3352 & -0.3056 & {     } & {$\kappa_{1,\text{year}_{6}}$} & 0.0637 & 0.0146 & 0.0364 & 0.0917 \\ 
			& {$\beta_{1,\text{age}_{4}}$} & -0.0378 & 0.0088 & -0.0553 & -0.0207 & {     } & {$\kappa_{1,\text{year}_{7}}$} & 0.0748 & 0.0142 & 0.0485 & 0.1053 \\ 
			& {$\beta_{1,\text{age}_{5}}$} & 0.3721 & 0.0088 & 0.3539 & 0.3890 & {     } & {$\kappa_{1,\text{year}_{8}}$} & 0.1005 & 0.0164 & 0.0699 & 0.1361 \\ 
			& {$\beta_{1,\text{age}_{6}}$} & 0.7774 & 0.0059 & 0.7657 & 0.7893 & {     } & {$\kappa_{1,\text{year}_{9}}$} & 0.1094 & 0.0168 & 0.0775 & 0.1448 \\ 
			& {$\beta_{1,\text{age}_{7}}$} & 0.9769 & 0.0065 & 0.9640 & 0.9889 & {     } & {$\kappa_{1,\text{year}_{10}}$} & 0.1122 & 0.0198 & 0.0746 & 0.1508 \\ 
			& {$\beta_{1,\text{age}_{8}}$} & 1.1270 & 0.0069 & 1.1140 & 1.1400 & {     } & {$\kappa_{1,\text{year}_{11}}$} & 0.1249 & 0.0196 & 0.0935 & 0.1699 \\ 
			& {$\beta_{2,\text{region}_{1}}$} & -0.5047 & 0.0110 & -0.5255 & -0.4826 & {     } & {$\kappa_{1,\text{year}_{12}}$} & 0.1298 & 0.0213 & 0.0920 & 0.1771 \\ 
			& {$\beta_{2,\text{region}_{2}}$} & 0.1296 & 0.0052 & 0.1195 & 0.1400 & {     } & {$\kappa_{1,\text{year}_{13}}$} & 0.1469 & 0.0230 & 0.1062 & 0.1950 \\ 
			& {$\beta_{2,\text{region}_{3}}$} & -0.1027 & 0.0062 & -0.1150 & -0.0901 & {     } & {$\kappa_{1,\text{year}_{14}}$} & 0.1562 & 0.0252 & 0.1121 & 0.2073 \\ 
			& {$\beta_{2,\text{region}_{4}}$} & 0.2865 & 0.0075 & 0.2714 & 0.3011 & {     } & {$\kappa_{1,\text{year}_{15}}$} & 0.1463 & 0.0263 & 0.0988 & 0.2029 \\ 
			& {$\beta_{2,\text{region}_{5}}$} & 0.2578 & 0.0075 & 0.2427 & 0.2722 & {     } & {$\kappa_{1,\text{year}_{16}}$} & 0.1647 & 0.0285 & 0.1161 & 0.2254 \\ 
			& {$\beta_{2,\text{region}_{6}}$} & 0.1515 & 0.0066 & 0.1383 & 0.1636 & {     } & {$\kappa_{1,\text{year}_{17}}$} & 0.1515 & 0.0295 & 0.0982 & 0.2175 \\ 
			& {$\beta_{2,\text{region}_{7}}$} & -0.8694 & 0.0108 & -0.8904 & -0.8468 & {     } & {$\kappa_{1,\text{year}_{18}}$} & 0.1456 & 0.0309 & 0.0915 & 0.2124 \\ 
			& {$\beta_{2,\text{region}_{8}}$} & 0.3276 & 0.0070 & 0.3142 & 0.3420 & {     } & {$\kappa^{*}_{1,\text{year}_{1}}$} & 0.1509 & 0.0498 & 0.0554 & 0.2516 \\ 
			& {$\beta_{2,\text{region}_{9}}$} & 0.3238 & 0.0084 & 0.3077 & 0.3407 & {     } & {$\kappa^{*}_{1,\text{year}_{2}}$} & 0.1573 & 0.0645 & 0.0349 & 0.2834 \\ 
			& {$\beta_{3,\text{deprivation}_{1}}$} & 3.0050 & 0.0314 & 2.9370 & 3.0660 & {     } & {$\kappa^{*}_{1,\text{year}_{3}}$} & 0.1619 & 0.0760 & 0.0097 & 0.3099 \\ 
			& {$\beta_{3,\text{deprivation}_{2}}$} & 1.2720 & 0.0134 & 1.2420 & 1.3000 & {     } & {$\kappa^{*}_{1,\text{year}_{4}}$} & 0.1665 & 0.0862 & -0.0043 & 0.3388 \\ 
			& {$\beta_{3,\text{deprivation}_{3}}$} & -0.1210 & 0.0051 & -0.1306 & -0.1108 & {     } & {$\kappa^{*}_{1,\text{year}_{5}}$} & 0.1715 & 0.0960 & -0.0191 & 0.3537 \\ 
			& {$\beta_{3,\text{deprivation}_{4}}$} & -1.4170 & 0.0151 & -1.4450 & -1.3850 & {     } & {$\kappa^{*}_{1,\text{year}_{6}}$} & 0.1760 & 0.1051 & -0.0347 & 0.3808 \\ 
			& {$\beta_{3,\text{deprivation}_{5}}$} & -2.7390 & 0.0281 & -2.7960 & -2.6770 & {     } & {$\kappa^{*}_{1,\text{year}_{7}}$} & 0.1813 & 0.1155 & -0.0516 & 0.4082 \\ 
			& {$\beta_{4}$} & 1.7610 & 0.0214 & 1.7170 & 1.8040 & {     } & {$\kappa^{*}_{1,\text{year}_{8}}$} & 0.1860 & 0.1237 & -0.0570 & 0.4335 \\ 
			& {$\beta_{5,\text{deprivation}_{1},\text{age}_{1}}$} & 0.1282 & 0.0135 & 0.1013 & 0.1543 & {     } & {$\kappa^{*}_{1,\text{year}_{9}}$} & 0.1911 & 0.1310 & -0.0692 & 0.4480 \\ 
			& {$\beta_{5,\text{deprivation}_{2},\text{age}_{1}}$} & 0.1466 & 0.0127 & 0.1216 & 0.1720 & {     } & {$\kappa^{*}_{1,\text{year}_{10}}$} & 0.1961 & 0.1403 & -0.0899 & 0.4747 \\ 
			& {$\beta_{5,\text{deprivation}_{3},\text{age}_{1}}$} & 0.1108 & 0.0105 & 0.0901 & 0.1320 & {     } & {$\kappa^{*}_{1,\text{year}_{11}}$} & 0.2001 & 0.1494 & -0.1095 & 0.4897 \\ 
			& {$\beta_{5,\text{deprivation}_{4},\text{age}_{1}}$} & 0.0657 & 0.0090 & 0.0483 & 0.0828 & {     } & {$\kappa^{*}_{1,\text{year}_{12}}$} & 0.2075 & 0.1567 & -0.1180 & 0.5154 \\ 
			& {$\beta_{5,\text{deprivation}_{5},\text{age}_{1}}$} & -0.0059 & 0.0086 & -0.0226 & 0.0115 & {     } & {$\kappa^{*}_{1,\text{year}_{13}}$} & 0.2133 & 0.1635 & -0.1248 & 0.5335 \\ 
			& {$\beta_{5,\text{deprivation}_{1},\text{age}_{2}}$} & -0.0708 & 0.0083 & -0.0867 & -0.0545 & {     } & {$\kappa^{*}_{1,\text{year}_{14}}$} & 0.2185 & 0.1692 & -0.1257 & 0.5457 \\ 
			& {$\beta_{5,\text{deprivation}_{2},\text{age}_{2}}$} & -0.1486 & 0.0093 & -0.1670 & -0.1302 & {     } & {$\kappa^{*}_{1,\text{year}_{15}}$} & 0.2232 & 0.1751 & -0.1326 & 0.5643 \\ 
			& {$\beta_{5,\text{deprivation}_{3},\text{age}_{2}}$} & -0.2260 & 0.0118 & -0.2487 & -0.2026 & {     } & {$\kappa^{*}_{1,\text{year}_{16}}$} & 0.2287 & 0.1814 & -0.1341 & 0.5802 \\ 
			& {$\beta_{5,\text{deprivation}_{4},\text{age}_{2}}$} & 0.0638 & 0.0159 & 0.0316 & 0.0938 & {     } & {$\kappa^{*}_{1,\text{year}_{17}}$} & 0.2351 & 0.1881 & -0.1330 & 0.5934 \\ 
			& {$\beta_{5,\text{deprivation}_{5},\text{age}_{2}}$} & 0.0445 & 0.0135 & 0.0164 & 0.0705 & {     } & {$\kappa^{*}_{1,\text{year}_{18}}$} & 0.2402 & 0.1940 & -0.1444 & 0.6009 \\ 
			& {$\beta_{5,\text{deprivation}_{1},\text{age}_{3}}$} & 0.0278 & 0.0117 & 0.0051 & 0.0503 & {     } & {$\kappa_{2,\text{year}_{2}}$} & 0.0003 & 0.0138 & -0.0270 & 0.0277 \\ 
			& {$\beta_{5,\text{deprivation}_{2},\text{age}_{3}}$} & 0.0391 & 0.0099 & 0.0191 & 0.0594 & {     } & {$\kappa_{2,\text{year}_{3}}$} & 0.0003 & 0.0137 & -0.0256 & 0.0280 \\ 
			& {$\beta_{5,\text{deprivation}_{3},\text{age}_{3}}$} & 0.0052 & 0.0096 & -0.0134 & 0.0240 & {     } & {$\kappa_{2,\text{year}_{4}}$} & -0.0205 & 0.0126 & -0.0448 & 0.0063 \\ 
			& {$\beta_{5,\text{deprivation}_{4},\text{age}_{3}}$} & -0.0179 & 0.0091 & -0.0359 & -0.0008 & {     } & {$\kappa_{2,\text{year}_{5}}$} & -0.0299 & 0.0119 & -0.0521 & -0.0057 \\ 
			& {$\beta_{5,\text{deprivation}_{5},\text{age}_{3}}$} & -0.0710 & 0.0099 & -0.0904 & -0.0517 & {     } & {$\kappa_{2,\text{year}_{6}}$} & -0.0348 & 0.0119 & -0.0563 & -0.0086 \\ 
			& {$\beta_{5,\text{deprivation}_{1},\text{age}_{4}}$} & -0.0916 & 0.0121 & -0.1157 & -0.0680 & {     } & {$\kappa_{2,\text{year}_{7}}$} & -0.0387 & 0.0140 & -0.0686 & -0.0129 \\ 
			& {$\beta_{5,\text{deprivation}_{2},\text{age}_{4}}$} & 0.0056 & 0.0176 & -0.0276 & 0.0434 & {     } & {$\kappa_{2,\text{year}_{8}}$} & -0.0480 & 0.0118 & -0.0716 & -0.0270 \\ 
			& {$\beta_{5,\text{deprivation}_{3},\text{age}_{4}}$} & 0.0044 & 0.0146 & -0.0230 & 0.0339 & {     } & {$\kappa_{2,\text{year}_{9}}$} & -0.0440 & 0.0123 & -0.0695 & -0.0218 \\ 
			& {$\beta_{5,\text{deprivation}_{4},\text{age}_{4}}$} & -0.0242 & 0.0126 & -0.0484 & 0.0015 & {     } & {$\kappa_{2,\text{year}_{10}}$} & -0.0556 & 0.0118 & -0.0773 & -0.0290 \\ 
			& {$\beta_{5,\text{deprivation}_{5},\text{age}_{4}}$} & 0.0020 & 0.0100 & -0.0178 & 0.0212 & {     } & {$\kappa_{2,\text{year}_{11}}$} & -0.0511 & 0.0119 & -0.0750 & -0.0251 \\ 
			& {$\beta_{5,\text{deprivation}_{1},\text{age}_{5}}$} & -0.0077 & 0.0098 & -0.0275 & 0.0116 & {     } & {$\kappa_{2,\text{year}_{12}}$} & -0.0532 & 0.0129 & -0.0774 & -0.0275 \\ 
			& {$\beta_{5,\text{deprivation}_{2},\text{age}_{5}}$} & 0.0100 & 0.0092 & -0.0082 & 0.0279 & {     } & {$\kappa_{2,\text{year}_{13}}$} & -0.0680 & 0.0125 & -0.0938 & -0.0437 \\ 
			& {$\beta_{5,\text{deprivation}_{3},\text{age}_{5}}$} & 0.0041 & 0.0096 & -0.0150 & 0.0230 & {     } & {$\kappa_{2,\text{year}_{14}}$} & -0.0798 & 0.0125 & -0.1044 & -0.0552 \\ 
			& {$\beta_{5,\text{deprivation}_{4},\text{age}_{5}}$} & 0.0057 & 0.0116 & -0.0168 & 0.0295 & {     } & {$\kappa_{2,\text{year}_{15}}$} & -0.0855 & 0.0123 & -0.1082 & -0.0605 \\ 
			& {$\beta_{5,\text{deprivation}_{5},\text{age}_{5}}$} & -0.0834 & 0.0183 & -0.1206 & -0.0491 & {     } & {$\kappa_{2,\text{year}_{16}}$} & -0.0804 & 0.0126 & -0.1049 & -0.0542 \\ 
			& {$\beta_{5,\text{deprivation}_{1},\text{age}_{6}}$} & -0.0724 & 0.0165 & -0.1048 & -0.0397 & {     } & {$\kappa_{2,\text{year}_{17}}$} & -0.0810 & 0.0122 & -0.1039 & -0.0552 \\ 
			& {$\beta_{5,\text{deprivation}_{2},\text{age}_{6}}$} & -0.0474 & 0.0127 & -0.0739 & -0.0223 & {     } & {$\kappa_{2,\text{year}_{18}}$} & -0.0847 & 0.0126 & -0.1094 & -0.0588 \\ 
			& {$\beta_{5,\text{deprivation}_{3},\text{age}_{6}}$} & -0.0418 & 0.0103 & -0.0617 & -0.0216 & {     } & {$\kappa^{*}_{2,\text{year}_{1}}$} & -0.0952 & 0.0395 & -0.1749 & -0.0175 \\ 
			& {$\beta_{5,\text{deprivation}_{4},\text{age}_{6}}$} & 0.0048 & 0.0096 & -0.0143 & 0.0236 & {     } & {$\kappa^{*}_{2,\text{year}_{2}}$} & -0.1065 & 0.0551 & -0.2142 & 0.0033 \\ 
			& {$\beta_{5,\text{deprivation}_{5},\text{age}_{6}}$} & 0.0338 & 0.0095 & 0.0158 & 0.0525 & {     } & {$\kappa^{*}_{2,\text{year}_{3}}$} & -0.1165 & 0.0656 & -0.2447 & 0.0140 \\ 
			& {$\beta_{5,\text{deprivation}_{1},\text{age}_{7}}$} & 0.0784 & 0.0099 & 0.0592 & 0.0985 & {     } & {$\kappa^{*}_{2,\text{year}_{4}}$} & -0.1271 & 0.0781 & -0.2823 & 0.0249 \\ 
			& {$\beta_{5,\text{deprivation}_{2},\text{age}_{7}}$} & 0.1280 & 0.0129 & 0.1028 & 0.1531 & {     } & {$\kappa^{*}_{2,\text{year}_{5}}$} & -0.1371 & 0.0882 & -0.3114 & 0.0427 \\ 
			& {$\beta_{5,\text{deprivation}_{3},\text{age}_{7}}$} & -0.1142 & 0.0209 & -0.1592 & -0.0742 & {     } & {$\kappa^{*}_{2,\text{year}_{6}}$} & -0.1489 & 0.0966 & -0.3392 & 0.0406 \\ 
			& {$\beta_{5,\text{deprivation}_{4},\text{age}_{7}}$} & -0.1231 & 0.0189 & -0.1605 & -0.0868 & {     } & {$\kappa^{*}_{2,\text{year}_{7}}$} & -0.1596 & 0.1049 & -0.3703 & 0.0389 \\ 
			& {$\beta_{5,\text{deprivation}_{5},\text{age}_{7}}$} & -0.0670 & 0.0145 & -0.0934 & -0.0366 & {     } & {$\kappa^{*}_{2,\text{year}_{8}}$} & -0.1691 & 0.1131 & -0.4038 & 0.0427 \\ 
			& {$\beta_{5,\text{deprivation}_{1},\text{age}_{8}}$} & -0.0651 & 0.0120 & -0.0878 & -0.0422 & {     } & {$\kappa^{*}_{2,\text{year}_{9}}$} & -0.1796 & 0.1216 & -0.4359 & 0.0547 \\ 
			& {$\beta_{5,\text{deprivation}_{2},\text{age}_{8}}$} & 0.0036 & 0.0104 & -0.0173 & 0.0228 & {     } & {$\kappa^{*}_{2,\text{year}_{10}}$} & -0.1911 & 0.1281 & -0.4500 & 0.0554 \\ 
			& {$\beta_{5,\text{deprivation}_{3},\text{age}_{8}}$} & 0.0448 & 0.0109 & 0.0238 & 0.0667 & {     } & {$\kappa^{*}_{2,\text{year}_{11}}$} & -0.2015 & 0.1339 & -0.4754 & 0.0511 \\ 
			& {$\beta_{5,\text{deprivation}_{4},\text{age}_{8}}$} & 0.1371 & 0.0114 & 0.1163 & 0.1594 & {     } & {$\kappa^{*}_{2,\text{year}_{12}}$} & -0.2113 & 0.1400 & -0.5031 & 0.0546 \\ 
			& {$\beta_{5,\text{deprivation}_{5},\text{age}_{8}}$} & 0.1839 & 0.0134 & 0.1570 & 0.2094 & {     } & {$\kappa^{*}_{2,\text{year}_{13}}$} & -0.2212 & 0.1449 & -0.5205 & 0.0590 \\ 
			& {$\beta_{6,\text{region}_{1}}$} & 0.0010 & 0.0075 & -0.0133 & 0.0164 & {     } & {$\kappa^{*}_{2,\text{year}_{14}}$} & -0.2310 & 0.1517 & -0.5447 & 0.0617 \\ 
			& {$\beta_{6,\text{region}_{2}}$} & -0.0040 & 0.0053 & -0.0140 & 0.0061 & {     } & {$\kappa^{*}_{2,\text{year}_{15}}$} & -0.2423 & 0.1571 & -0.5715 & 0.0563 \\ 
			& {$\beta_{6,\text{region}_{3}}$} & 0.0030 & 0.0058 & -0.0085 & 0.0143 & {     } & {$\kappa^{*}_{2,\text{year}_{16}}$} & -0.2542 & 0.1613 & -0.5719 & 0.0402 \\ 
			& {$\beta_{6,\text{region}_{4}}$} & 0.0201 & 0.0068 & 0.0069 & 0.0332 & {     } & {$\kappa^{*}_{2,\text{year}_{17}}$} & -0.2656 & 0.1661 & -0.6038 & 0.0482 \\ 
			& {$\beta_{6,\text{region}_{5}}$} & 0.0074 & 0.0058 & -0.0041 & 0.0190 & {     } & {$\kappa^{*}_{2,\text{year}_{18}}$} & -0.2760 & 0.1699 & -0.6196 & 0.0458 \\ 
			& {$\beta_{6,\text{region}_{6}}$} & 0.0285 & 0.0063 & 0.0161 & 0.0406 & {     } & {$\sigma^2$} & 0.0020 & 0.0004 & 0.0014 & 0.0028 \\ 
			& {$\beta_{6,\text{region}_{7}}$} & 0.0185 & 0.0062 & 0.0062 & 0.0309 & {     } & {$\sigma^2_{\psi_{\kappa_2}}$} & 0.0001 & 0.0000 & 0.0000 & 0.0002 \\ 
			& {$\beta_{6,\text{region}_{8}}$} & -0.0340 & 0.0058 & -0.0454 & -0.0228 & {     } & {$\sigma^2_{\psi_{\kappa_1}}$} & 0.0001 & 0.0000 & 0.0000 & 0.0002 \\ 
			& {$\beta_{6,\text{region}_{9}}$} & -0.0404 & 0.0066 & -0.0533 & -0.0275 & {     } & {$\sigma^2_{\kappa_1}$} & 0.0014 & 0.0005 & 0.0008 & 0.0028 \\ 
			& {$\beta_{7}$} & -0.3810 & 0.0207 & -0.4265 & -0.3425 & {     } & {$\sigma^2_{\kappa_2}$} & 0.0014 & 0.0005 & 0.0007 & 0.0027 \\ 
			& {$\kappa_{1,\text{year}_{2}}$} & 0.0152 & 0.0101 & -0.0037 & 0.0341 & {     } & &  &  &  &  \\ 
			\hline
		\end{tabular}
	}
\end{table}

%----------------------------------------------------------------------------------------
% PARAMETER ESTIMATES - BREAST  CANCER, FEMALES 
%----------------------------------------------------------------------------------------	

% latex table generated in R 4.1.2 by xtable 1.8-4 package
% Fri Nov 17 14:24:20 2023
\begin{table}[H]
	\centering
	%\captionsetup{labelformat=empty}
	\caption{{{Estimated coefficients for breast cancer mortality based on \eqref{eq:FemaleBreastLocationPrmtr}. }}}
	\addtolength{\tabcolsep}{-4pt}
	\resizebox{0.85\columnwidth}{!}{
		\begin{tabular}{llrrrr|llrrrr}
			\hline
			Covariate& Parameter & Mean & SD & \%2.5 & \%97.5 &Covariate& Parameter & Mean & SD & \%2.5 & \%97.5 \\ 
			\hline
			& {$\beta_{0}$} & -7.4490 & 0.0231 & -7.4920 & -7.3980 & {     } & {$\kappa_{1,\text{year}_{10}}$} & -0.1507 & 0.0171 & -0.1860 & -0.1184 \\ 
			& {$\beta_{1,\text{age}_{1}}$} & -1.8890 & 0.0184 & -1.9250 & -1.8530 & {     } & {$\kappa_{1,\text{year}_{11}}$} & -0.1618 & 0.0178 & -0.1985 & -0.1304 \\ 
			& {$\beta_{1,\text{age}_{2}}$} & -1.2540 & 0.0141 & -1.2810 & -1.2260 & {     } & {$\kappa_{1,\text{year}_{12}}$} & -0.1756 & 0.0182 & -0.2114 & -0.1395 \\ 
			& {$\beta_{1,\text{age}_{3}}$} & -0.8577 & 0.0234 & -0.9029 & -0.8071 & {     } & {$\kappa_{1,\text{year}_{13}}$} & -0.1973 & 0.0189 & -0.2351 & -0.1589 \\ 
			& {$\beta_{1,\text{age}_{4}}$} & -0.4689 & 0.0230 & -0.5098 & -0.4189 & {     } & {$\kappa_{1,\text{year}_{14}}$} & -0.2132 & 0.0194 & -0.2525 & -0.1760 \\ 
			& {$\beta_{1,\text{age}_{5}}$} & -0.1101 & 0.0100 & -0.1297 & -0.0896 & {     } & {$\kappa_{1,\text{year}_{15}}$} & -0.2164 & 0.0202 & -0.2582 & -0.1770 \\ 
			& {$\beta_{1,\text{age}_{6}}$} & 0.0867 & 0.0098 & 0.0679 & 0.1056 & {     } & {$\kappa_{1,\text{year}_{16}}$} & -0.2238 & 0.0206 & -0.2666 & -0.1841 \\ 
			& {$\beta_{1,\text{age}_{7}}$} & 0.2495 & 0.0102 & 0.2295 & 0.2691 & {     } & {$\kappa_{1,\text{year}_{17}}$} & -0.2360 & 0.0211 & -0.2807 & -0.1962 \\ 
			& {$\beta_{1,\text{age}_{8}}$} & 0.4826 & 0.0104 & 0.4631 & 0.5027 & {     } & {$\kappa_{1,\text{year}_{18}}$} & -0.2448 & 0.0225 & -0.2911 & -0.2019 \\ 
			& {$\beta_{1,\text{age}_{9}}$} & 0.9480 & 0.0244 & 0.8951 & 0.9916 & {     } & {$\kappa^{*}_{1,\text{year}_{1}}$} & -0.2530 & 0.0310 & -0.3150 & -0.1953 \\ 
			& {$\beta_{1,\text{age}_{10}}$} & 1.2420 & 0.0240 & 1.1890 & 1.2840 & {     } & {$\kappa^{*}_{1,\text{year}_{2}}$} & -0.2608 & 0.0373 & -0.3341 & -0.1875 \\ 
			& {$\beta_{1,\text{age}_{11}}$} & 1.5710 & 0.0242 & 1.5170 & 1.6170 & {     } & {$\kappa^{*}_{1,\text{year}_{3}}$} & -0.2688 & 0.0440 & -0.3582 & -0.1805 \\ 
			& {$\beta_{2,\text{region}_{1}}$} & -0.0316 & 0.0112 & -0.0540 & -0.0099 & {     } & {$\kappa^{*}_{1,\text{year}_{4}}$} & -0.2758 & 0.0502 & -0.3781 & -0.1785 \\ 
			& {$\beta_{2,\text{region}_{2}}$} & -0.0144 & 0.0081 & -0.0302 & 0.0013 & {     } & {$\kappa^{*}_{1,\text{year}_{5}}$} & -0.2839 & 0.0558 & -0.4041 & -0.1789 \\ 
			& {$\beta_{2,\text{region}_{3}}$} & -0.0298 & 0.0091 & -0.0479 & -0.0117 & {     } & {$\kappa^{*}_{1,\text{year}_{6}}$} & -0.2923 & 0.0597 & -0.4121 & -0.1759 \\ 
			& {$\beta_{2,\text{region}_{4}}$} & 0.0234 & 0.0092 & 0.0053 & 0.0413 & {     } & {$\kappa^{*}_{1,\text{year}_{7}}$} & -0.2997 & 0.0645 & -0.4307 & -0.1720 \\ 
			& {$\beta_{2,\text{region}_{5}}$} & 0.0305 & 0.0084 & 0.0143 & 0.0470 & {     } & {$\kappa^{*}_{1,\text{year}_{8}}$} & -0.3074 & 0.0693 & -0.4450 & -0.1677 \\ 
			& {$\beta_{2,\text{region}_{6}}$} & 0.0283 & 0.0081 & 0.0123 & 0.0441 & {     } & {$\kappa^{*}_{1,\text{year}_{9}}$} & -0.3154 & 0.0740 & -0.4592 & -0.1708 \\ 
			& {$\beta_{2,\text{region}_{7}}$} & -0.0101 & 0.0085 & -0.0267 & 0.0066 & {     } & {$\kappa^{*}_{1,\text{year}_{10}}$} & -0.3230 & 0.0776 & -0.4727 & -0.1717 \\ 
			& {$\beta_{2,\text{region}_{8}}$} & 0.0189 & 0.0074 & 0.0045 & 0.0330 & {     } & {$\kappa^{*}_{1,\text{year}_{11}}$} & -0.3302 & 0.0823 & -0.4918 & -0.1688 \\ 
			& {$\beta_{2,\text{region}_{9}}$} & -0.0152 & 0.0083 & -0.0320 & 0.0006 & {     } & {$\kappa^{*}_{1,\text{year}_{12}}$} & -0.3384 & 0.0870 & -0.5060 & -0.1723 \\ 
			& {$\beta_{3}$} & -0.1079 & 0.0181 & -0.1409 & -0.0673 & {     } & {$\kappa^{*}_{1,\text{year}_{13}}$} & -0.3466 & 0.0910 & -0.5243 & -0.1722 \\ 
			& {$\kappa_{1,\text{year}_{2}}$} & -0.0137 & 0.0129 & -0.0391 & 0.0131 & {     } & {$\kappa^{*}_{1,\text{year}_{14}}$} & -0.3550 & 0.0949 & -0.5470 & -0.1732 \\ 
			& {$\kappa_{1,\text{year}_{3}}$} & -0.0396 & 0.0135 & -0.0673 & -0.0135 & {     } & {$\kappa^{*}_{1,\text{year}_{15}}$} & -0.3631 & 0.0989 & -0.5630 & -0.1711 \\ 
			& {$\kappa_{1,\text{year}_{4}}$} & -0.0540 & 0.0146 & -0.0835 & -0.0265 & {     } & {$\kappa^{*}_{1,\text{year}_{16}}$} & -0.3716 & 0.1029 & -0.5788 & -0.1732 \\ 
			& {$\kappa_{1,\text{year}_{5}}$} & -0.0585 & 0.0151 & -0.0882 & -0.0286 & {     } & {$\kappa^{*}_{1,\text{year}_{17}}$} & -0.3793 & 0.1073 & -0.6001 & -0.1722 \\ 
			& {$\kappa_{1,\text{year}_{6}}$} & -0.0723 & 0.0150 & -0.1019 & -0.0426 & {     } & {$\kappa^{*}_{1,\text{year}_{18}}$} & -0.3870 & 0.1116 & -0.6129 & -0.1692 \\ 
			& {$\kappa_{1,\text{year}_{7}}$} & -0.0954 & 0.0156 & -0.1256 & -0.0632 & {     } & {$\psi_{\kappa_1}$} & -0.0081 & 0.0037 & -0.0156 & -0.0013 \\ 
			& {$\kappa_{1,\text{year}_{8}}$} & -0.1065 & 0.0159 & -0.1373 & -0.0743 & {     } & {$\sigma^2$} & 0.0038 & 0.0004 & 0.0031 & 0.0046 \\ 
			& {$\kappa_{1,\text{year}_{9}}$} & -0.1269 & 0.0158 & -0.1578 & -0.0961 & {     } & {$\sigma^2_{\kappa_1}$} & 0.0004 & 0.0002 & 0.0002 & 0.0009 \\ 
			\hline
		\end{tabular}
	}
\end{table}

\newpage
\section{Age-specific fitted and projected cancer mortality} \label{sec:AgeSpecificFittedProjectedRates}

\subsection{Female lung cancer mortality}\label{SecApp:FiguresLungCODFemales_v2}

\begin{figure}[H]
	\centering
	\includegraphics[width=0.8\textwidth, angle =0]{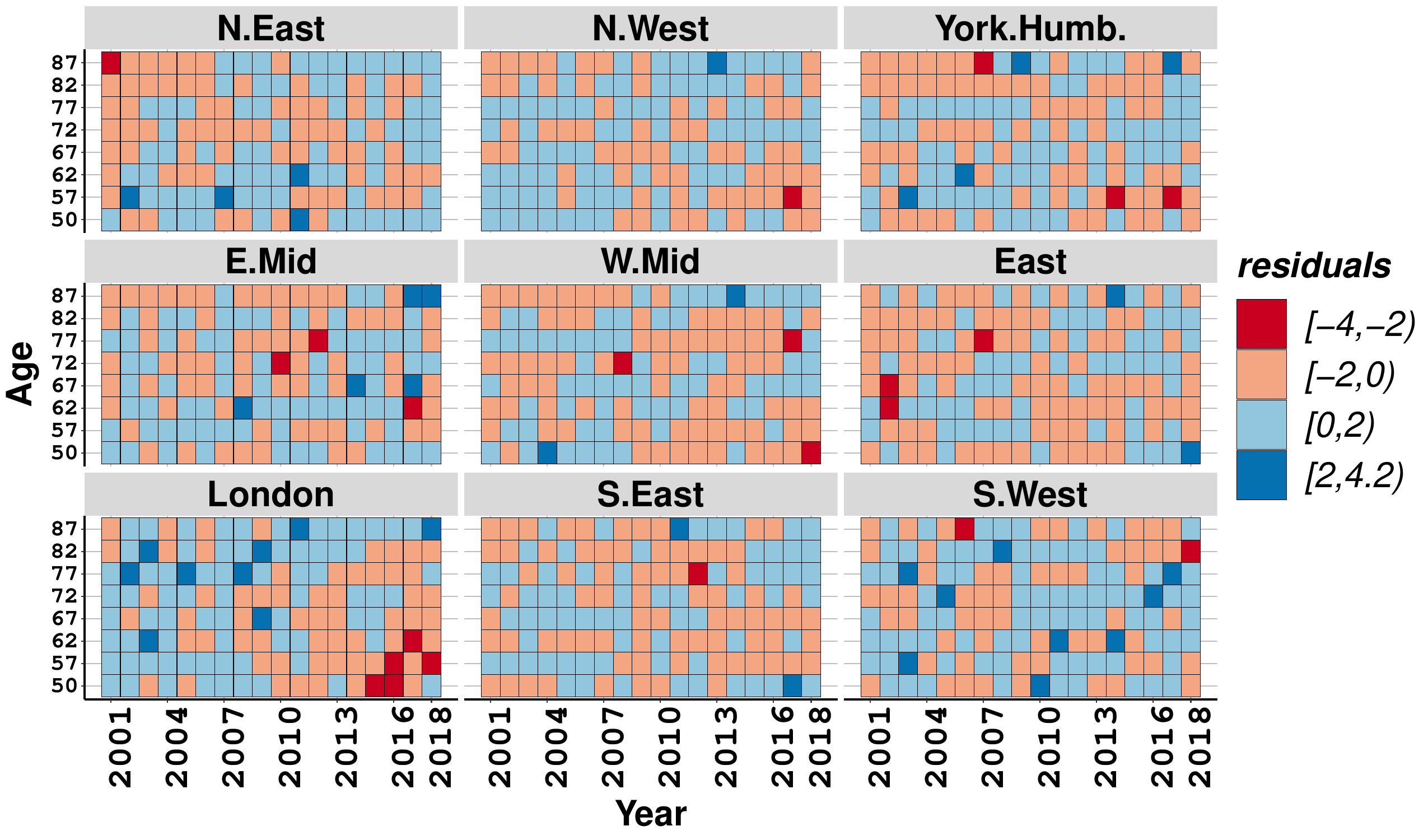}
	\caption{Heat map of Pearson residuals for female lung cancer mortality in regions of England, deprivation quintile 1 (most deprived), based on \eqref{eq:FemaleLungCODLocationPrmtr2}: orange/light blue cells indicate areas with good fit, while red/dark blue cells indicate areas with poor fit. Note that there is a small number of residuals greater than 4, and these are included in the last category.}
	%	\label{fig:LungCOD_heatmap_baselinemodel_female}
\end{figure}

\begin{figure}[H]
	\centering
	\includegraphics[width=0.8\textwidth, angle =0]{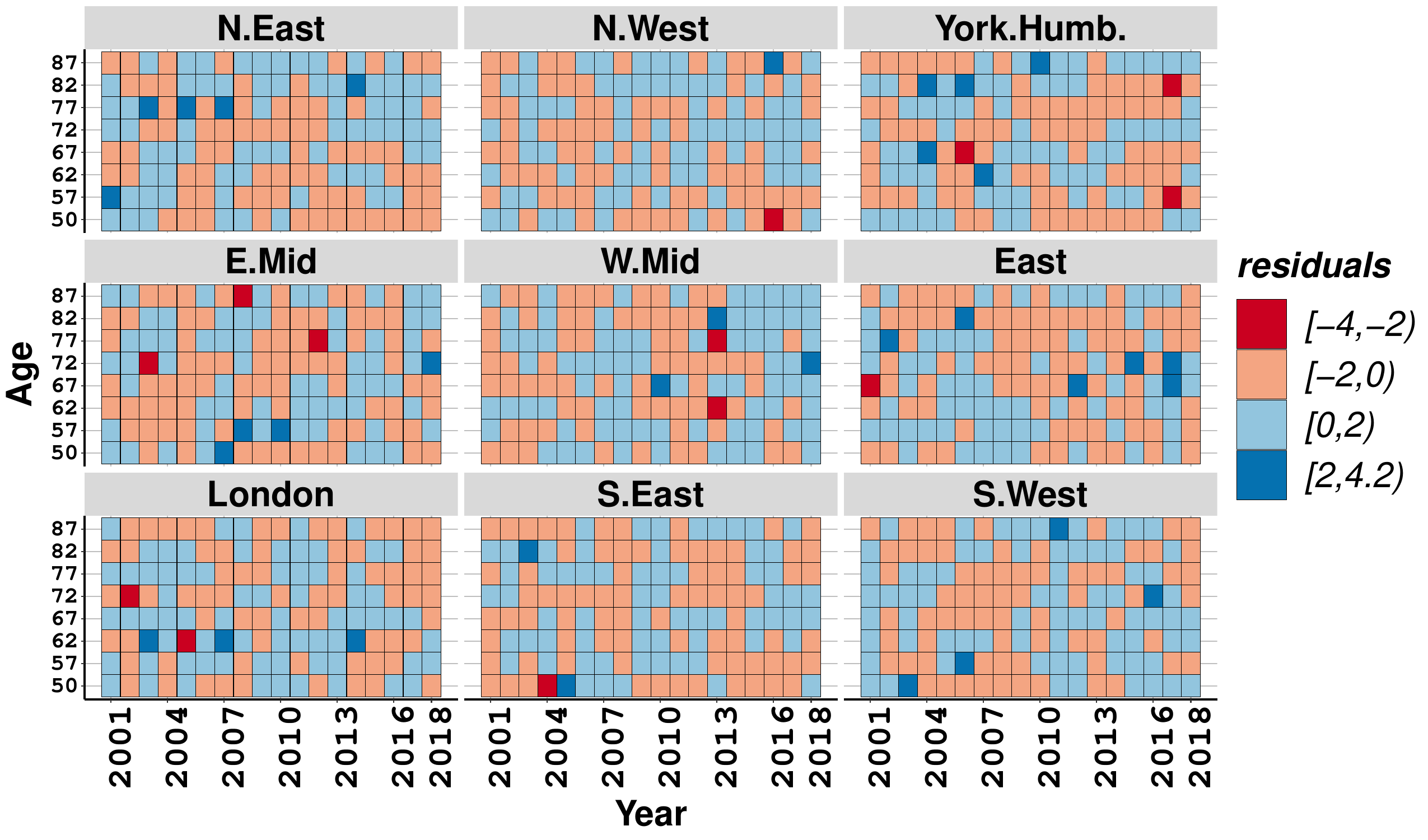}
	\caption{Heat map of Pearson residuals for female lung cancer mortality in regions of England, deprivation quintile 5 (least deprived), based on \eqref{eq:FemaleLungCODLocationPrmtr2}: orange/light blue cells indicate areas with good fit, while red/dark blue cells indicate areas with poor fit. Note that there is a small number of residuals greater than 4, and these are included in the last category.}
	%	\label{fig:LungCOD_heatmap_baselinemodel_female}
\end{figure}

\begin{figure}[H]
	\centering
	\includegraphics[width=0.8\textwidth, angle =0]{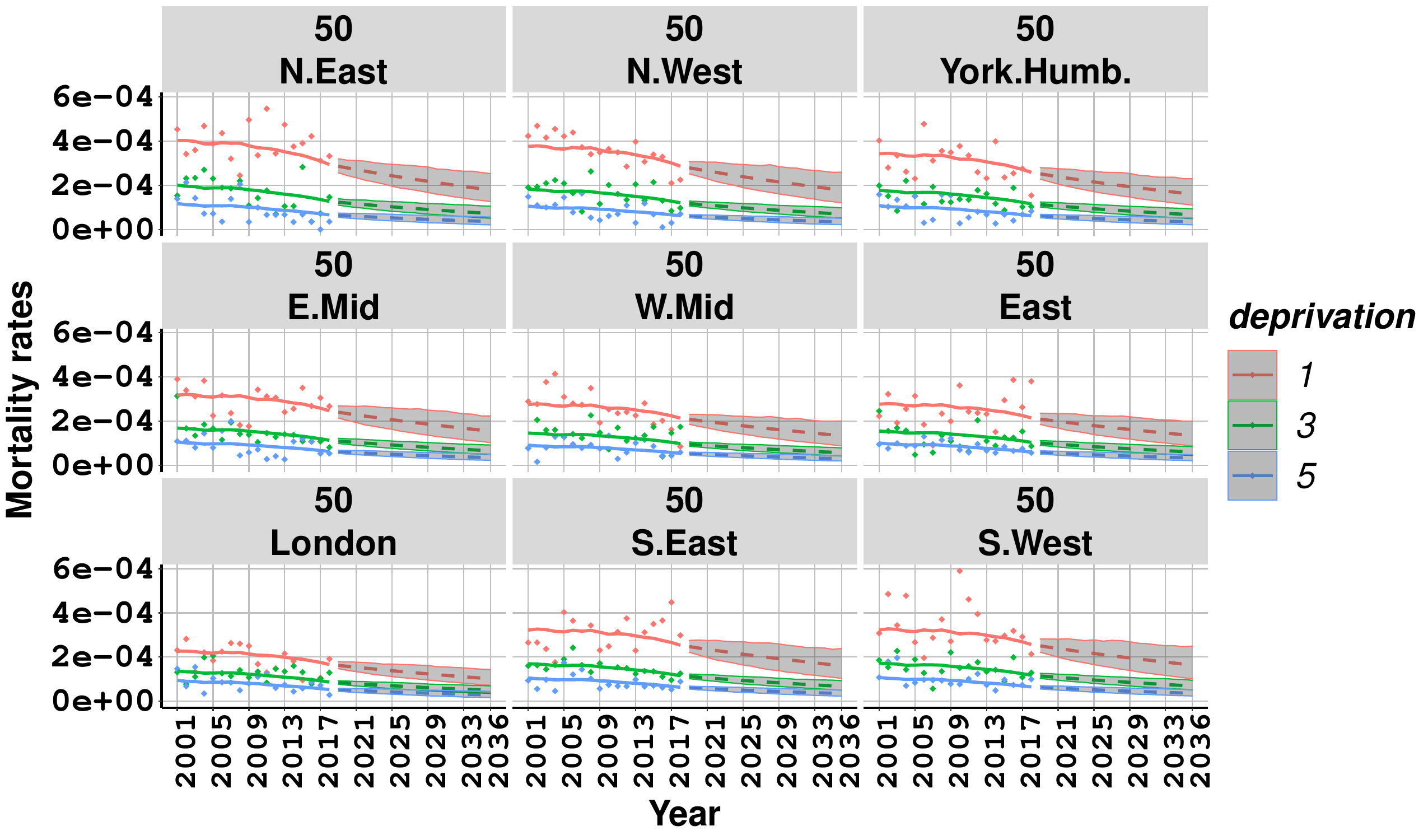}
	\caption{Lung cancer mortality, females, ages at death 50, in selected deprivation quintiles 1 (most deprived), 3, and 5 (least deprived) in regions of England based on \eqref{eq:FemaleLungCODLocationPrmtr2}: observed rates (dots), fitted rates (lines), projected rates (dashed lines) with 95\% credible intervals for the projected rates.}
\end{figure}

%\vspace{-1.5cm} % Adjust the value to reduce the space

\begin{figure}[H]
	\centering
	\includegraphics[width=0.8\textwidth, angle =0]{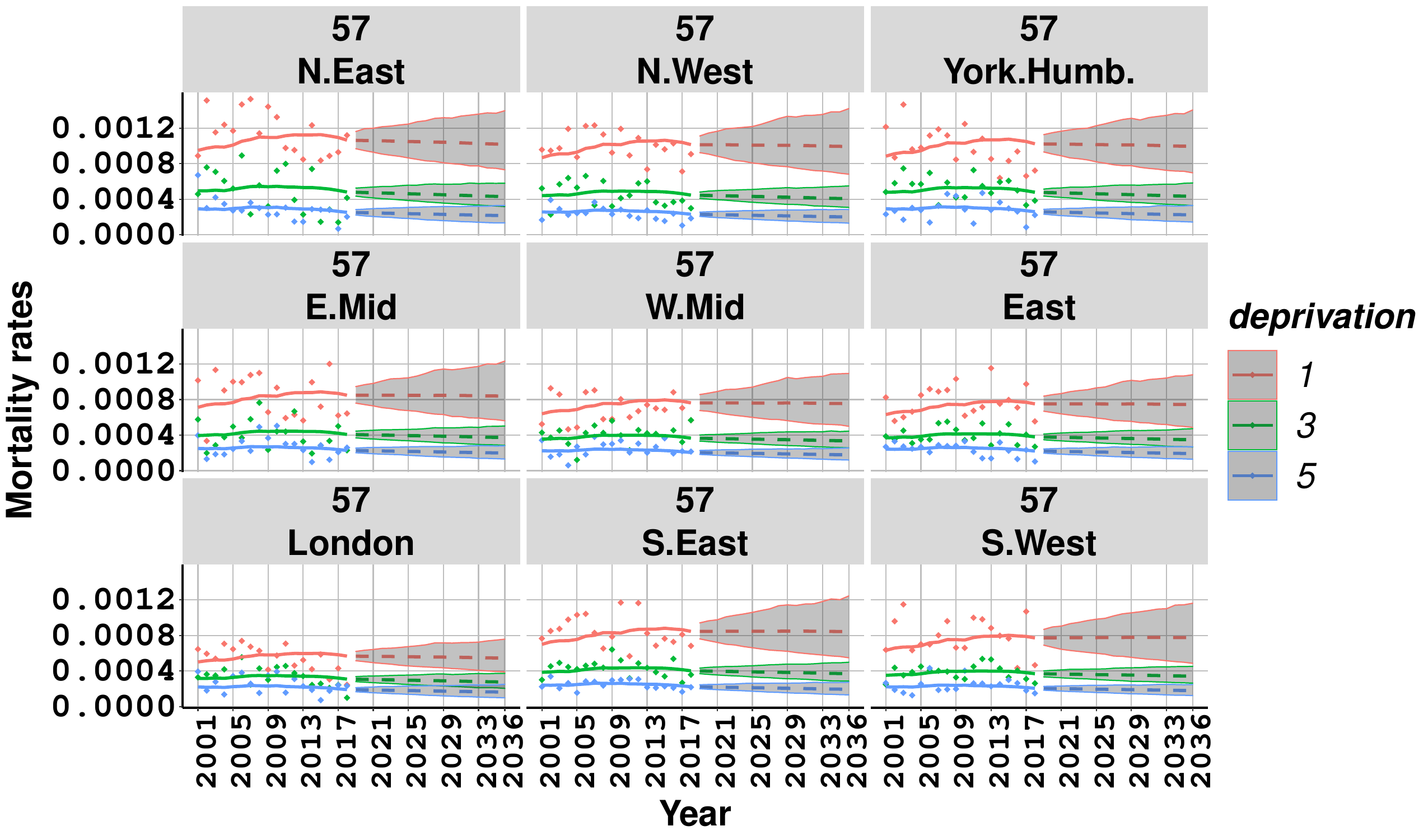}
	\caption{Lung cancer mortality, females, ages at death 57, in selected deprivation quintiles 1 (most deprived), 3, and 5 (least deprived) in regions of England based on \eqref{eq:FemaleLungCODLocationPrmtr2}: observed rates (dots), fitted rates (lines), projected rates (dashed lines) with 95\% credible intervals for the projected rates.}
\end{figure}

\begin{figure}[H]
	\centering
	\includegraphics[width=0.8\textwidth, angle =0]{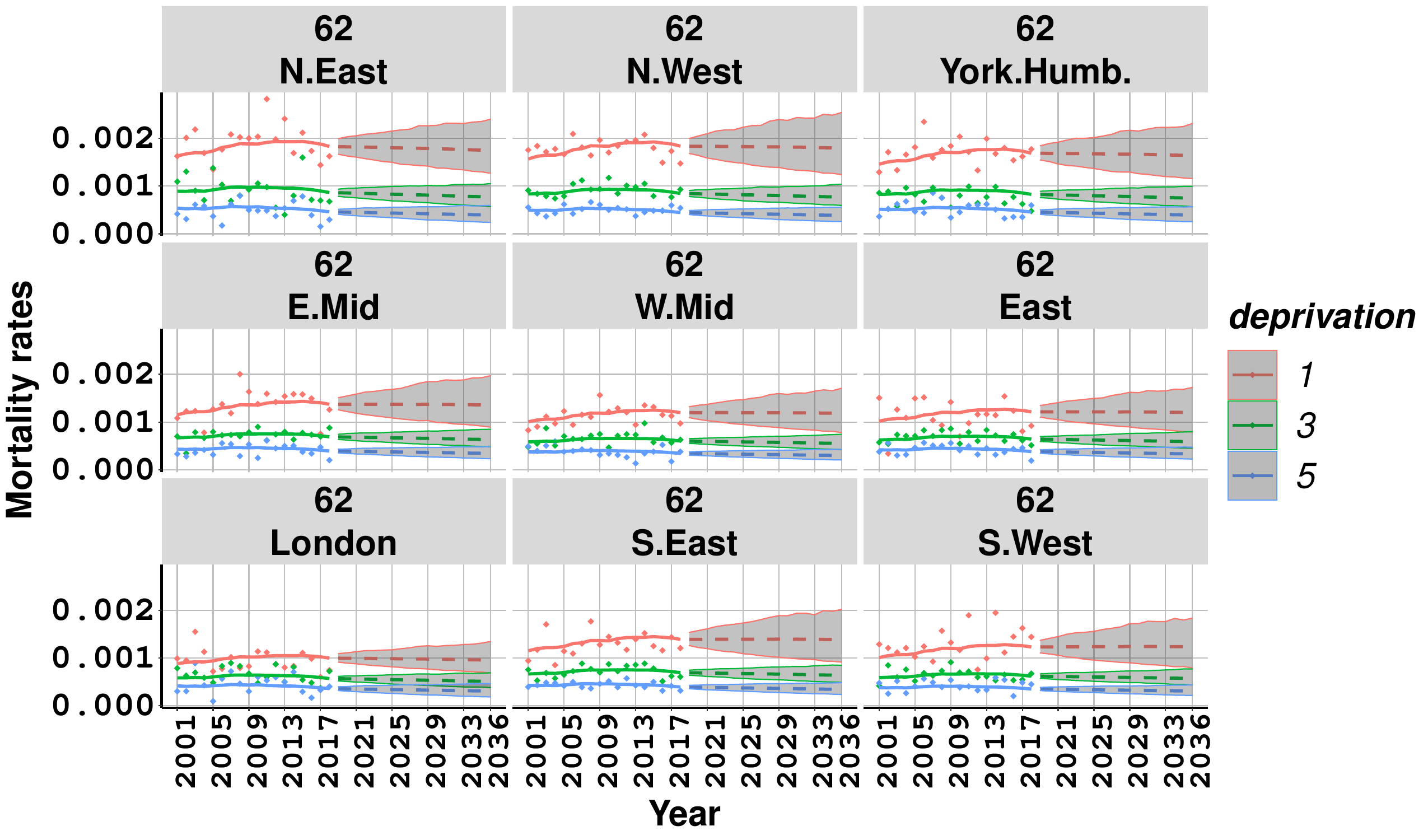}
	\caption{Lung cancer mortality, females, ages at death 62, in selected deprivation quintiles 1 (most deprived), 3, and 5 (least deprived) in regions of England based on \eqref{eq:FemaleLungCODLocationPrmtr2}: observed rates (dots), fitted rates (lines), projected rates (dashed lines) with 95\% credible intervals for the projected rates.}
\end{figure}

\begin{figure}[H]
	\centering
	\includegraphics[width=0.8\textwidth, angle =0]{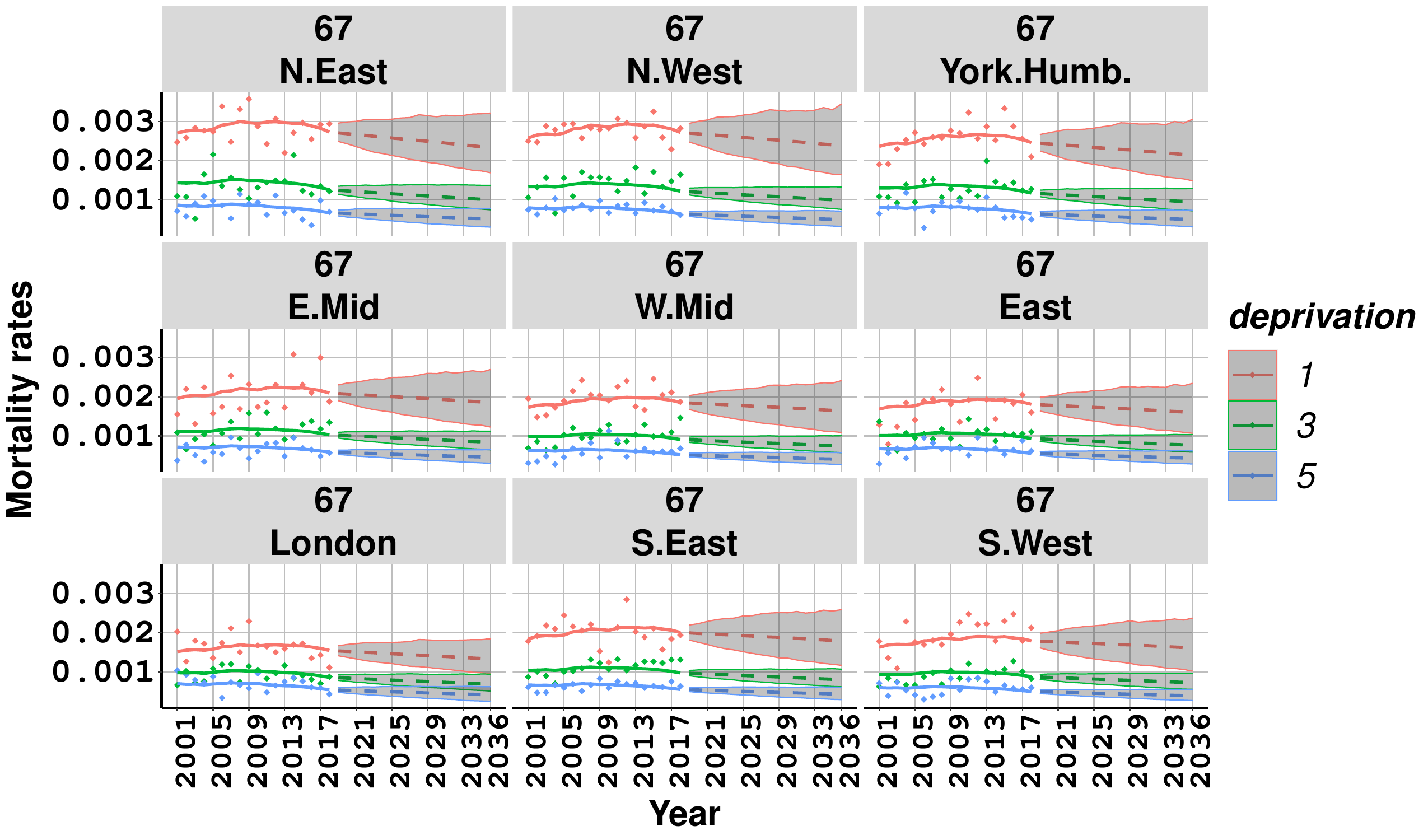}
	\caption{Lung cancer mortality, females, ages at death 67, in selected deprivation quintiles 1 (most deprived), 3, and 5 (least deprived) in regions of England based on \eqref{eq:FemaleLungCODLocationPrmtr2}: observed rates (dots), fitted rates (lines), projected rates (dashed lines) with 95\% credible intervals for the projected rates.}
\end{figure}

\begin{figure}[H]
	\centering
	\includegraphics[width=0.8\textwidth, angle =0]{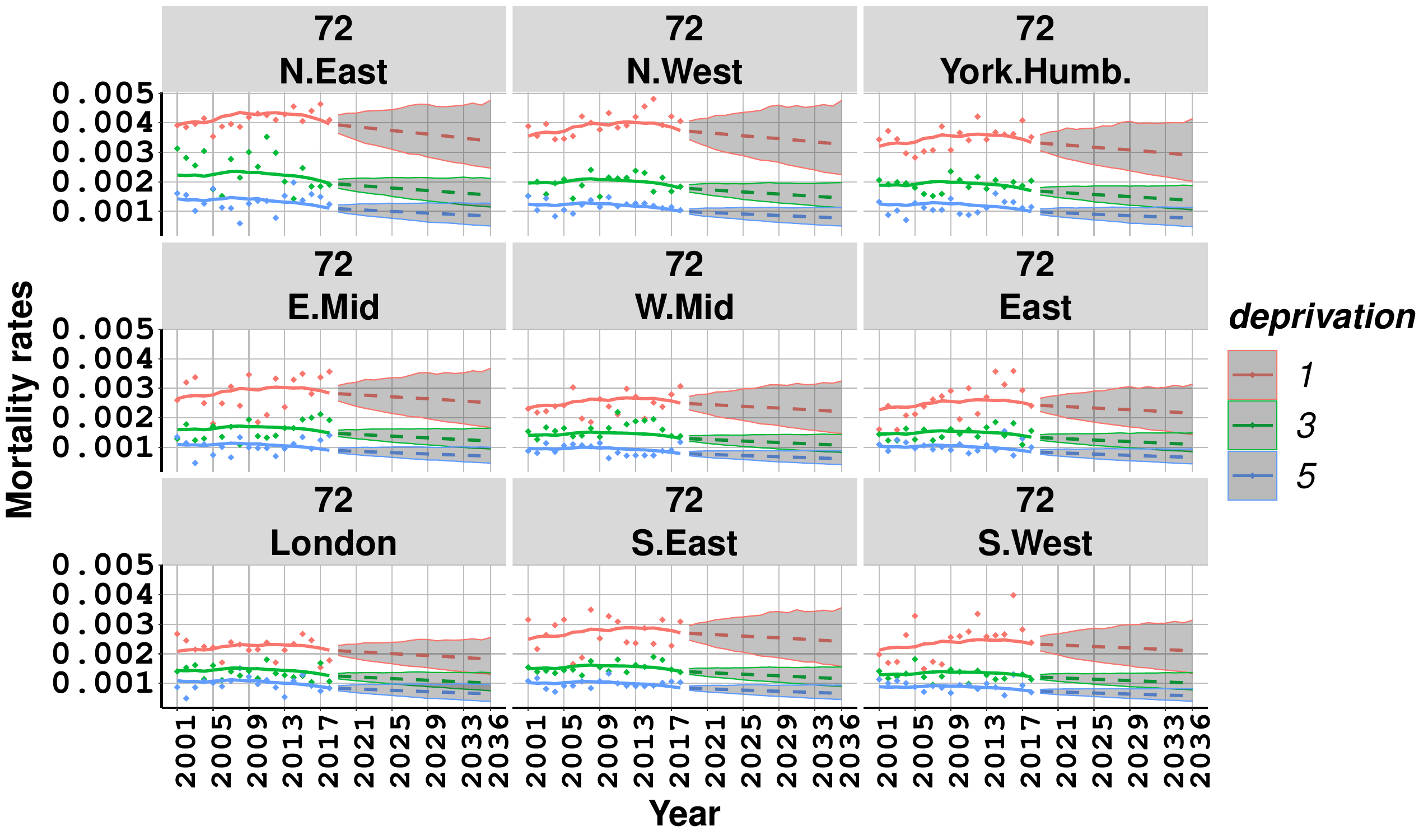}
	\caption{Lung cancer mortality, females, ages at death 72, in selected deprivation quintiles 1 (most deprived), 3, and 5 (least deprived) in regions of England based on \eqref{eq:FemaleLungCODLocationPrmtr2}: observed rates (dots), fitted rates (lines), projected rates (dashed lines) with 95\% credible intervals for the projected rates.}
\end{figure}

\begin{figure}[H]
	\centering
	\includegraphics[width=0.8\textwidth, angle =0]{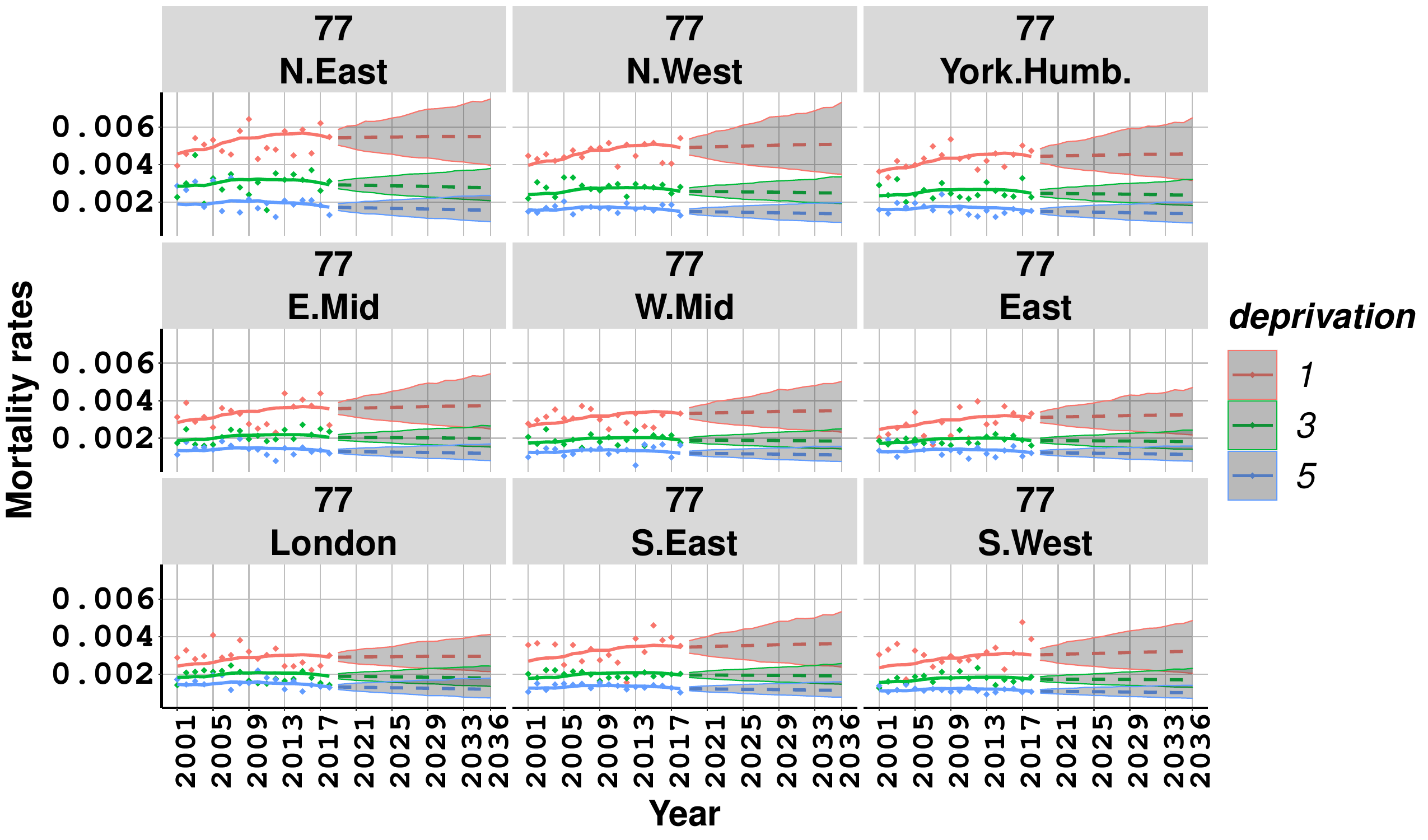}
	\caption{Lung cancer mortality, females, ages at death 77, in selected deprivation quintiles 1 (most deprived), 3, and 5 (least deprived) in regions of England based on \eqref{eq:FemaleLungCODLocationPrmtr2}: observed rates (dots), fitted rates (lines), projected rates (dashed lines) with 95\% credible intervals for the projected rates.}
\end{figure}

\begin{figure}[H]
	\centering
	\includegraphics[width=0.8\textwidth, angle =0]{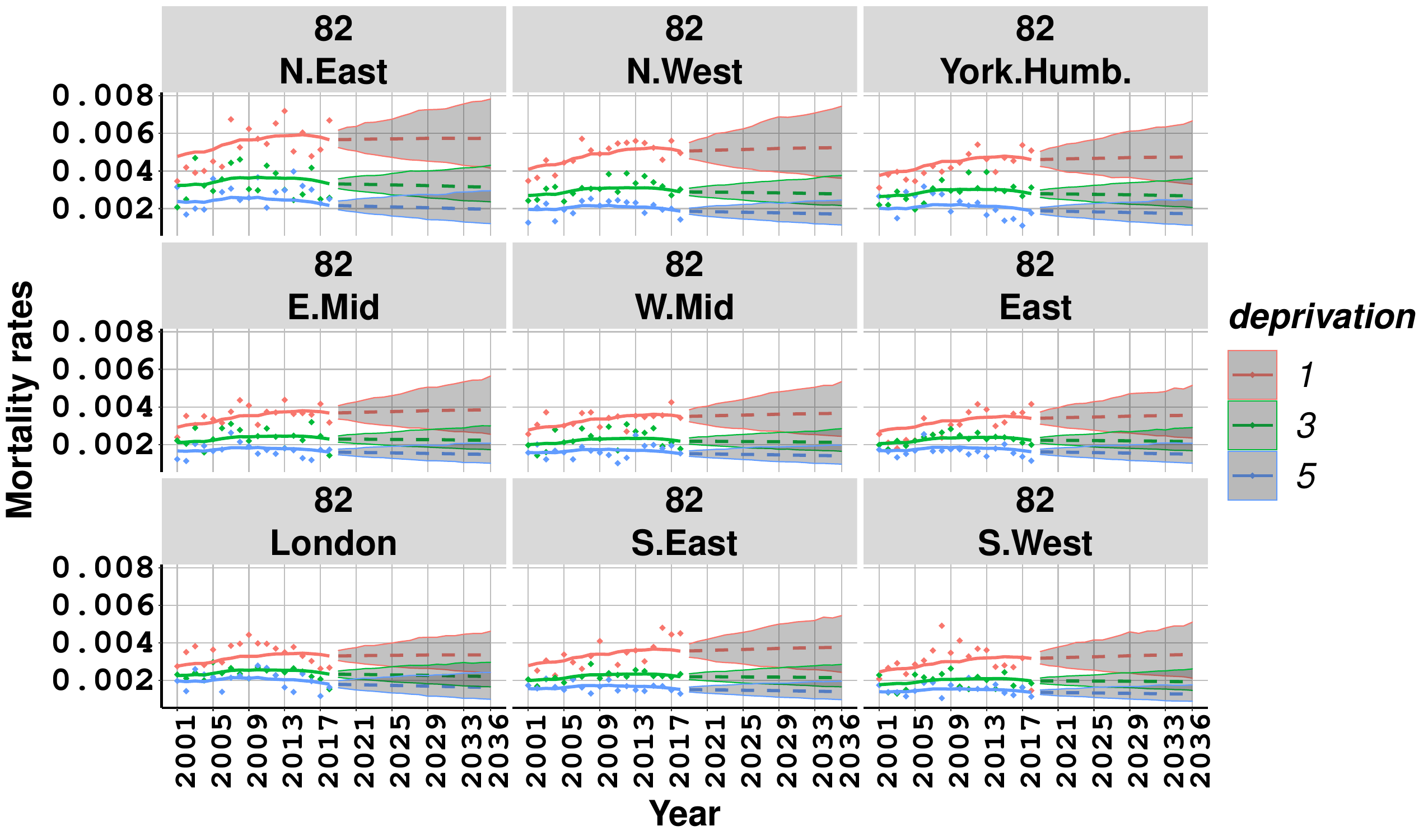}
	\caption{Lung cancer mortality, females, ages at death 82, in selected deprivation quintiles 1 (most deprived), 3, and 5 (least deprived)in regions of England based on \eqref{eq:FemaleLungCODLocationPrmtr2}: observed rates (dots), fitted rates (lines), projected rates (dashed lines) with 95\% credible intervals for the projected rates.}
\end{figure}

\begin{figure}[H]
	\centering
	\includegraphics[width=0.8\textwidth, angle =0]{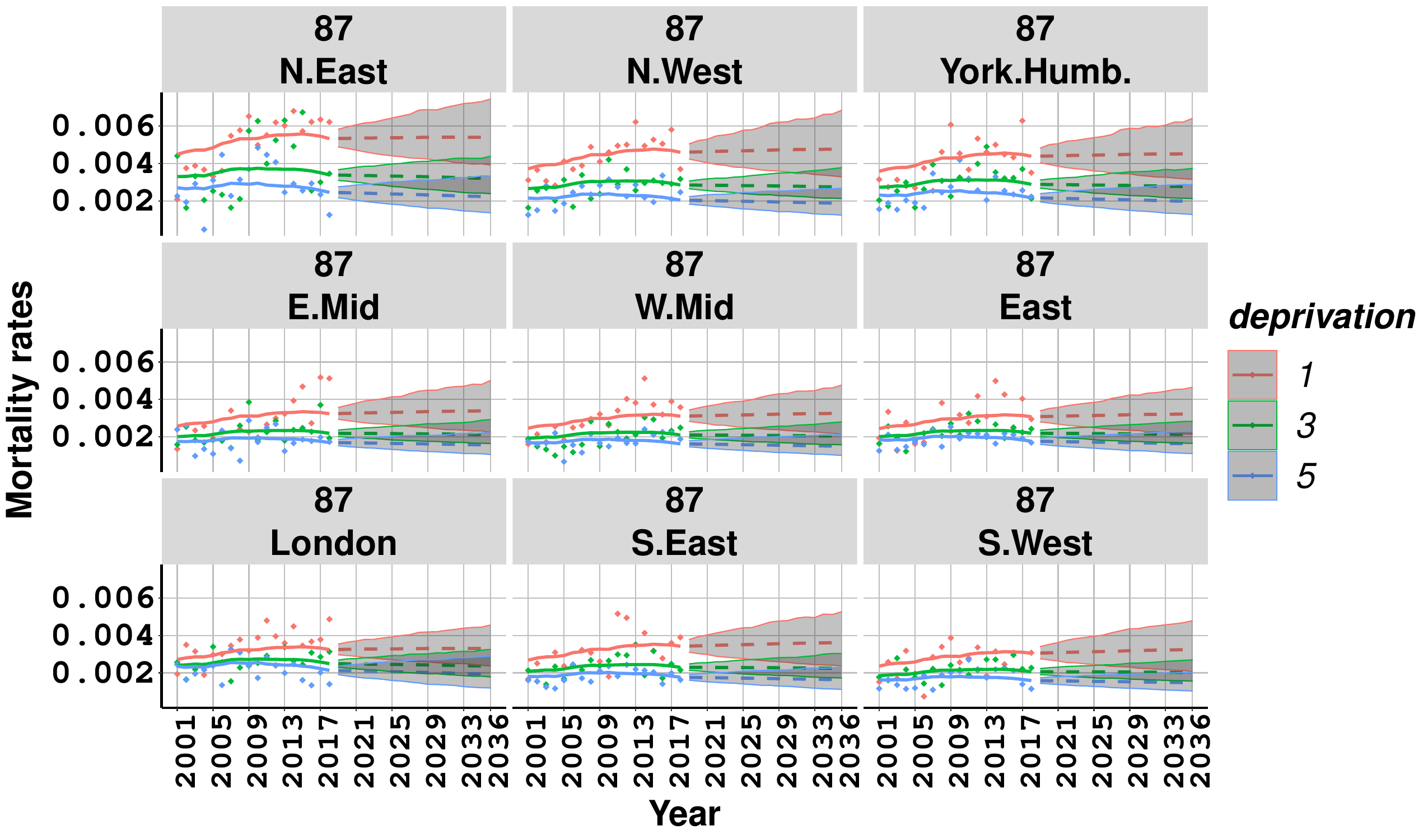}
	\caption{Lung cancer mortality, females, ages at death 87, in selected deprivation quintiles 1 (most deprived), 3, and 5 (least deprived) in regions of England based on \eqref{eq:FemaleLungCODLocationPrmtr2}: observed rates (dots), fitted rates (lines), projected rates (dashed lines) with 95\% credible intervals for the projected rates.}
\end{figure}

\subsection{Male lung cancer mortality}\label{SecApp:FiguresLungCODMales_v2}

%% To reduce spacing between the title and figure
%\vspace{-3pt}

\begin{figure}[H]
	\centering
	\includegraphics[width=.8\textwidth, angle =0]{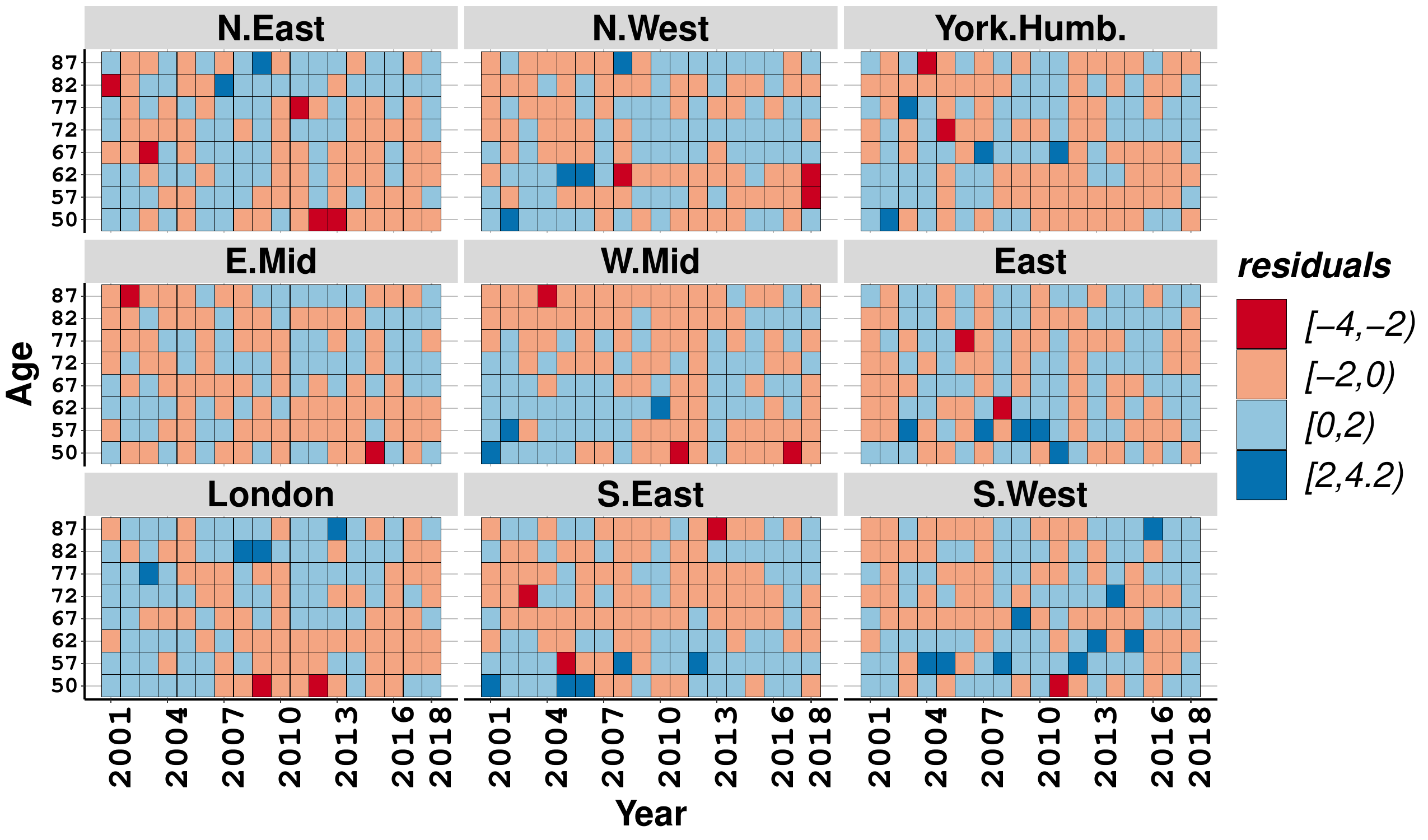}
	\caption{Heat map of Pearson residuals for male lung cancer mortality in regions of England, deprivation quintile 1 (most deprived), based on \eqref{eq:MaleLungCODLocationPrmtr2}: orange/light blue cells indicate areas with good fit, while red/dark blue cells indicate areas with poor fit. Note that there is a small number of residuals greater than 4, and these are included in the last category.}
	%	\label{fig:LungCOD_heatmap_baselinemodel_female}
\end{figure}

\begin{figure}[H]
	\centering
	\includegraphics[width=.8\textwidth, angle =0]{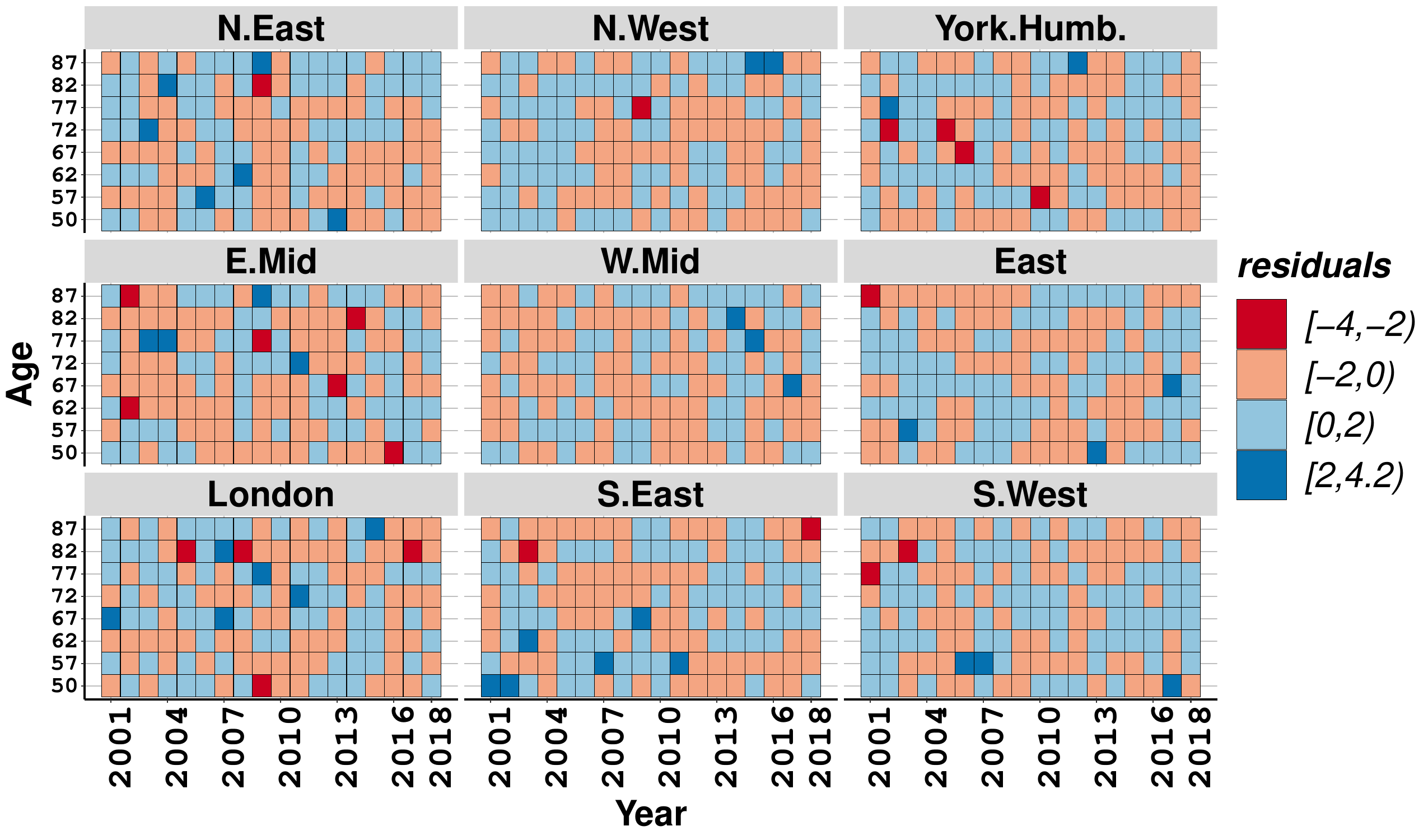}
	\caption{Heat map of Pearson residuals for male lung cancer mortality in regions of England, deprivation quintile 5 (least deprived), based on \eqref{eq:MaleLungCODLocationPrmtr2}: orange/light blue cells indicate areas with good fit, while red/dark blue cells indicate areas with poor fit. Note that there is a small number of residuals greater than 4, and these are included in the last category.}
	%	\label{fig:LungCOD_heatmap_baselinemodel_female}
\end{figure}

\begin{figure}[H]
	\centering
	\includegraphics[width=0.8\textwidth, angle =0]{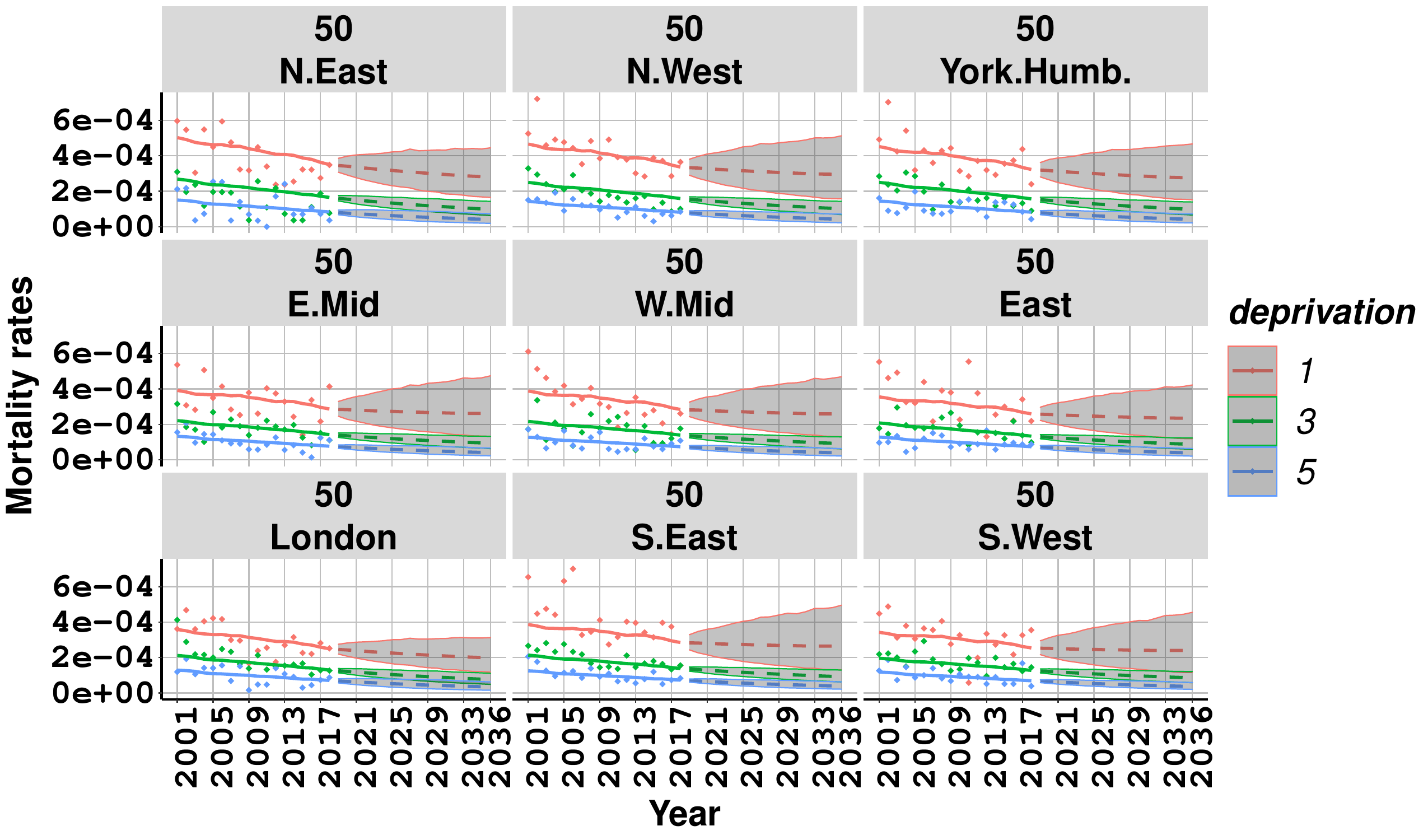}
	\caption{Lung cancer mortality, males, ages at death 50, in selected deprivation quintiles 1 (most deprived), 3, and 5 (least deprived) in regions of England based on \eqref{eq:MaleLungCODLocationPrmtr2}: observed rates (dots), fitted rates (lines), projected rates (dashed lines) with 95\% credible intervals for the projected rates.}
\end{figure}

\begin{figure}[H]
	\centering
	\includegraphics[width=0.8\textwidth, angle =0]{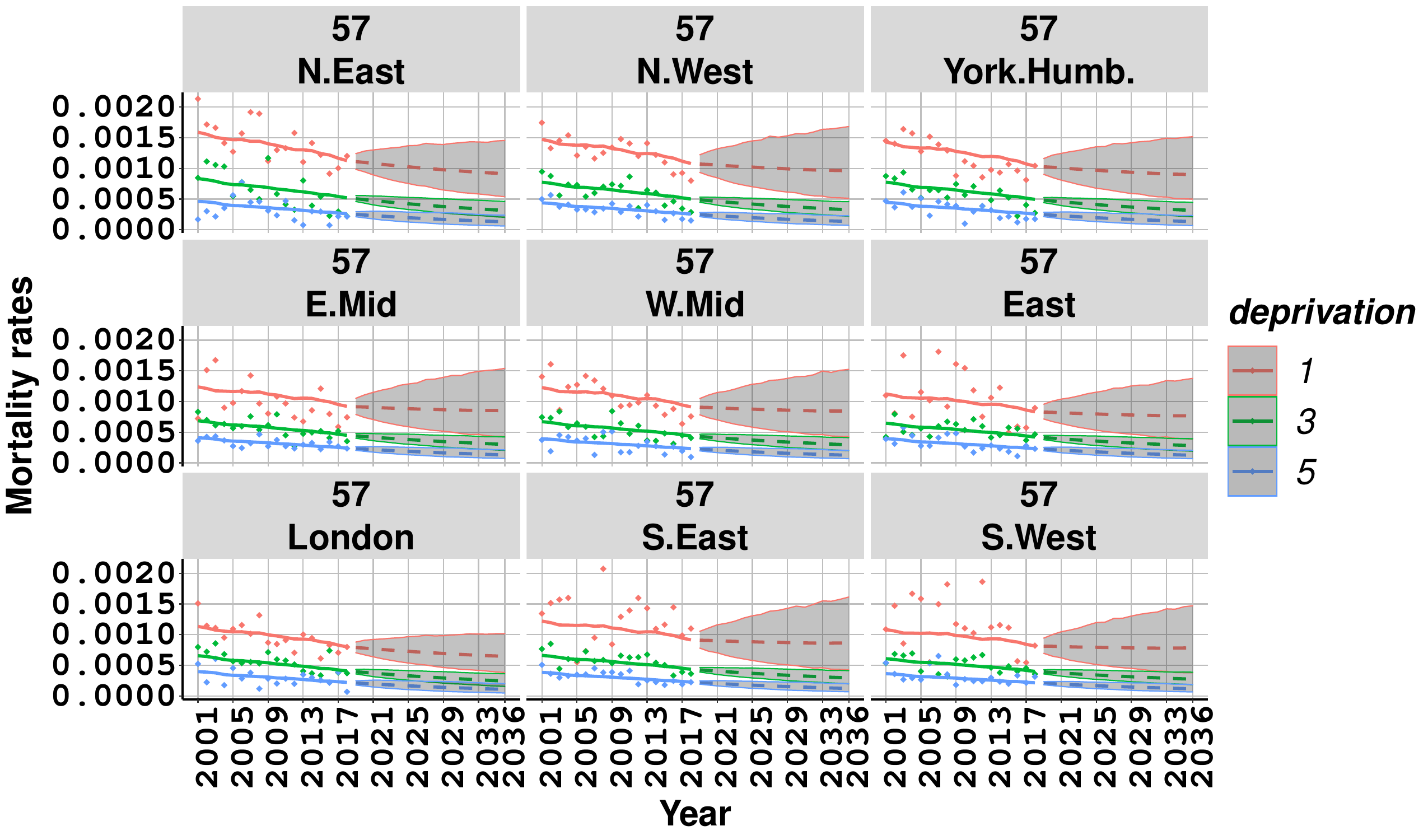}
	\caption{Lung cancer mortality, males, ages at death 57, in selected deprivation quintiles 1 (most deprived), 3, and 5 (least deprived) in regions of England based on \eqref{eq:MaleLungCODLocationPrmtr2}: observed rates (dots), fitted rates (lines), projected rates (dashed lines) with 95\% credible intervals for the projected rates.}
\end{figure}

\begin{figure}[H]
	\centering
	\includegraphics[width=0.8\textwidth, angle =0]{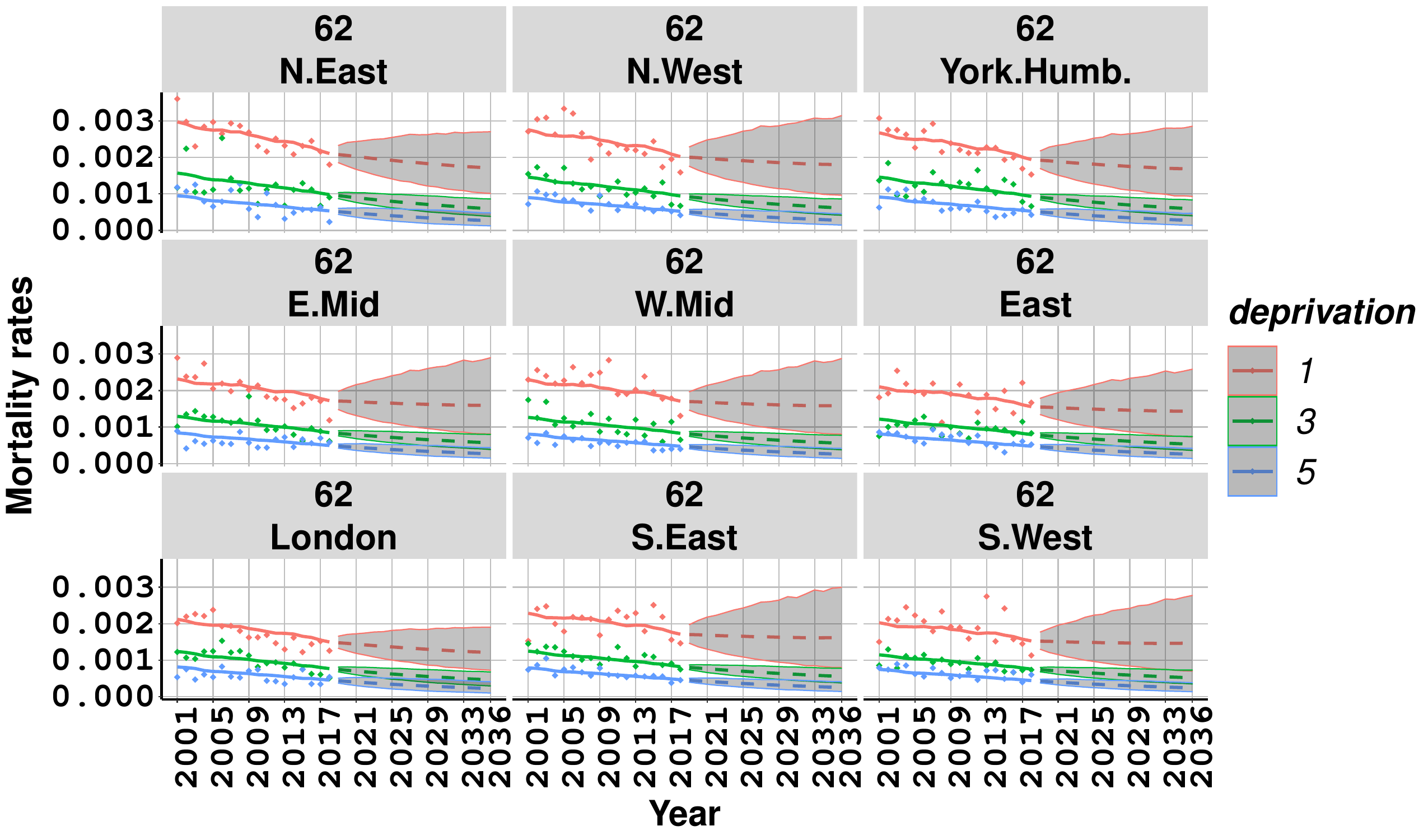}
	\caption{Lung cancer mortality, males, ages at death 62, in selected deprivation quintiles 1 (most deprived), 3, and 5 (least deprived) in regions of England based on \eqref{eq:MaleLungCODLocationPrmtr2}: observed rates (dots), fitted rates (lines), projected rates (dashed lines) with 95\% credible intervals for the projected rates.}
\end{figure}

\begin{figure}[H]
	\centering
	\includegraphics[width=0.8\textwidth, angle =0]{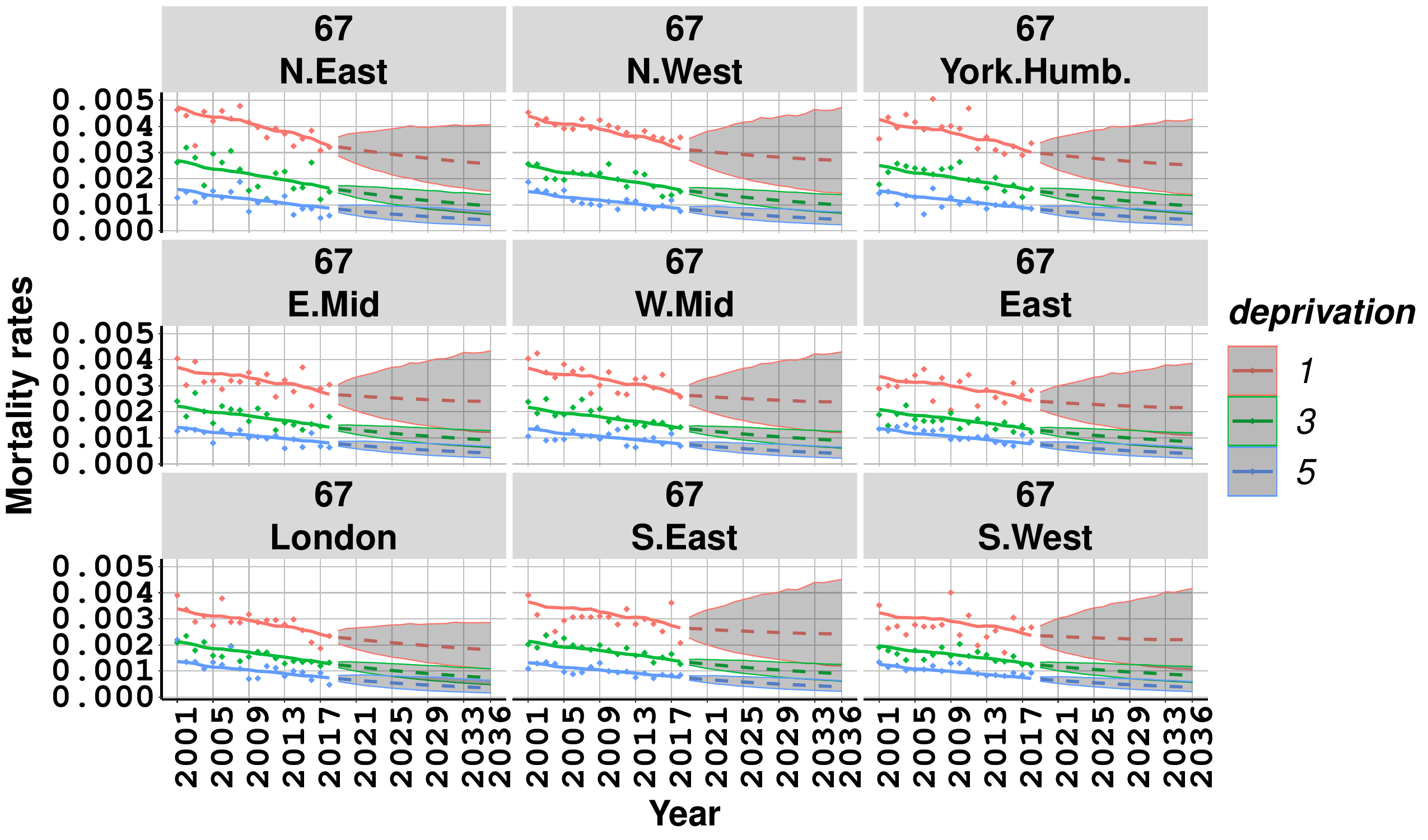}
	\caption{Lung cancer mortality, males, ages at death 67, in selected deprivation quintiles 1 (most deprived), 3, and 5 (least deprived)in regions of England based on \eqref{eq:MaleLungCODLocationPrmtr2}: observed rates (dots), fitted rates (lines), projected rates (dashed lines) with 95\% credible intervals for the projected rates.}
\end{figure}

\begin{figure}[H]
	\centering
	\includegraphics[width=0.8\textwidth, angle =0]{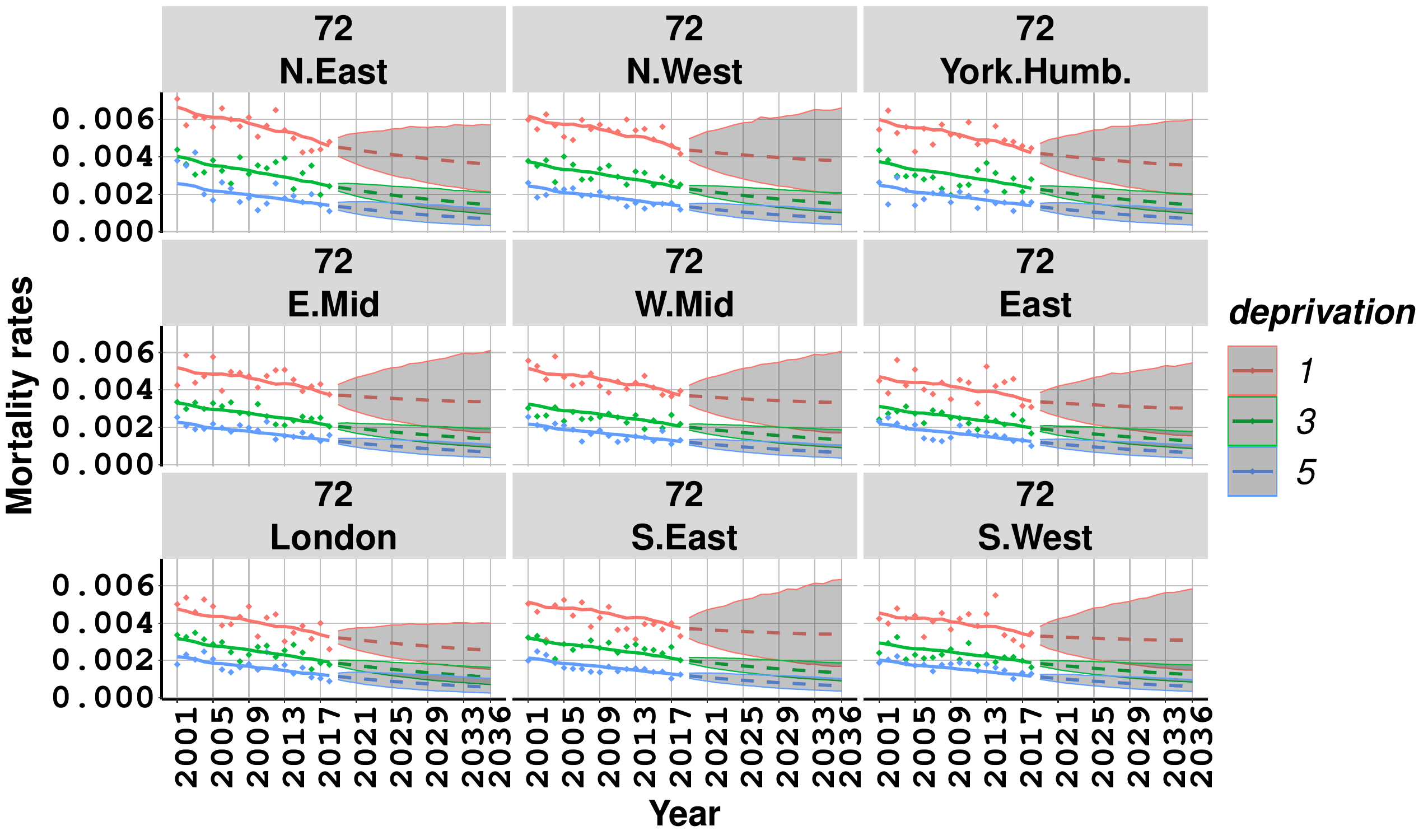}
	\caption{Lung cancer mortality, males, ages at death 72, in selected deprivation quintiles 1 (most deprived), 3, and 5 (least deprived) in regions of England based on \eqref{eq:MaleLungCODLocationPrmtr2}: observed rates (dots), fitted rates (lines), projected rates (dashed lines) with 95\% credible intervals for the projected rates.}
\end{figure}

\begin{figure}[H]
	\centering
	\includegraphics[width=0.8\textwidth, angle =0]{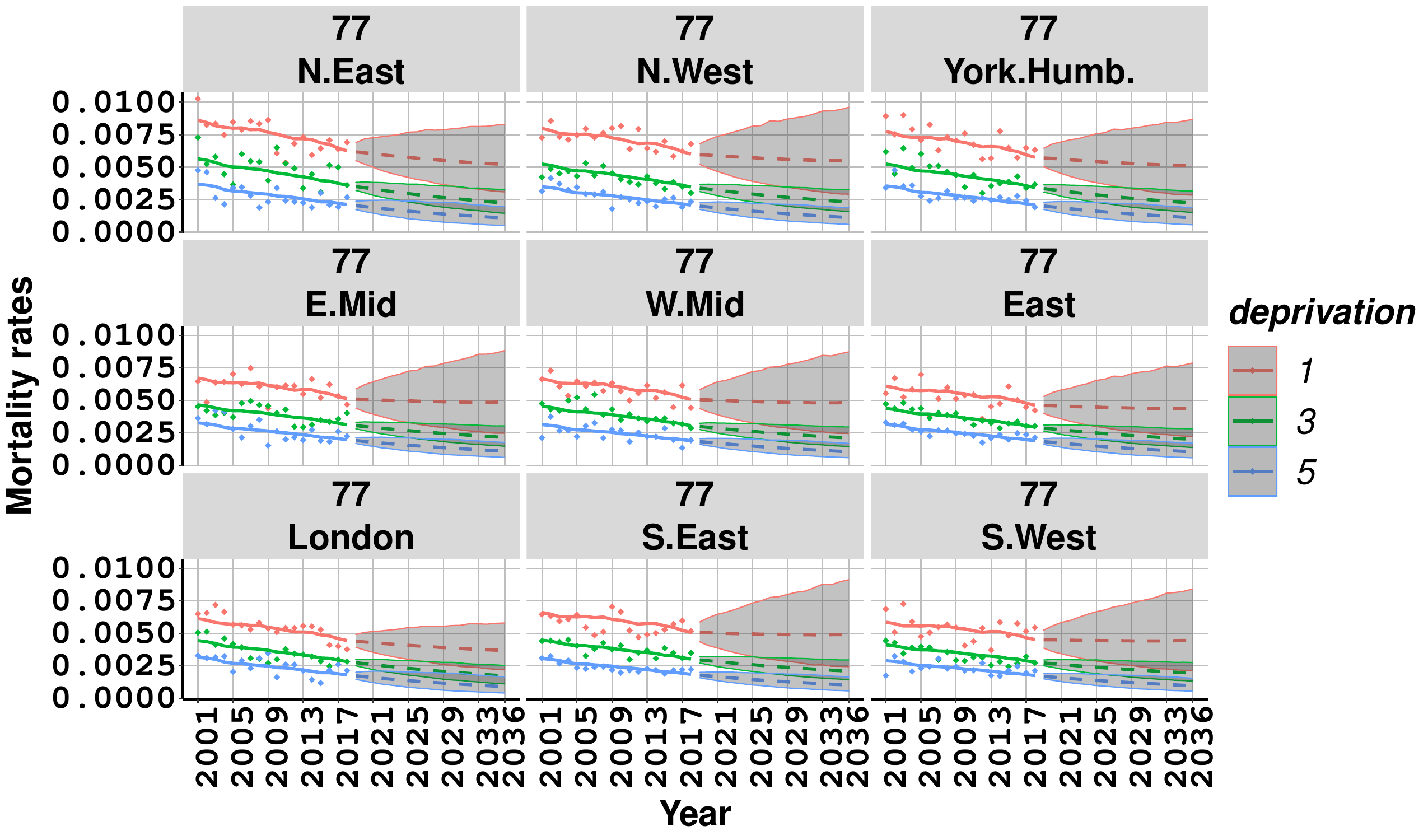}
	\caption{Lung cancer mortality, males, ages at death 77, in selected deprivation quintiles 1 (most deprived), 3, and 5 (least deprived) in regions of England based on \eqref{eq:MaleLungCODLocationPrmtr2}: observed rates (dots), fitted rates (lines), projected rates (dashed lines) with 95\% credible intervals for the projected rates.}
\end{figure}

\begin{figure}[H]
	\centering
	\includegraphics[width=0.8\textwidth, angle =0]{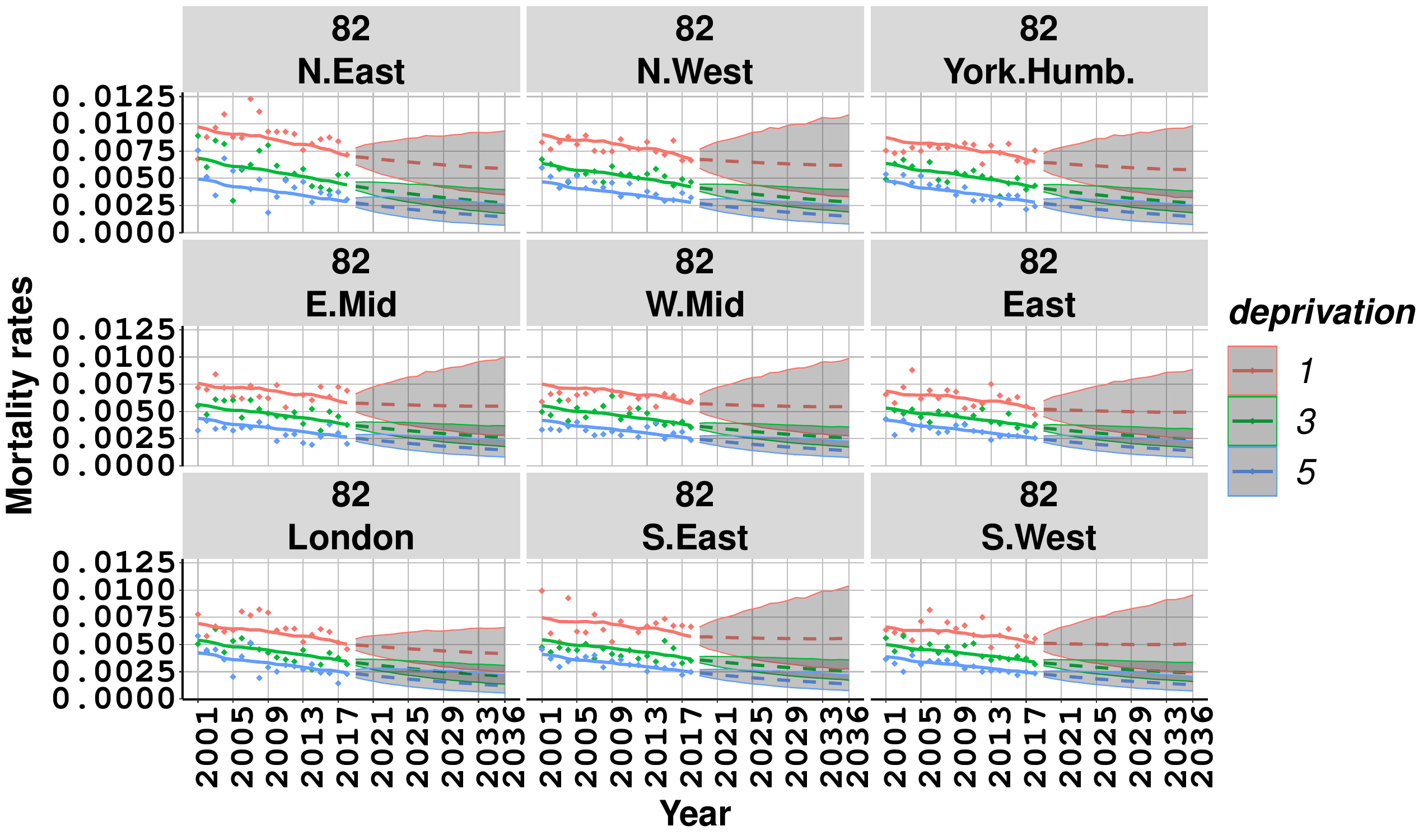}
	\caption{Lung cancer mortality, males, ages at death 82, in selected deprivation quintiles 1 (most deprived), 3, and 5 (least deprived) in regions of England based on \eqref{eq:MaleLungCODLocationPrmtr2}: observed rates (dots), fitted rates (lines), projected rates (dashed lines) with 95\% credible intervals for the projected rates.}
\end{figure}

\begin{figure}[H]
	\centering
	\includegraphics[width=0.8\textwidth, angle =0]{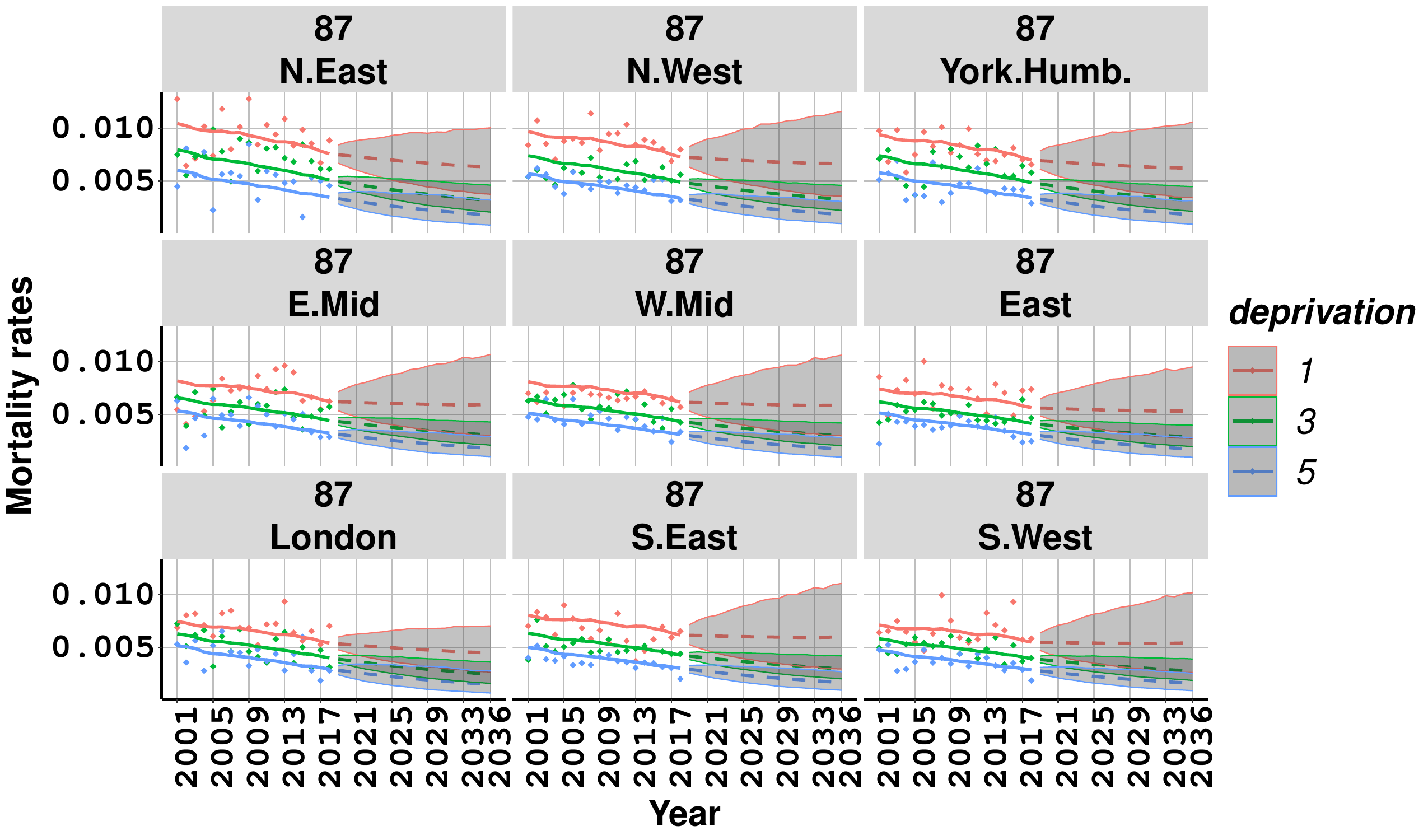}
	\caption{Lung cancer mortality, males, ages at death 87, in selected deprivation quintiles 1 (most deprived), 3, and 5 (least deprived) in regions of England based on \eqref{eq:MaleLungCODLocationPrmtr2}: observed rates (dots), fitted rates (lines), projected rates (dashed lines) with 95\% credible intervals for the projected rates.}
\end{figure}

\subsection{Breast cancer mortality}
\label{Sec:FiguresBreastCOD}

\begin{figure}[H]
	\centering
	\includegraphics[width=.8\textwidth, angle =0]{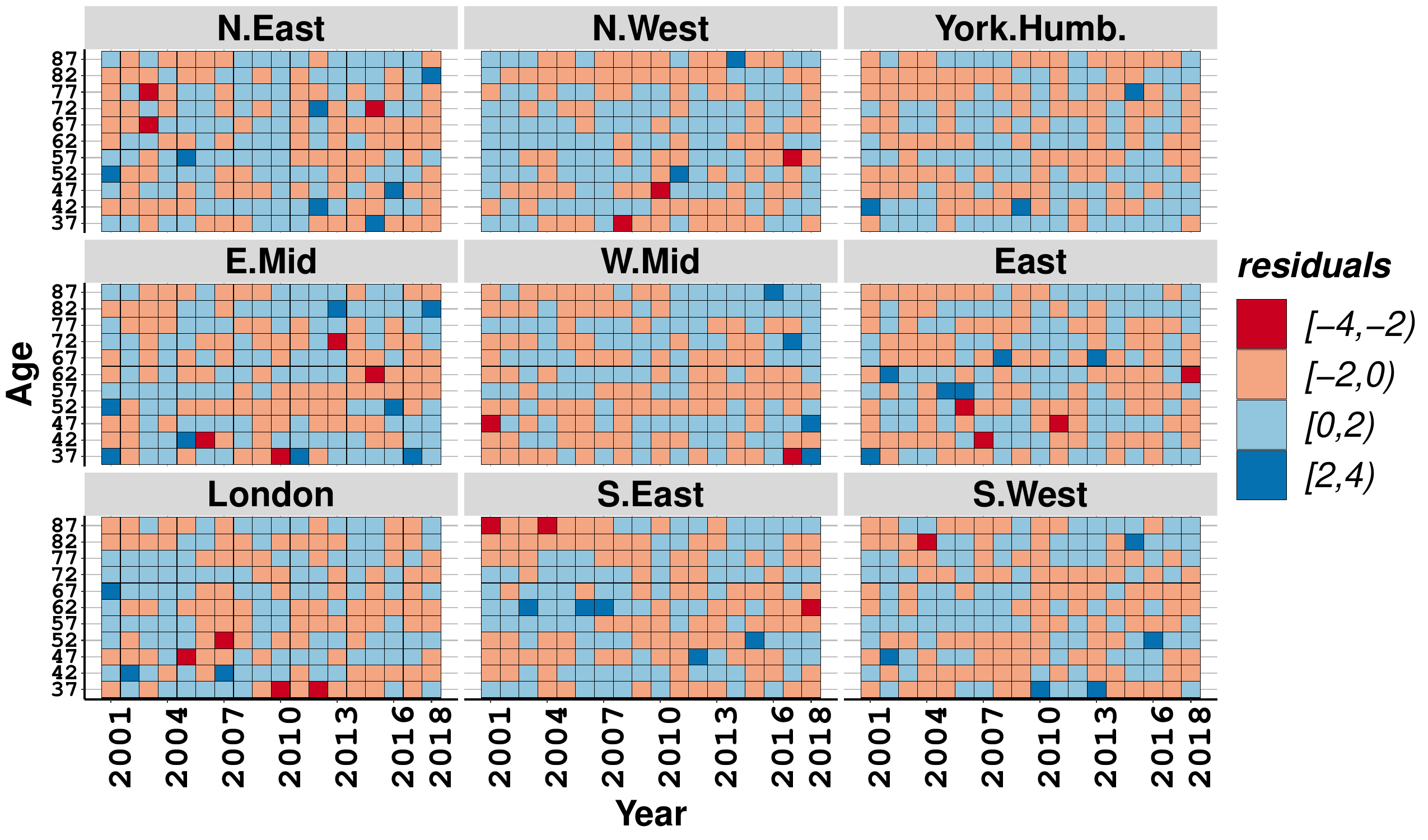}
	\caption{Heat map of Pearson residuals for breast cancer mortality in regions of England based on \eqref{eq:FemaleBreastLocationPrmtr}: orange/light blue cells indicate areas with good fit, while red/dark blue cells indicate areas with poor fit. Note that there is a small number of residuals greater than 4, and these are included in the last category.}
\end{figure}

\begin{figure}[H]
	\centering
	\includegraphics[width=.8\textwidth, angle =0]{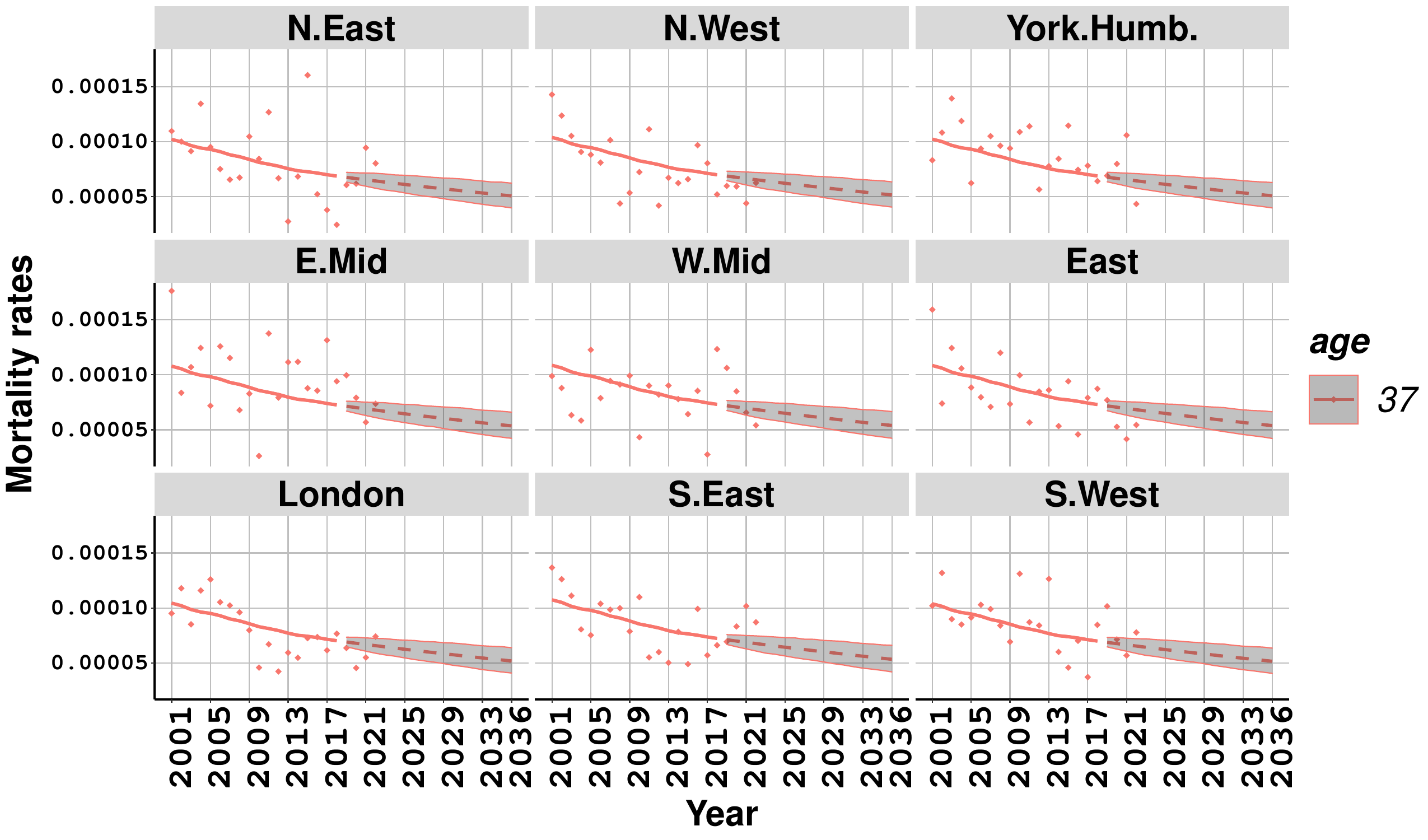}
	\caption{Breast cancer mortality, females, age at death 37, in regions of England based on \eqref{eq:FemaleBreastLocationPrmtr}: observed rates (dots), fitted rates (lines), projected rates (dashed lines) with 95\% credible intervals for the projected rates.}
\end{figure}

\begin{figure}[H]
	\centering
	\includegraphics[width=.8\textwidth, angle =0]{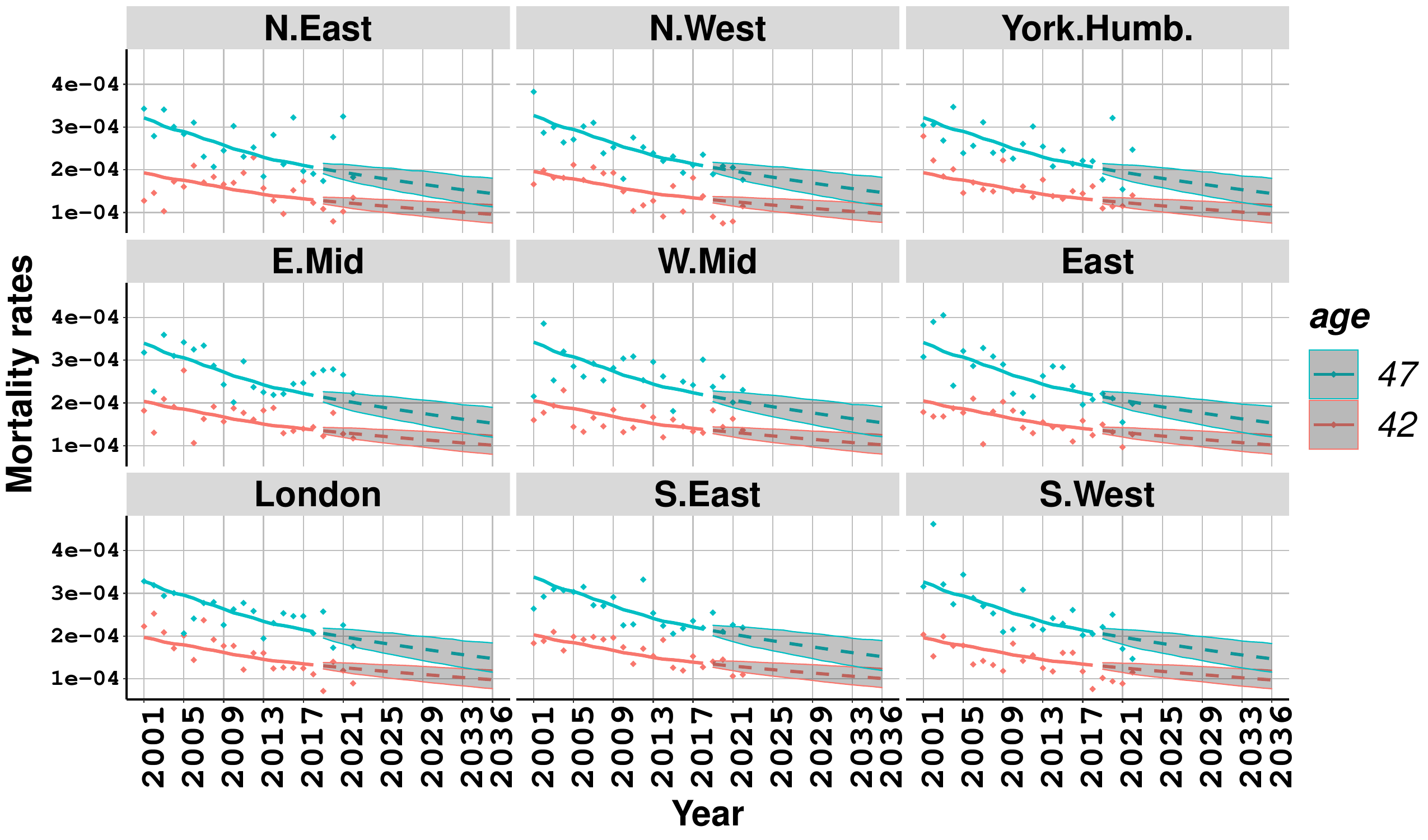}
	\caption{Breast cancer mortality, females, ages at death 42 and 47, in regions of England based on \eqref{eq:FemaleBreastLocationPrmtr}: observed rates (dots), fitted rates (lines), projected rates (dashed lines) with 95\% credible intervals for the projected rates.}
\end{figure}

\begin{figure}[H] 
	\centering
	\includegraphics[width=.8\textwidth, angle =0]{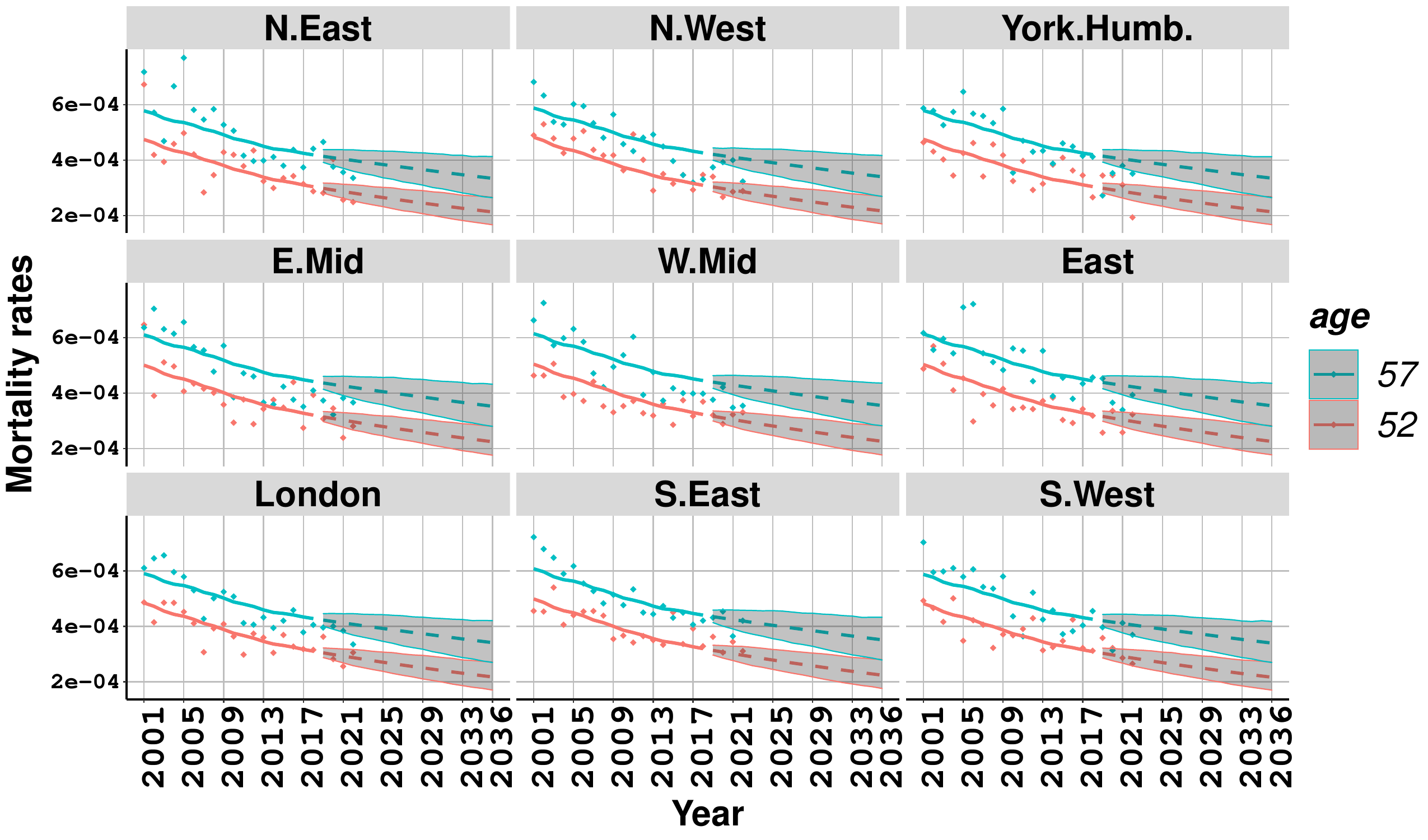}
	\caption{Breast cancer mortality, females, ages at death 52 and 57, in regions of England based on \eqref{eq:FemaleBreastLocationPrmtr}: observed rates (dots), fitted rates (lines), projected rates (dashed lines) with 95\% credible intervals for the projected rates.}
\end{figure}

\begin{figure}[H]
	\centering
	\includegraphics[width=.8\textwidth, angle =0]{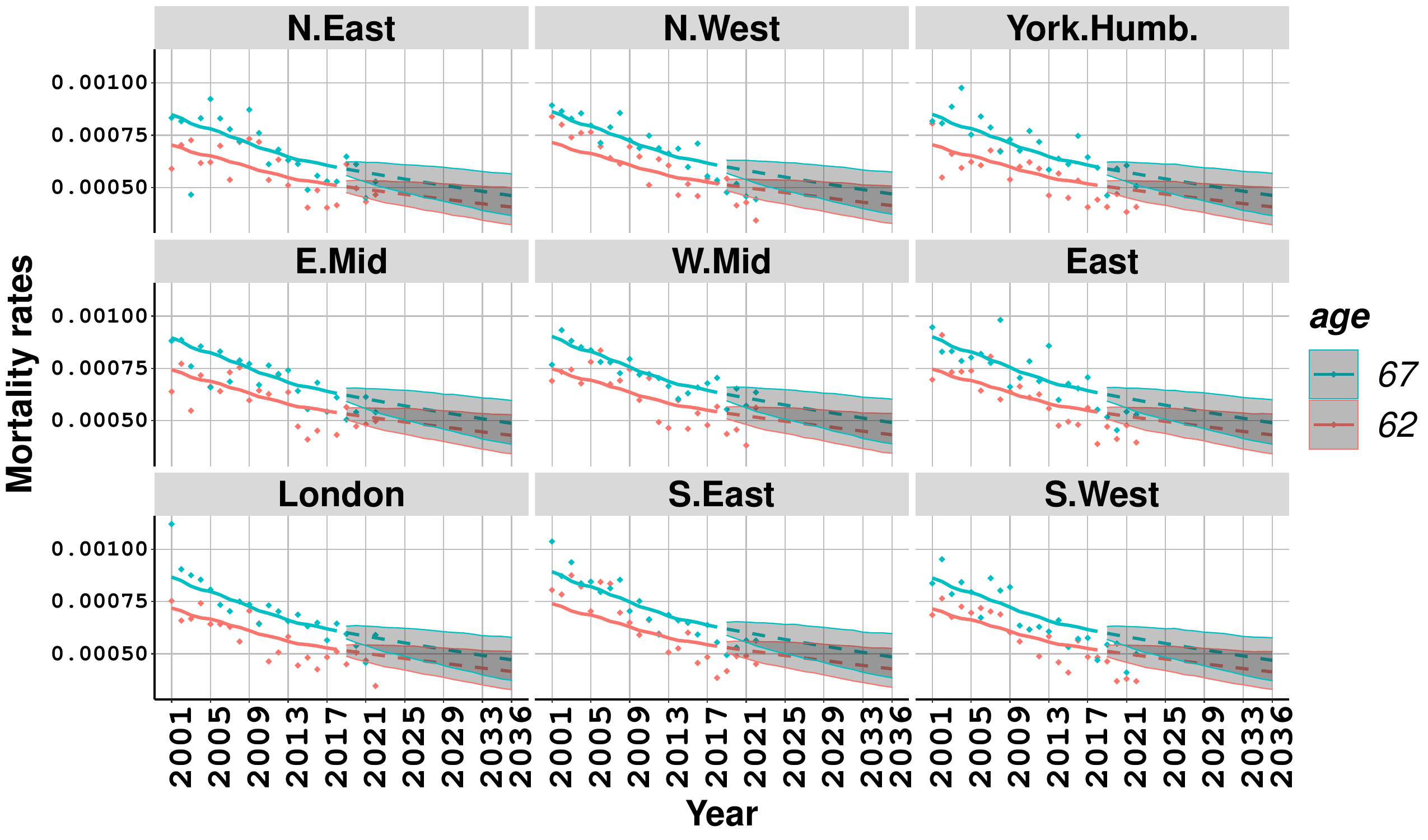}
	\caption{Breast cancer mortality, females, ages at death 62 and 67, in regions of England based on \eqref{eq:FemaleBreastLocationPrmtr}: observed rates (dots), fitted rates (lines), projected rates (dashed lines) with 95\% credible intervals for the projected rates.}
\end{figure}

\begin{figure}[H]
	\centering
	\includegraphics[width=.8\textwidth, angle =0]{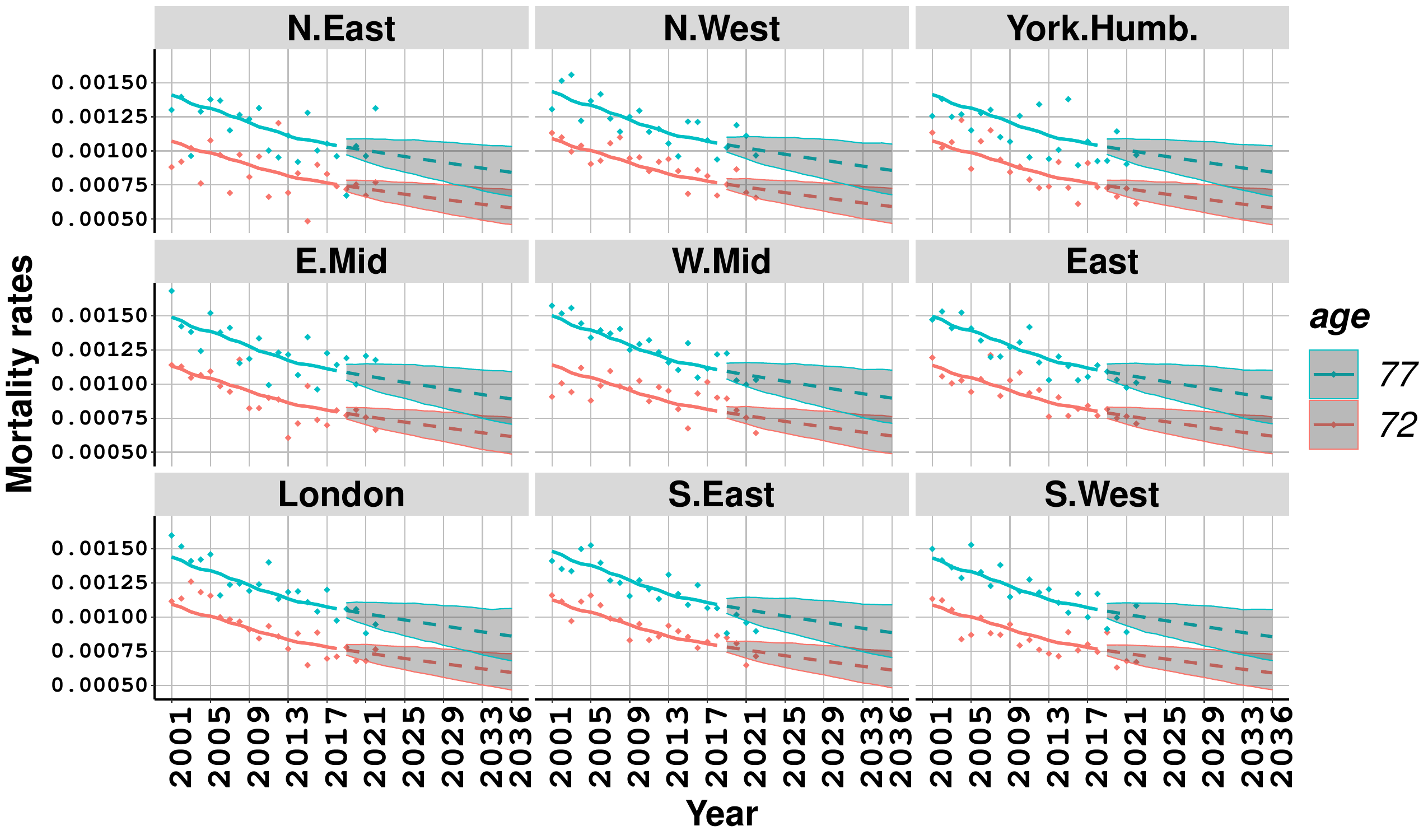}
	\caption{Breast cancer mortality, females, ages at death 72 and 77, in regions of England based on \eqref{eq:FemaleBreastLocationPrmtr}: observed rates (dots), fitted rates (lines), projected rates (dashed lines) with 95\% credible intervals for the projected rates.}
\end{figure}

\begin{figure}[H]
	\centering
	\includegraphics[width=.8\textwidth, angle =0]{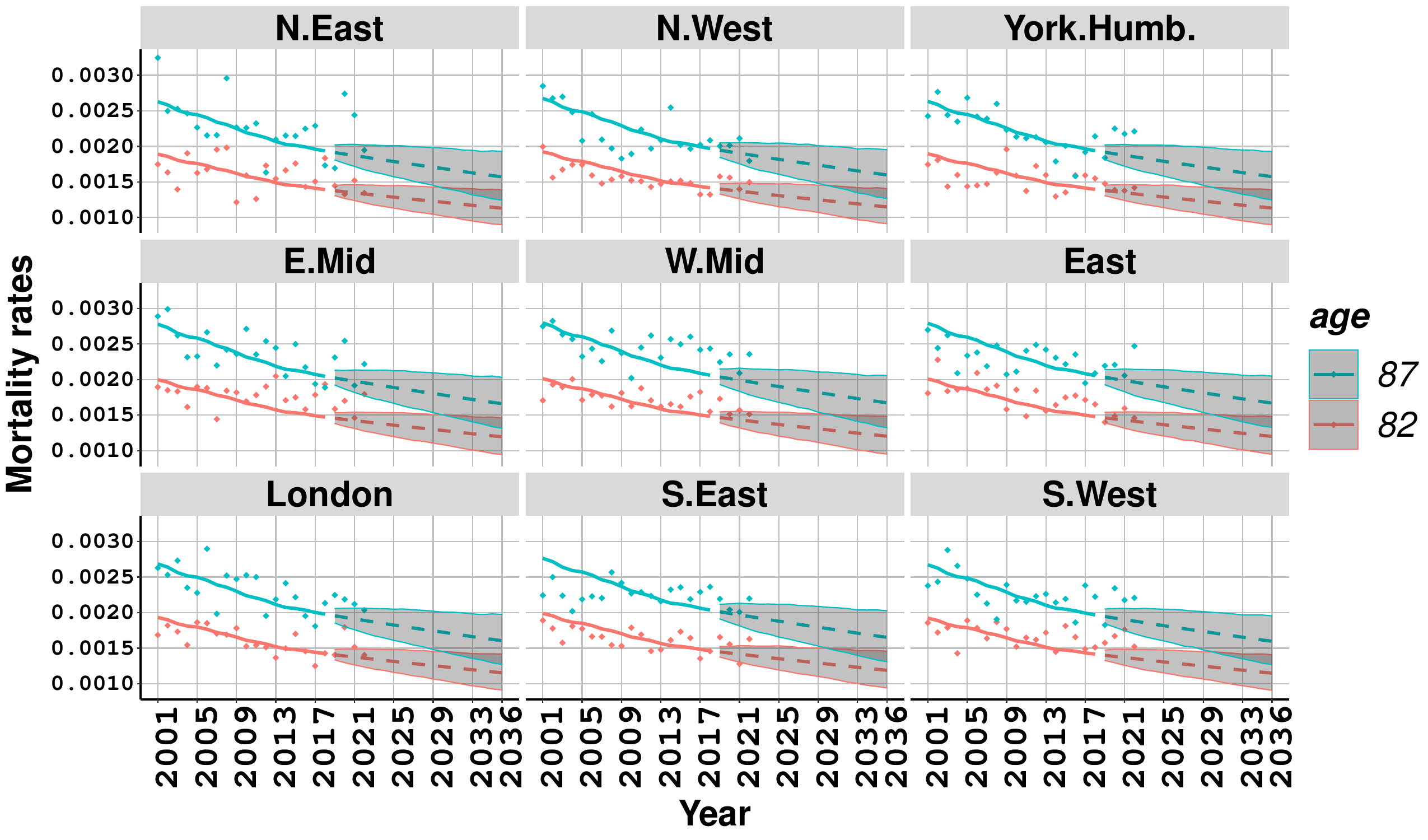}
	\caption{Breast cancer mortality, females, ages at death 82 and 87, in regions of England based on \eqref{eq:FemaleBreastLocationPrmtr}: observed rates (dots), fitted rates (lines), projected rates (dashed lines) with 95\% credible intervals for the projected rates.}
\end{figure}

\subsection{Excess lung cancer deaths in men}\label{sec:ExcessLCMen}

%% To reduce spacing between the title and figure
%\vspace{-4pt}

\begin{figure}[H]
	\centering
	\subfloat[ Excess deaths by region \label{fig:ExcessDeathsMenRegionSmokv2}]{\includegraphics[width=0.5\textwidth]{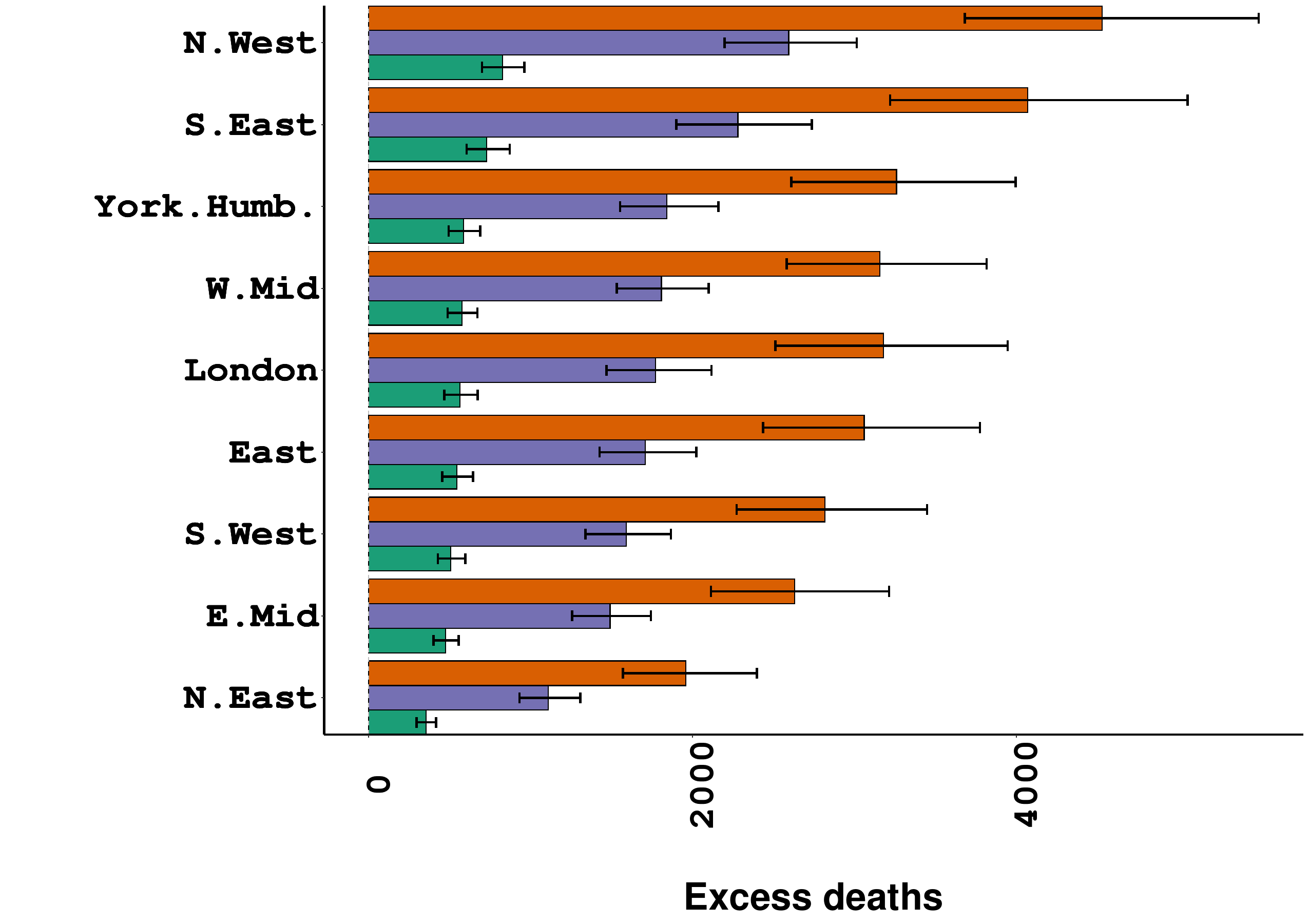}}
	\hfill
	\subfloat[ Excess deaths by deprivation quintiles \label{fig:ExcessDeathsMenDeprivSmokv2}]{\includegraphics[width=0.5\textwidth]{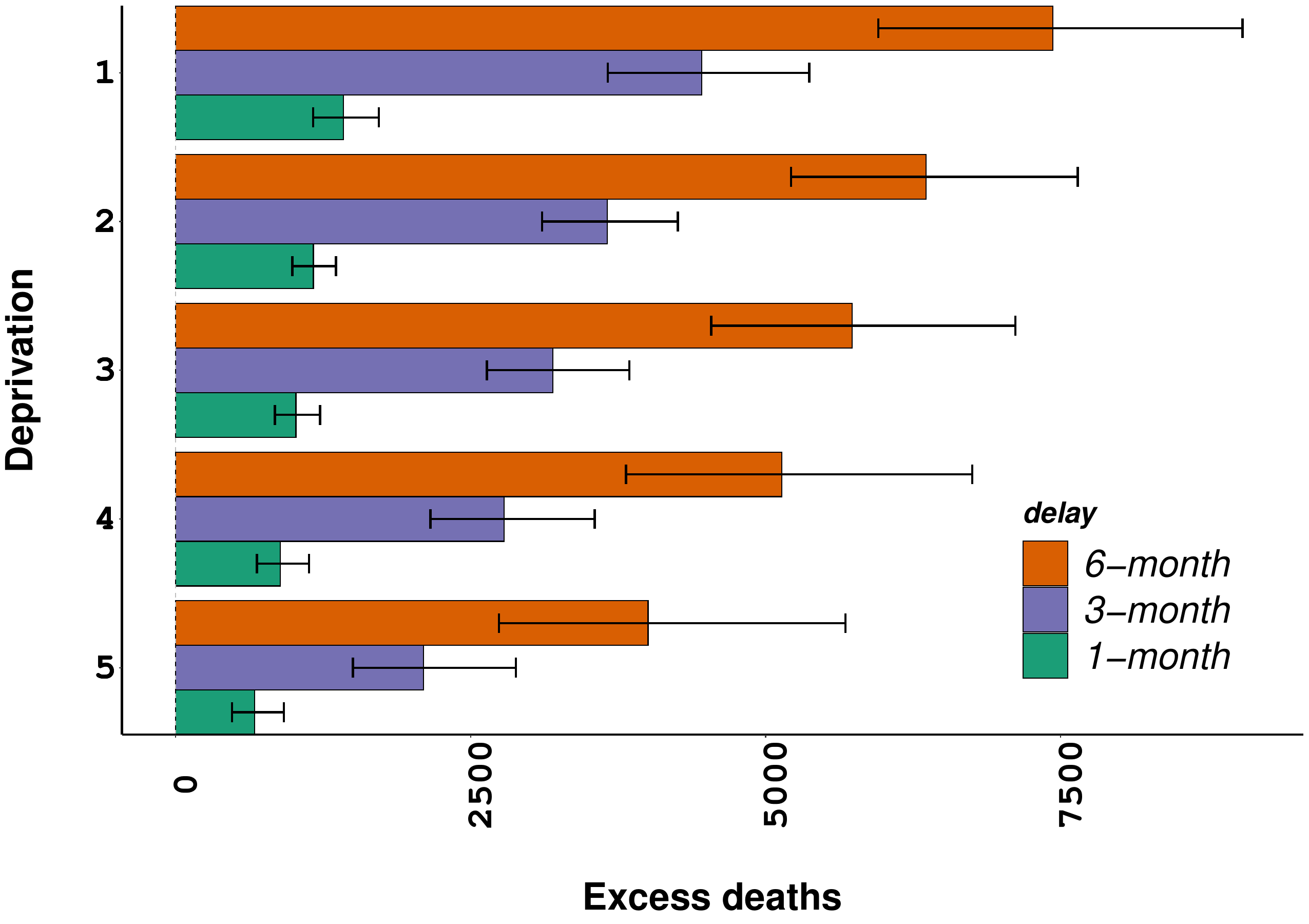}}
	\caption{Total lung cancer excess deaths in men, $\text{ED}^{\text{lung}}_{\text{men},d}$ and $\text{ED}^{\text{lung}}_{\text{men},r}$, respectively, in different deprivation quintiles and regions of England from 2020 to 2036, over 17 years, with 95\% credible intervals, based on \eqref{eq:MaleLungCODLocationPrmtr2}.}
	\label{fig:ExcessDeathsMenRegionDepSmok} 	
\end{figure}

\begin{figure}[H]
	\centering
	\subfloat[1-month delay \label{fig:ExcessRate1monthMaleSmok}]{\includegraphics[width=0.5\textwidth]{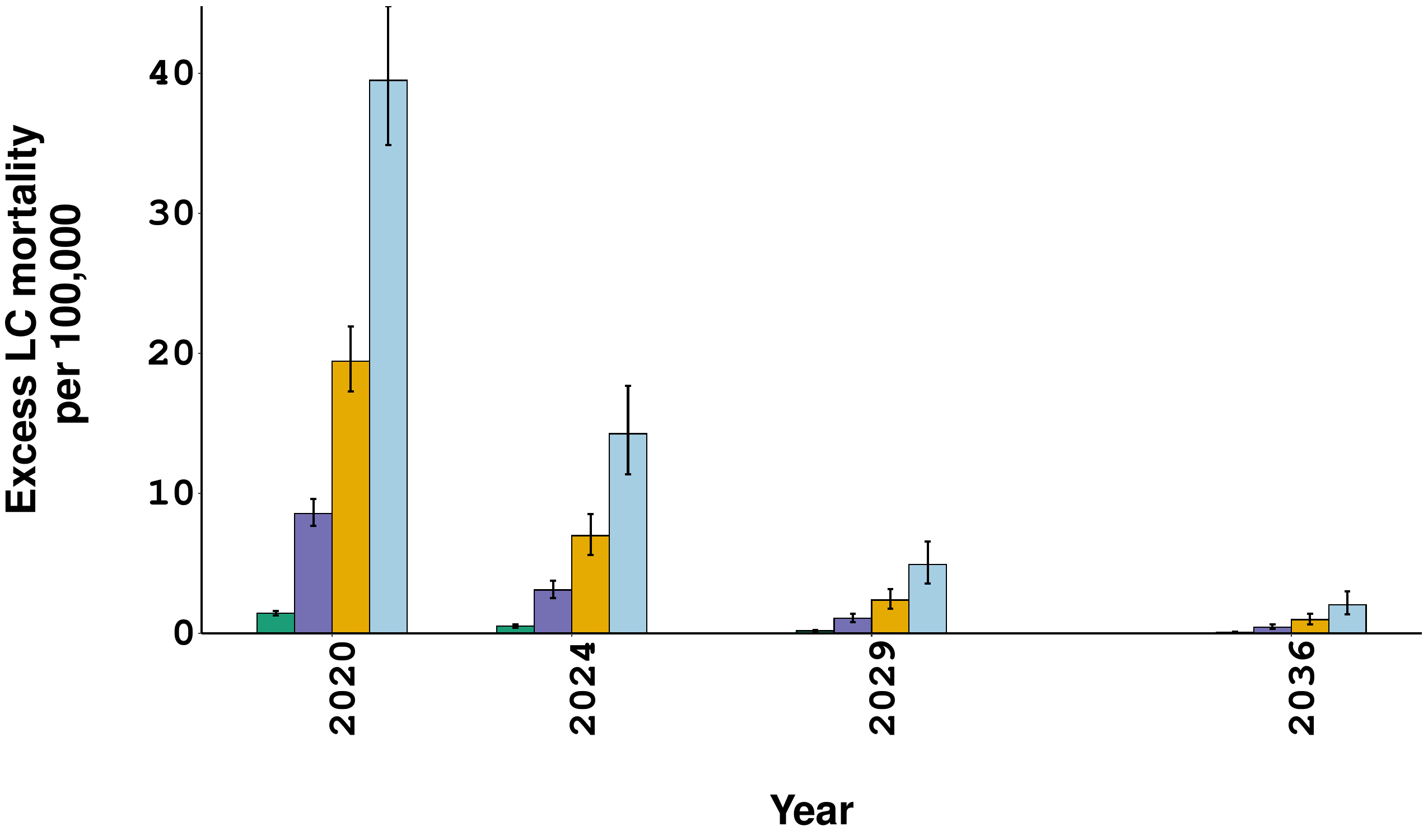}}
	\hfill
	\subfloat[ 6-month delay \label{fig:ExcessRate1monthMaleSmokv2}]{\includegraphics[width=0.5\textwidth]{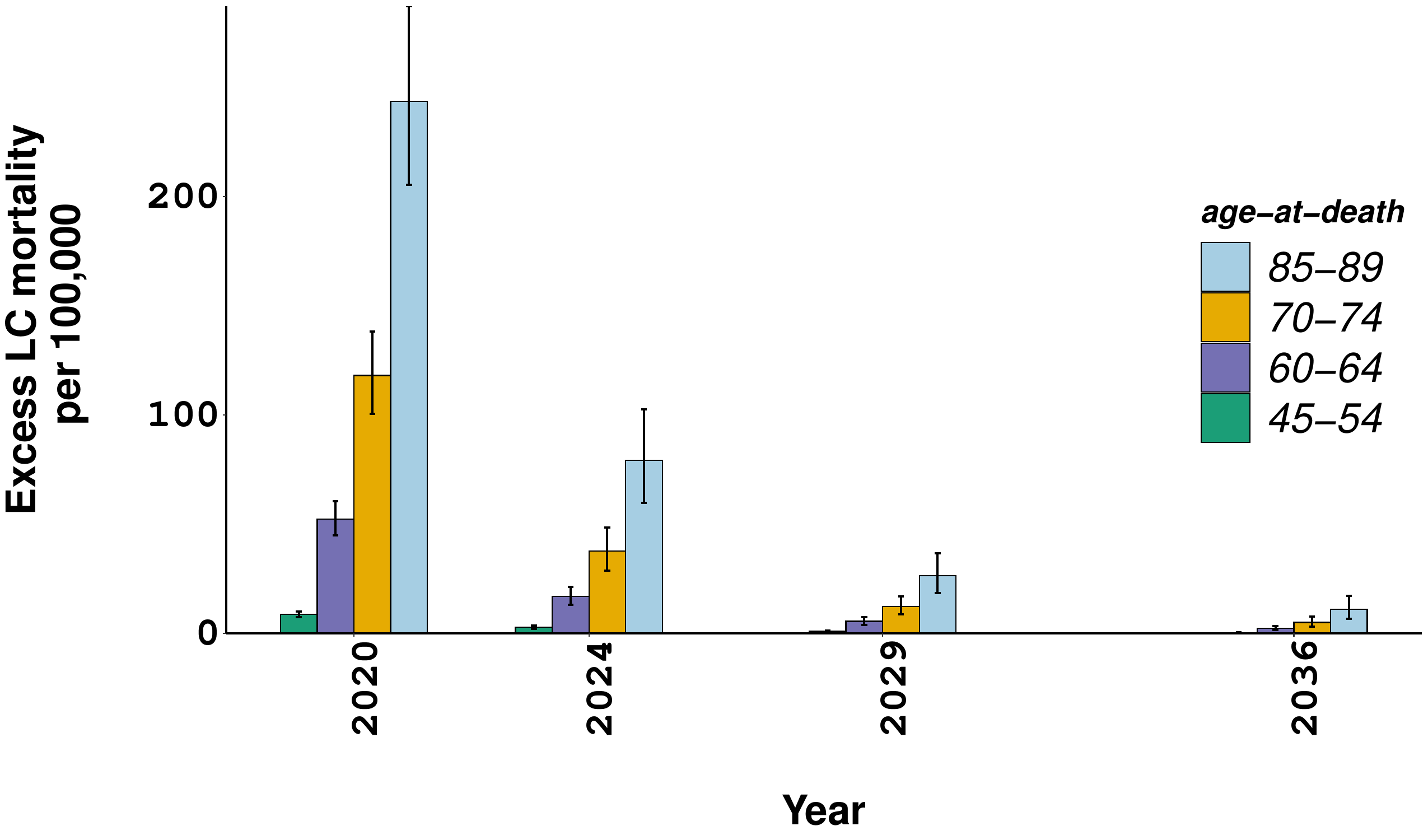}}
	\caption{Lung cancer excess mortality, per 100,000 men, $\text{EAM}^{\text{lung}}_{a, \text{men},t}$, by age-at-death in England from 2020 to 2036 based on \eqref{eq:MaleLungCODLocationPrmtr2}. Note that differences in lung cancer excess mortality by age in intermediate years are negligible. } 	
	\label{fig:ExcessRateMaleAgeSmok}
\end{figure}

%\vspace{-0.9cm} % Adjust the value to reduce the space

\begin{figure}[H]
	\centering
	\subfloat[1-month delay \label{fig:ExcessRate1monthMenRegSmok}]{\includegraphics[width=0.5\textwidth]{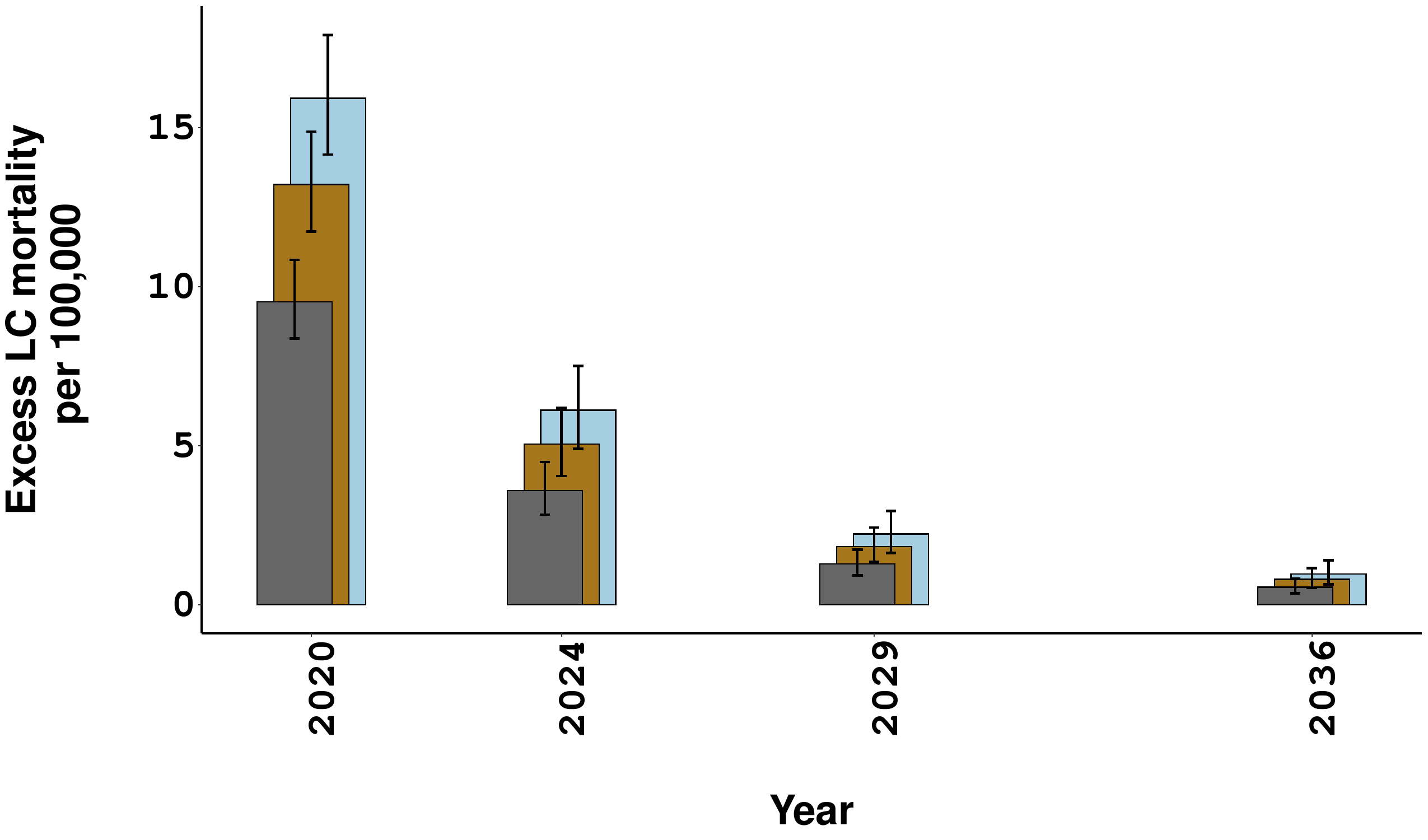}}
	\hfill
	\subfloat[ 6-month delay \label{fig:ExcessRate6monthMenReg2Smok}]{\includegraphics[width=0.5\textwidth]{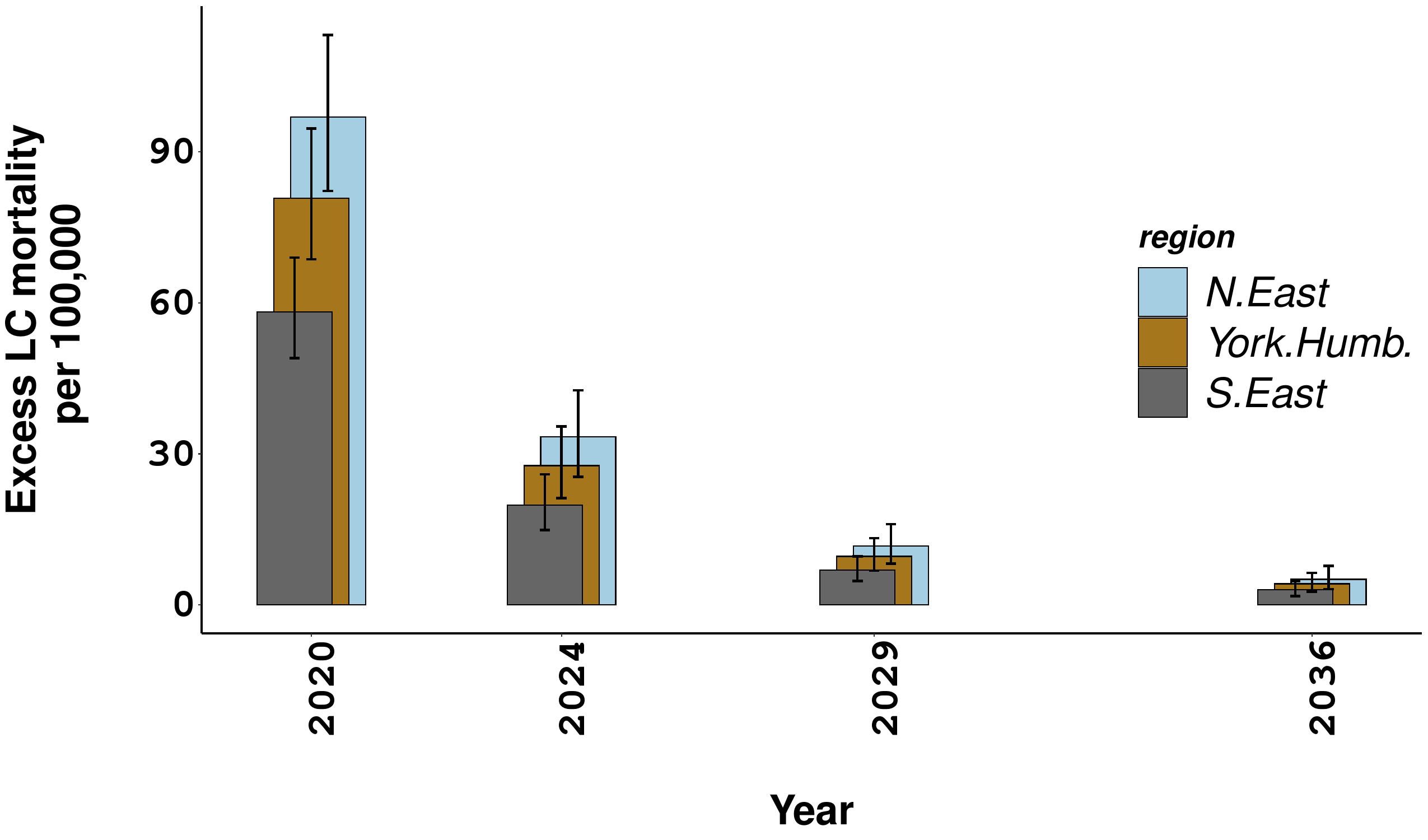}}
	\caption{Lung cancer excess mortality, per 100,000 men, $\text{ERM}^{\text{lung}}_{\text{men},r, t}$, by selected regions of England from 2020 to 2036, with 95\% credible intervals, based on \eqref{eq:MaleLungCODLocationPrmtr2}. Note that differences in lung cancer excess mortality in other regions in a given year, and differences in intermediate years, are negligible. } 	
	\label{fig:ExcessRateMenRegionSmok}
\end{figure}

\begin{figure}[H]
	\centering
	\subfloat[1-month delay \label{fig:ExcessRate1monthMenDepSmok}]{\includegraphics[width=0.5\textwidth]{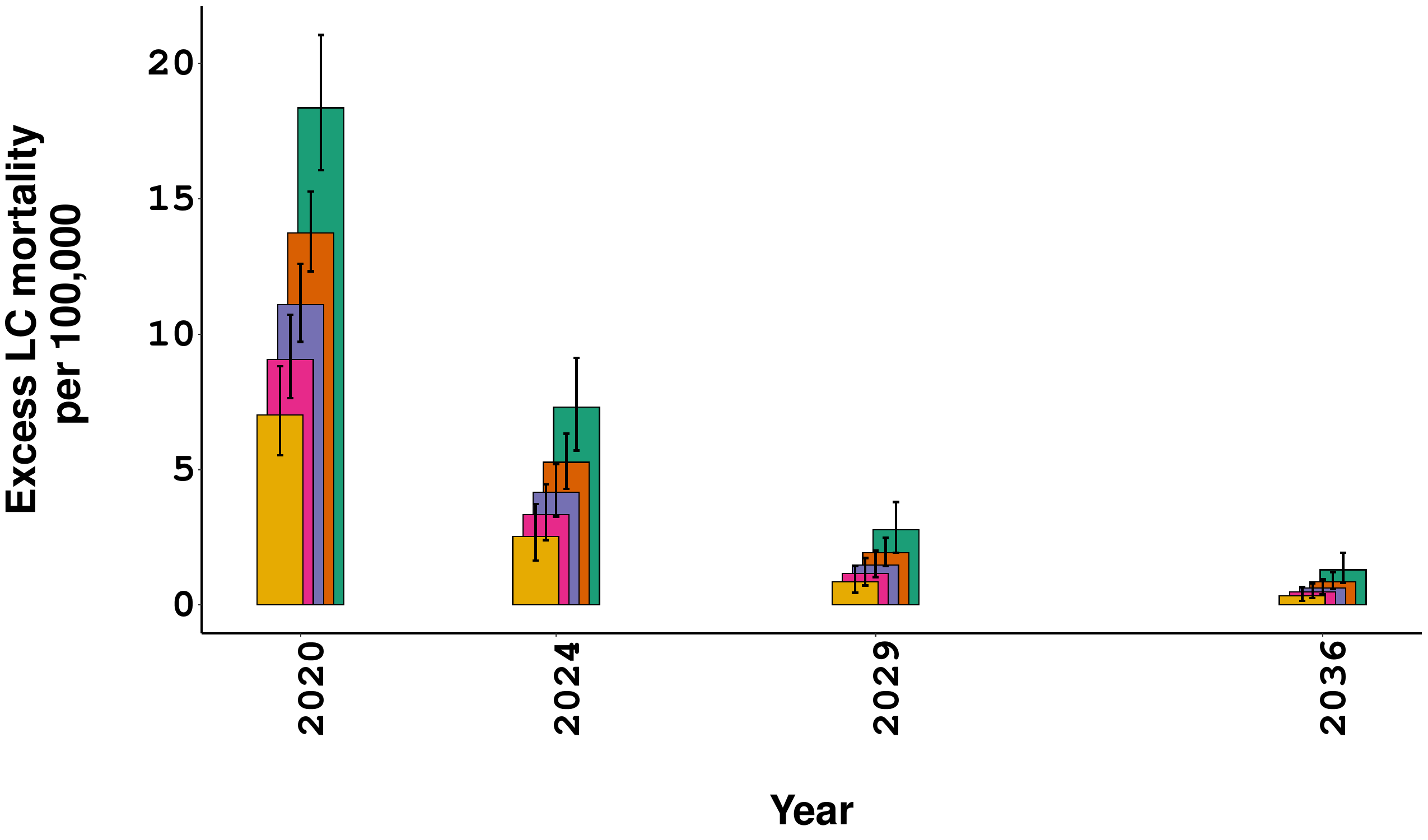}}
	\hfill
	\subfloat[ 6-month delay \label{fig:ExcessRate6monthMenDep2Smok}]{\includegraphics[width=0.5\textwidth]{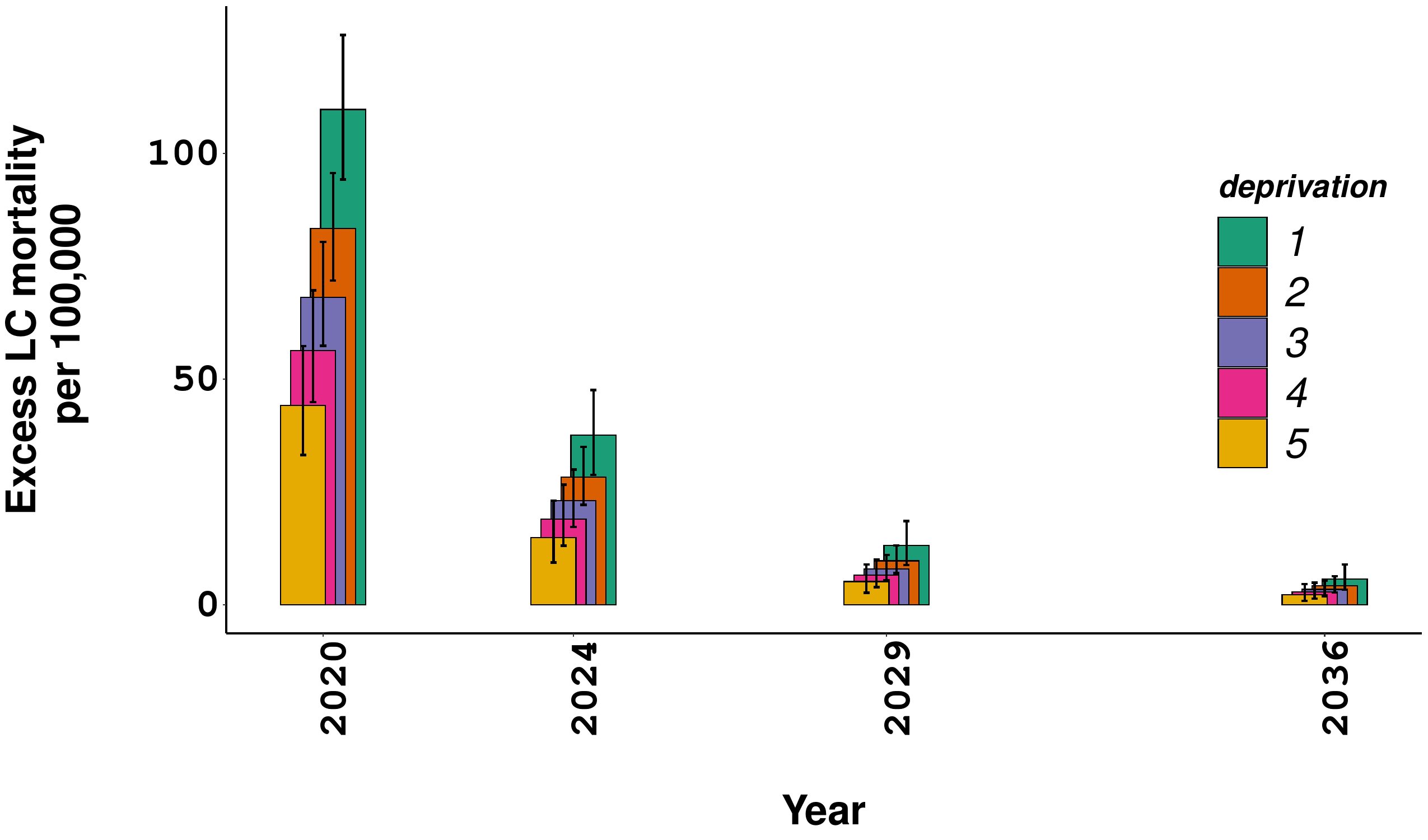}}
	\caption{Lung cancer excess mortality, per 100,000 men, $\text{EDM}^{\text{lung}}_{\text{men},d, t}$, by deprivation quintiles in England from 2020 to 2036, with 95\% credible intervals, based on \eqref{eq:MaleLungCODLocationPrmtr2}. Note that differences in lung cancer excess mortality by deprivation in intermediate years are negligible.} 	
	\label{fig:ExcessRateMenDepSmok}
\end{figure}

\end{document}